\documentclass[12pt,preprint,aps]{revtex4}
\usepackage{amsmath}
\usepackage{amsfonts}
\usepackage{amssymb}
\usepackage{graphicx}
\begin{document}

\preprint{hep-ph/0104040} 
\preprint{IFIC/01-16}

\title{Broken R-parity, stop decays, and neutrino physics}
\author{D.~Restrepo} 
\author{W.~Porod} 
\author{J.~W.~F.~Valle} 
\affiliation{Inst.~de F{\'\i}sica Corpuscular (IFIC), CSIC - U. de Val{\`e}ncia, \\ 
Edificio Institutos de Paterna, Apartado de Correos  22085\\ 
E-46071--Val{\`e}ncia, Spain \\
}

\begin{abstract}
  
  We discuss the phenomenology of the lightest stop in models where
  R-parity is broken by bilinear superpotential terms. In this class
  of models we consider scenarios where the R-parity breaking two-body
  decay ${\tilde t}_1 \to \tau^+ \, b $ competes with the leading
  three-body decays such as ${\tilde t}_1 \to W^+ \, b \, {\tilde
    \chi}^0_1$.  We demonstrate that the R--parity violating decay can
  be sizable and in some parts of the parameter space even the
  dominant one.  Moreover we discuss the expectations for ${\tilde
    t}_1 \to \mu^+ \, b $ and ${\tilde t}_1 \to e^+ \, b $. The recent
  results from solar and atmospheric neutrinos suggest that these are
  as important as the $\tau^+ b$ mode.  The ${\tilde t}_1 \to l^+ \,
  b$ decays are of particular interest for hadron colliders, as they
  may allow a full mass reconstruction of the lighter stop.  Moreover
  these decay modes allow cross checks on the neutrino mixing angle
  involved in the solar neutrino puzzle complementary to those
  possible using neutralino decays. For the so--called small mixing
  angle or SMA solution ${\tilde t}_1 \to e^+ \, b$ should be
  negligible, while for the large mixing angle type solutions
  \texttt{all} ${\tilde t}_1 \to l^+ \, b$ decays should have
  comparable magnitude.

\end{abstract}
\maketitle
\section{Introduction}
\label{sec:introduction}

The search for supersymmetry (SUSY) \cite{Nilles:1984ge,Haber:1985rc}
plays an important r{\^o}le in the experimental program of present and
at future colliders, e.g.~the Tevatron, LHC, or an $e^+ e^-$ linear
collider.  Therefore many phenomenological studies have been carried
out in recent years (see
e.g.~\cite{phen1,Carena:1998uy,Accomando:1998wt,Allanach:1999bf} and
references therein) focusing mainly on the minimal supersymmetric
standard model (MSSM) \cite{Tata:1995zj}.  However, neither gauge
invariance nor supersymmetry require the conservation of R-parity.
Indeed, there is considerable theoretical and phenomenological
interest in studying possible implications of alternative
scenarios~\cite{beyond} in which R-parity is
broken~\cite{Aulakh:1982yn,Hall:1984id,rpold,arca}.  These theories
are of particular interest as they lead to a pattern of neutrino
masses and mixing angles~\cite{Romao:2000up} which can account for the
observed anomalies in solar and atmospheric neutrinos
~\cite{Gonzalez-Garcia:2001sq}.  In general the violation of R-parity
could arise explicitly~\cite{expl} as a residual effect of some larger
unified theory~\cite{Hall:1984id}, or spontaneously, through nonzero
vacuum expectation values (vev's) for scalar
neutrinos~\cite{Aulakh:1982yn,rpold,arca,sponsearch}.  In realistic
spontaneous R-parity breaking models there is an $SU(2) \otimes U(1) $
singlet sneutrino vacuum expectation value (vev) characterizing the
scale of R-parity violation \cite{MASIpot3,MASI,ROMA,ZR} which is
expected to be in the order of 1~TeV.

There are two generic cases of spontaneous R-parity breaking models to
consider.
In the absence of any additional gauge symmetry, these models lead to
the existence of a physical massless Nambu-Goldstone boson, called
majoron (J) which is {\sl the lightest SUSY particle}, massless and
therefore stable. 
As in the standard case in R-parity breaking models the lightest SUSY
particle (LSP) is in general a neutralino. However, it now decays
mostly into visible states, therefore diluting the missing momentum
signal and bringing in increased multiplicity events which arise
mainly from three-body decays such as $ \tilde{\chi}^{0}_{1} \to f
\bar{f} \nu$, where $f$ denotes a charged fermion~\footnote{Note that
  some of the neutralino decay modes such as $ \tilde{\chi}^{0}_{1}
  \to 3 \nu$ and $ \tilde{\chi}^{0}_{1} \to \nu J$ are experimentally
  ``invisible'' and maintain the missing momentum signal unchanged.
This last decay conserves R-parity since the majoron has a large R-odd
singlet sneutrino component.}.
If lepton number is part of the gauge symmetry and R-parity is
spontaneously broken then there is an additional gauge boson which
gets mass via the Higgs mechanism, and there is no physical Goldstone
boson~\cite{ZR}.  In this case R-parity violating effects relevant for
collider physics are conveniently parameterized by adding bilinear
terms $\epsilon_i L_i H_2$ to the MSSM superpotential and
corresponding terms for the soft SUSY breaking part of the Lagrangian.
Bilinear R-parity violation may also be assumed ab initio as the
fundamental theory. For example, it may be the only violation
permitted by higher Abelian flavour symmetries\cite{Mira:2000gg}.

Owing to the large top Yukawa coupling the stops have a quite
different phenomenology compared to those of the first two generations
of up--type squarks (see e.g.~\cite{Bartl97a,Bartl:1997wt} and
references therein).  The large Yukawa coupling implies a large mixing
between ${\tilde t}_L$ and ${\tilde t}_R$ \cite{Ellis83} and large
couplings to the higgsino components of neutralinos and charginos.
The large top quark mass also implies the existence of scenarios where
all MSSM two-body decay modes of ${\tilde t}_1$ are kinematically
forbidden at the tree-level (e.g. ${\tilde t}_1 \to t \, {\tilde
  \chi}^0_i, b \, {\tilde \chi}^+_j, t \, \tilde g$). In such case
higher order decays of ${\tilde t}_1$ become relevant
\cite{Hikasa:1987db,Porod:1997at,Porod:1999yp,Djouadi:2000bx}:
${\tilde t}_1 \to c \, {\tilde \chi}^0_{1,2}$, 
${\tilde t}_1 \to W^+ \, b \, {\tilde \chi}^0_1$,
${\tilde t}_1 \to H^+ \, b \, {\tilde \chi}^0_1$,
${\tilde t}_1 \to b \, {\tilde l}^+_i \, \nu_l$,
${\tilde t}_1 \to b \, {\tilde \nu}_l \, l^+$, 
%
where $l$ denotes $e,\mu,\tau$. Also 4-body decays may become
important if the 3-body decays are kinematically forbidden
\cite{Boehm:2000tr}.  In
\cite{Porod:1997at,Porod:1999yp,Bartl:2000kw,Djouadi:2000bx} it has
been shown that in the MSSM the three-body decay modes are in general
much more important than the two body decay mode.  Recently it has
been demonstrated that not only LSP decays but also the light stop can
be a good candidate for observing R-parity violation, even if its
magnitude is as small as indicated by the solutions to the present
neutrino anomalies
\cite{Allanach:1999bf,Bartl:1996gz,Diaz:1999ge,Datta:2000yc}.  It has
been demonstrated that there exists a large parameter region where the
R-parity violating decay
\begin{equation*}
{\tilde t}_1 \to b \, \tau
\end{equation*}
is much more important than the $\not\!\!R_p$ conserving decays
\begin{equation*}
{\tilde t}_1 \to c \, {\tilde \chi}^0_{1,2} 
\end{equation*}
in scenarios where two-body decay modes are possible.  It is therefore
natural to ask if there exist scenarios where the decay ${\tilde t}_1
\to b \, \tau$ is as important as the three--body decays.  Note that
in the R-parity violating models under consideration the neutral
(charged) Higgs--bosons mix with the neutral (charged) sleptons
\cite{deCampos:1995av,Akeroyd:1998iq}.  These states are denoted by
$S^0_i$, $P^0_j$, and $S^\pm_k$ for the neutral scalars, pseudoscalars
and charged scalars, respectively.  Therefore in the R-parity
violating case one has the following three-body decay modes:
\begin{eqnarray}
{\tilde t}_1 &\to& W^+ \, b \, {\tilde \chi}^0_1 \nonumber\\
{\tilde t}_1 &\to& S^+_k \, b \, {\tilde \chi}^0_1 \nonumber\\
{\tilde t}_1 &\to& S^+_k \, b \, \nu_3 \nonumber\\
{\tilde t}_1 &\to& b \, S^0_i \, \tau^+ \,,\nonumber  \\
{\tilde t}_1 &\to& b \, P^0_j \, \tau^+ \, \nonumber\\
{\tilde t}_1 &\to& b \, {\tilde l}^+_i \, \nu_l \,, \nonumber\\
{\tilde t}_1 &\to& b \, {\tilde \nu}_l \, l^+ \hspace{1cm} (l=e,\mu) \, .
\nonumber
\end{eqnarray}
We will show that there exist regions in parameter space where
${\tilde t}_1 \to b \, \tau^+$ is sizeable and even the most important
decay mode.  In particular we will consider a mass range for the light
stop ${\tilde t}_1$, where it is difficult for the LHC to discover it
in the MSSM due to the large top background \cite{ulrike}.  In
contrast to the existing LSP decay studies in $\not\!\!R_p$
models~\cite{Bartl2000rp,Porod:2000hv} which are mainly sensitive to
the atmospheric neutrino anomaly parameters, the stop decay processes
considered here are very sensitive to the solar neutrino parameters
and therefore gives valuable complementary information.

The paper is organized in the following way: in the next section we
will introduce the model.  In Sect.~\ref{sec:numerical-analisis}
numerical results for stop decays are presented. We first explore the
extent to which the decay $\tilde t_1\to b\,\tau$ can be sizeable when
compared with the 3--body decay modes.  Moreover, we discuss the
connections between the decay modes ${\tilde t}_1 \to b \, l^+$ and
neutrino physics, in particular we discuss a possible test of the
solution to the solar neutrino puzzle. In Sect.~\ref{sec:con} we
present our conclusions. The appendixes contain complete formulas for
the total widths of the three-body decay modes as well as for the
couplings.

\section{The model}
\label{sec:themodel}

The supersymmetric Lagrangian is specified by the superpotential $W$
given by
\begin{equation*}  
W=\varepsilon_{ab}\left[ 
 h_U^{ij}\widehat Q_i^a\widehat U_j\widehat H_2^b 
+h_D^{ij}\widehat Q_i^b\widehat D_j\widehat H_1^a 
+h_E^{ij}\widehat L_i^b\widehat R_j\widehat H_1^a 
-\mu\widehat H_1^a\widehat H_2^b 
\right] + \varepsilon_{ab}\epsilon_i\widehat L_i^a\widehat H_2^b\,,
\end{equation*}
where $i,j=1,2,3$ are generation indices, $a,b=1,2$ are $SU(2)$
indices, and $\varepsilon$ is a completely antisymmetric $2\times2$
matrix, with $\varepsilon_{12}=1$. The symbol ``hat'' over each letter
indicates a superfield, with $\widehat Q_i$, $\widehat L_i$, $\widehat
H_1$, and $\widehat H_2$ being $SU(2)$ doublets with hypercharges
$1/3$, $-1$, $-1$, and $1$ respectively, and $\widehat U$, $\widehat
D$, and $\widehat R$ being $SU(2)$ singlets with hypercharges
$-\frac43$, $\frac23$, and $2$ respectively. The couplings $h_U$,
$h_D$ and $h_E$ are $3\times 3$ Yukawa matrices, and $\mu$ and
$\epsilon_i$ are parameters with units of mass.
 
Supersymmetry breaking is parameterized by the standard set of soft
supersymmetry breaking terms 
\begin{eqnarray} 
V_{soft}&=& 
M_Q^{ij2}\widetilde Q^{a*}_i\widetilde Q^a_j+M_U^{ij2} 
\widetilde U^*_i\widetilde U_j+M_D^{ij2}\widetilde D^*_i 
\widetilde D_j+M_L^{ij2}\widetilde L^{a*}_i\widetilde L^a_j+ 
M_R^{ij2}\widetilde R^*_i\widetilde R_j \nonumber\\ 
&&\!\!\!\!+m_{H_1}^2 H^{a*}_1 H^a_1+m_{H_2}^2 H^{a*}_2 H^a_2\nonumber\\
&&\!\!\!\!- \left[{\textstyle\frac12} M_3\lambda_3\lambda_3+{\textstyle\frac12} M\lambda_2\lambda_2 
+{\textstyle\frac12} M'\lambda_1\lambda_1+h.c.\right] 
\nonumber\\ 
&&\!\!\!\!+\varepsilon_{ab}\left[ 
A_U^{ij}h_U^{ij}\widetilde Q_i^a\widetilde U_j H_2^b 
+A_D^{ij}h_D^{ij}\widetilde Q_i^b\widetilde D_j H_1^a 
+A_E^{ij}h_E^{ij}\widetilde L_i^b\widetilde R_j H_1^a\right.
\nonumber\\ 
&&\!\!\!\!\left.-B\mu H_1^a H_2^b+B_i\epsilon_i\widetilde L_i^a H_2^b\right] 
\,,\label{eq:1}
\end{eqnarray} 

Note that, in the presence of soft supersymmetry breaking terms the
bilinear terms proportional to the $\epsilon_i$ can not be rotated
away except for the very special case $B_i = B$ and only if the scalar
masses are adjusted in a special way.  Such 'fine-tuned' assumptions
at a low scale are, from our point of view, very unnatural. If
realized at a high scale such as the unification or GUT-scale, then
the trilinear R-parity breaking couplings introduced as a result of
the rotation would re-introduce bilinear terms at the electroweak scale
due to the structure of the corresponding RGEs \cite{rbvRGE}. In
contrast, bilinear terms are closed under RGE evolution from the high
scale to the electroweak scale~\cite{Diaz:2000is}.

In order to compare the $\tilde t_1\to b\, \tau$ decay mode with the
3-body MSSM modes it is sufficient for us to consider a 1-generation
$\not\!\!R_p$ model. Note however, that for the detailed connection of stop
decays with neutrino physics we must consider the complete model, and
we will do so in a second step when we discuss the connection with the
solar neutrino mixing.  
Finally, notice that we also allow for R-parity-conserving Flavour
Changing Neutral Currents (FCNC) effects, such as the process $\tilde
t_1\to c\,\tilde \chi_1^0$ involving the three generations of quarks
in both cases.  

Our 1-generation model is specified by the superpotential
\cite{BRpV,moreBRpV,otros,Ferrandis:1999ii}
\begin{equation} 
W=h_t\widehat Q_3\widehat U_3\widehat H_2
 +h_b\widehat Q_3\widehat D_3\widehat H_1
 +h_{\tau}\widehat L_3\widehat R_3\widehat H_1
 -\mu\widehat H_1\widehat H_2
 +\epsilon_3\widehat L_3\widehat H_2
\label{eq:2}
\end{equation}

For simplicity, in the remaining part of this section we adopt the
1-generation model when presenting the formulas for the mass matrices
which are needed in the subsequent sections. The complete formulas for
the 3-generation case are given in the second paper of
\cite{Romao:2000up} and have been used whenever required.
The electroweak symmetry is broken when the two Higgs doublets, $H_1$
and $H_2$, and left slepton doublet $\widetilde L_3$ acquire vacuum
expectation values. We introduce the notation:
\begin{equation*} 
H_1={{\frac1{\sqrt{2}}[\theta^0_1+v_1+i\varphi^0_1]}\choose{ 
H^-_1}}\,, \,\,
H_2={{H^+_2}\choose{\frac1{\sqrt{2}}[\theta^0_2+v_2+ i\varphi^0_2]}}\,, 
\widetilde L_3={{\frac1{\sqrt{2}} 
[\tilde\nu^R_{\tau}+v_3+i\tilde\nu^I_{\tau}]}\choose{\tilde\tau^-}}\,. 
\end{equation*} 
The mass of $W$ is given by $m_W^2={\textstyle\frac14} g^2v^2$, where
$v^2\equiv v_1^2+v_2^2+v_3^2\simeq(246 \; \rm{GeV})^2$. We define
$\tan\beta=v_2/\sqrt(v_d^2+v_3^2)$.
In addition to the above MSSM parameters, our model contains three new
parameters, $\epsilon_3$, $v_3$ and $B_3$, of which only two are
independent, because there is an additional tad-pole equation
\cite{BRpV}.  These may be chosen as $\epsilon_3$ and $v_3$.

The stop mass matrix is given by
\begin{equation*}
  \mathbf{M_{\tilde t}}^2=\left[ \begin{array}{cc}
      {M_Q^2}+\frac12v_2^2 {h_t}^2+\Delta_{UL}&
      \frac{h_t}{\sqrt2}
      \left( v_2{A_t}- \mu v_1 +\epsilon_3 v_3 \right) \\
      \frac{h_t}{\sqrt2}
      \left( v_2{A_t}- \mu v_1 +\epsilon_3 v_3 \right) &
      {M_U^2}+\frac12 v_2^2{h_t}^2+\Delta_{UR}
    \end{array} \right]
\end{equation*}
with $\Delta_{UL}=\frac18\big(g^2-\frac13{g'}^2\big)\big(v_1^2-v_2^2
+v_3^2\big)$ and $\Delta_{UR}=\frac16 {g'}^2(v_1^2-v_2^2+v_3^2)$.
The mass matrix for the sbottoms is given by
\begin{equation*}
  \mathbf{M_{\tilde b}}^2=
  \left[
    \begin{array}{cc}
      {M_Q^2}+\frac12v_1^2{h_b^2}+\Delta_{DL}&
      \frac{h_b}{\sqrt2}(v_1{A_D}-\mu v_2)\\
      \frac{h_b}{\sqrt2}(v_1{A_D} -\mu v_2)  &
      {M_D^2}+\frac12v_1^2{h_b^2}+\Delta_{DR}
    \end{array}
  \right]
\end{equation*}
where $\Delta_{DL}=-\frac18\big(g^2+\frac13{g'}^2\big)\big(v_1^2-v_2^2
+v_3^2\big)$, $\Delta_{DR}=-\frac1{12}
{g'}^2(v_1^2-v_2^2+v_3^2)$.
The  mass eigenstates are obtained by ($q=t,b$): 
\begin{equation*}
  \left[
    \begin{array}{c}
      \tilde q_1\\
      \tilde q_2
    \end{array}
  \right]=
  \left[
    \begin{array}{cc}
      \cos\theta_{\tilde q} & \sin\theta_{\tilde q}\\
      -\sin\theta_{\tilde q} & \cos\theta_{\tilde q}\\
    \end{array}
  \right] 
  \left[
    \begin{array}{c}
      \tilde q_L\\
      \tilde q_R
    \end{array}
  \right]=
  \mathcal{R}^{\tilde f}
  \left[
    \begin{array}{c}
      \tilde q_L\\
      \tilde q_R
    \end{array}
  \right]
\end{equation*}
with
\begin{equation*}
  \cos \theta_{\widetilde q} = \frac{- M^2_{\widetilde q_{12}}}{\sqrt{(M^2_{\widetilde q_{11}} -
      m^2_{\widetilde q_1})^2 + (M^2_{\widetilde q_{12}})^2}}, \qquad 
  \sin \theta_{\widetilde q} = \frac{M^2_{\widetilde q_{11}} - m^2_{\widetilde q_1}}
  {\sqrt{(M^2_{\widetilde q_{11}} - m^2_{\widetilde q_1})^2 + (M^2_{\widetilde q_{12}})^2}} \, .
\end{equation*}

The bilinear term in Eq.~(\ref{eq:2}) leads to a mixing between
the charginos and the $\tau$--lepton which induces the decay
${\tilde t}_1 \to b \, \tau$. The mass matrix is given by
\begin{equation*}
\mathbf{M_C}=
\begin{bmatrix}
  M & {\textstyle\frac1{\sqrt{2}}}gv_2 & 0 \\
  {\textstyle{\frac1{\sqrt{2}}}}gv_1 & \mu & 
  -{\textstyle{\frac1{\sqrt{2}}}}h_{\tau}v_3 \\
  {\textstyle{\frac1{\sqrt{2}}}}gv_3 & -\epsilon_3 &
  {\textstyle{\frac1{\sqrt{2}}}}h_{\tau}v_1
\end{bmatrix}
\end{equation*}
As in the MSSM, the chargino mass matrix is diagonalized by two
rotation matrices $\mathbf{U}$ and $\mathbf{V}$
\begin{equation*}
  \mathbf{U}^*\mathbf{M_C}\mathbf{V}^{-1}=\left[
    \begin{array}{ccc}
      m_{\tilde\chi^{\pm}_1} & 0 & 0 \\
      0 & m_{\tilde\chi^{\pm}_2} & 0 \\
      0 & 0 & m_{\tilde\chi^{\pm}_3}
    \end{array}
  \right]\,.
\end{equation*}
The lightest eigenstate of this mass matrix must be the tau lepton
($\tau^{\pm}=\tilde\chi^\pm_3$) and so the mass is constrained to be
$1.77703^{+0.30}_{-0.26}\,$ GeV\cite{Groom:2000in}~\footnote{Strictly
  speaking one must take into account that the tau Yukawa coupling
  becomes a function of the parameters in the mass matrix, as given
  in~\cite{Akeroyd:1998iq}. In practice the corrections are negligible
  for the case of interest.}

In our model, the one of the three neutrinos acquires mass at the tree
level due to a mixing between the neutralino sector and one of the
neutrinos~\cite{rpold,arca,Romao:1992ex}. The neutralino/neutrino mass
matrix is
\begin{equation*} 
  \mathbf{M_N}=
  \left[  
    \begin{array}{ccccc}  
      M^{\prime } & 0 & -\frac 12g^{\prime }v_1 & 
      \frac 12g^{\prime }v_2 & -\frac  
      12g^{\prime }v_3 \\   
      0 & M & \frac 12gv_1 & -\frac 12gv_2 & \frac 12gv_3 \\   
      -\frac 12g^{\prime }v_1 & \frac 12gv_1 & 0 & -\mu  & 0 \\   
      \frac 12g^{\prime }v_2 & -\frac 12gv_2 & -\mu  & 0 & \epsilon _3 \\   
      -\frac 12g^{\prime }v_3 & \frac 12gv_3 & 0 & \epsilon _3 & 0  
    \end{array}  
  \right] 
\end{equation*} 
and $M'$ is the $U(1)$ gaugino soft mass. This neutralino/neutrino
mass matrix is diagonalized by a $5\times 5$ rotation matrix
$\mathbf{N'}$ such that
\begin{equation*} 
  \mathbf{N'}^*\mathbf{M_N}\mathbf{N'}^{-1}=\mathrm{diag}(m_{\tilde\chi^0_1},
  m_{\tilde\chi^0_2}, m_{\tilde\chi^0_3},m_{\tilde\chi^0_4},m_{\tilde\chi^0_5})
\end{equation*} 
with $m_{\nu_3}=m_{\tilde\chi^0_5}$. Note that in
\cite{Porod:1997at,Porod:1999yp} a different basis for the neutralinos
was used.
Assuming small R-parity violating couplings the tree-level neutrino
mass is approximately given by
\begin{equation}
m_{\nu_3} \approx 
-\frac{(g^2M'+{g'}^2M){\mu'}^2}{
4 M M'{\mu'}^2-2(g^2M'+{g'}^2M){\mu'}v_2{v_1'} \cos\xi}{v_1'}^2\sin^2\xi
\label{eq:3}
\end{equation}
with
\begin{equation*}
\sin\xi = \frac{\epsilon_3 v_1 +\mu  v_3 }
               {\mu'v'_1} \hspace{1cm},
  \mu' = \sqrt{\mu^2+\epsilon_3^2} \, , \hspace{2cm}
  v'_1 = \sqrt{v_1^2+v_3^2} \, .
\end{equation*}
Notice that in mSUGRA models with bilinear R--parity
violation~\cite{Diaz:1999ge} $m_{\nu_3}$ is calculable through the RGE
evolution and one finds in this case cancellations up to two orders of
magnitude for the combination $\Lambda_3=\epsilon_3 v_1 +\mu v_3$. In
general models the smallness of $m_{\nu_3}$ requires relatively small
$\epsilon_3$ values from the start as might arise, for example, in the
models considered in ref.~\cite{Mira:2000gg}.
The remaining two neutrinos acquire mass radiatively.
Rigorous quantitative results were given in the second paper in
ref.~\cite{Romao:2000up}. Typically they are hierarchically lighter
than the heaviest neutrino, whose mass arises at the tree-level.  This
way one accounts for the observed hierarchy between the solar and the
atmospheric neutrino mass scales.
 
Similarly, the Higgs bosons mix with charged sleptons and the real
(imaginary) parts of the sneutrino mix with the scalar (pseudoscalar)
Higgs bosons.  We denote the scalar bosons by $S^0_i$, the
pseudo-scalar by $P^0_i$ and the charged bosons by $S^\pm_i$.  The
relevant formulas have all been given e.g.~in
\cite{Akeroyd:1998iq,Bartl2000rp}.
However, since we will take one of the pseudoscalar mass eigenvalues as
input, it is worth briefly repeating here the discussion of the
pseudoscalar bosons masses.  The pseudo-scalar mass matrix is given by:
\begin{equation} 
  \mathbf{ M^2_{P^0}}=
  \left[
    \begin{array}{ccc}
      B\mu{\frac{v_2}{v_1}}+\mu\epsilon_3{\frac{v_3}{v_1}}
      & B\mu & -\mu\epsilon_3 \\ 
      B\mu & B\mu{\frac{v_1}{v_2}}-B_3\epsilon_3{\frac{v_3}{v_2}}
      & -B_3\epsilon_3 \\ 
      -\mu\epsilon_3 & -B_3\epsilon_3 &  
      \mu\epsilon_3{\frac{v_1}{v_3}}-B_3\epsilon_3{\frac{v_2}{v_3}} 
    \end{array}
  \right] \, .
\label{eq:4} 
\end{equation} 
As expected, this matrix has zero determinant, since the neutral
Goldstone boson eaten by the $Z$ is one of the corresponding states.
Therefore, the masses of the two physical states are given by the
formula:
\begin{equation} 
  m_{2,3}={\frac12}{\rm Tr}\mathbf{ M} 
  \pm {\frac12}\sqrt{\left({\rm Tr}\mathbf{ M}\right)^2 
    -4(M_{11}M_{22}-M_{12}^2+M_{11}M_{33}-M_{13}^2+M_{22}M_{33}-M_{23}^2)}.
  \label{eq:5}
\end{equation}
Therefore we can easily take one these masses as input and calculate
$B\mu$ from it using
\begin{equation*}
  B\mu=\frac{-m_{P^0_2}^4v_1v_2v_3+B_3\epsilon_3\mu v_3(v_1^2+v_2^2+v_3^2)+
    \epsilon_3 m_{P^0_2}^2[\mu v_2(v_1^2+v_3^2)]-B_3v_1(v_2^2+v_3^2)}
  {-m_{P^0_2}^2(v_1^2+v_2^2)v_3+\epsilon_{3}(\mu v_1-B_3v_2)
    (v_1^2+v_2^2+v_3^2) }   
\end{equation*}
$B_3$ is obtained from the minimum equation for given $\epsilon_3$ and
$v_3$~\cite{Akeroyd:1998iq}. 

\section{Numerical Analyses}
\label{sec:numerical-analisis}

In this section we present our numerical results for the branching
ratios of the lighter stop $\tilde t_1$. Here we consider scenarios
where all two-body decays induced at tree-level are kinematically
forbidden except the $b \, l^+$ decays.  Before going into detail it
is useful to have some approximate formulas at
hand~\cite{Diaz:1999ge}:
\begin{align}
  \Gamma(\tilde t_1\to b\,\tau) \approx\,&
  \frac{g^2 |U_{32}|^2 h_b^2 \cos^2_{\theta_{\tilde t}} m_{\tilde t_1}}{16\pi} 
  \approx
  \frac{g^2 |\epsilon_3|^2 h_b^2 \cos^2_{\theta_{\tilde t}} m_{\tilde t_1}}{16\pi |\mu|^2} 
  \label{eq:6} \\
  \Gamma(\tilde t_1 \to c\tilde\chi_1^0)\approx\,& F h_b^4(
  \delta_{m_0^2}\cos\theta_{\tilde t}-\delta_A\sin\theta_{\tilde t})^2
  f_L^2 
   m_{\tilde t_1} 
  \left(
    1-\frac{m_{\tilde \chi_1^0}^2}{m_{\tilde t_1}^2}
  \right)^2\,
\label{eq:7} ,
\end{align}
where $F = \frac{g^2}{16\pi}\left(
 \log(m_{GUT}/m_Z) K_{cb}K_{tb}/16 \pi^2 \right)^2
 \sim 6 \times 10^{-7}$,
$f_L = \sqrt2(\tan\theta_W\,N_{11}+3N_{12}) / 6$ and
the parameter $\delta_{m_0^2}$ is given by
\begin{equation*}
\delta_{m_0^2}=\frac{M_Q^2+M_D^2+m_{H_1}^2+A_b^2}{m_{\tilde c_L}^2-m_{\tilde
t_L}^2}
\end{equation*}
For the minimal SUGRA models one finds $\delta_{m_0^2}={\cal O}(1)$ which is
basically independent of the initial conditions due to the $m_0$
dependence both in the numerator and in the denominator. Finally we
have
\begin{equation*}
\delta_A=\frac{m_t(A_b+\frac12A_t)}{m_{\tilde c_L}^2-m_{\tilde t_L}^2}
\end{equation*}
The complete formulas are given in \cite{Bartl:1996gz,Diaz:1999ge},
while for the three--body decays they are given in
Appendix~\ref{sec:appA}. They reduce to the ones given
in~\cite{Porod:1997at,Porod:1999yp} for vanishing R-parity violation
parameters.

We have fixed the parameters as in \cite{Porod:1999yp} to avoid colour
breaking minima, while in the top squark sector we have used
$m_{\tilde t_1}$, $\cos \theta_{\tilde t}$, $\tan \beta$, and $\mu$ as
input parameters.  For the sbottom sector we have fixed $M_{\tilde Q},
M_{\tilde D}$ and $A_b$ as input parameters whereas for the charged
scalars we took $m_{P^0_2}$, $M_{\widetilde E}, M_{\tilde L}$, and
$A_\tau$ as input~\footnote{ $m_{P^0_2}$ plays here the same role as
  $m_{A^0}$ in the MSSM case \cite{Porod:1999yp}.}.  In addition we
have chosen the R-parity violating parameters $\epsilon_3$ and $v_3$
in such a way that the heaviest neutrino mass is fixed with the help
of Eq.~(\ref{eq:3}).
For simplicity, we have also assumed that the soft SUSY breaking
parameters are equal for all generations.

In order to get a feeling for minimal size of branching ratios that
can be measured let us first shortly discuss the expected size for the
direct production of light stops at future colliders.
One expects for example at the LHC a production cross section of $\sim
35$~pb for 220 GeV stop mass.  Therefore, once the full luminosity has
been reached, one has to expect approximately 3.5 $10^6$ events per
year.
The corresponding stop production cross section at a future $e^+ e^-$
linear collider of $800$ c.m.s.~energy is of ${\cal O}(10-100 \mathrm{fb})$
\cite{Bartl:1997wt}.  For an integrated luminosity of 500 fb$^{-1}$
per year one can expect ${\cal O}(10^4)$ stop pairs per year.  This implies
that branching ratio as low as $10^{-3}$ can in principle be measured.

We consider first the simplest case of one generation model which, as
already mentioned, is sufficient to describe the relative importance
of the ${\tilde t}_1 \to \tau^+ \, b $ decay mode relative to the
possible 3-body decay modes
\begin{eqnarray}
{\tilde t}_1 &\to& W^+ \, b \, {\tilde \chi}^0_1 \nonumber\\
{\tilde t}_1 &\to& S^+_k \, b \, {\tilde \chi}^0_1 \nonumber\\
{\tilde t}_1 &\to& S^+_k \, b \, \nu_3 \nonumber\\
\label{eq:8}
{\tilde t}_1 &\to& b \, S^0_i \, \tau^+ \,,  \\
{\tilde t}_1 &\to& b \, P^0_j \, \tau^+ \, \nonumber\\
{\tilde t}_1 &\to& b \, {\tilde l}^+_i \, \nu_l \,, \nonumber\\
{\tilde t}_1 &\to& b \, {\tilde \nu}_l \, l^+ \hspace{1cm} (l=e,\mu) \, .
\nonumber
\end{eqnarray}
In general the important final states are those that conserve
R-parity. For example, for the case of decays involving $S^+_k$ the
most important are those in which the scalars mainly a stau.  Due to
the fact that the existing bounds on the MSSM sneutrinos are below 100
GeV there exists the possibility that the sneutrino has nearly the
same mass as one of the Higgs boson.  Similarly it could be that the
charged MSSM boson has nearly the same mass as one of the staus. This
implies large mixing effects even for small R-parity breaking
parameters \cite{deCampos:1995av,Akeroyd:1998iq,Porod:2000pw}.  We
therefore have used the complete formulas for the 3-body decay modes
which are presented in the Appendix. The latter include R-parity
violating decays such as $\tilde t_1 \to W^+ b \nu_3$. In addition to
the above mentioned decays there is also ${\tilde t}_1 \to b \, Z^0 \,
\tau^+$.  This decay mode is kinematically suppressed compared to
${\tilde t}_1 \to b \, \tau^+$ and there is no possible enhancement
due to a mixing with an R-parity conserving final state.  Therefore it
can be safely neglected.

\begin{figure}[htp]
 \setlength{\unitlength}{1mm}
 \begin{picture}(142,160)
 \put(5,175){\mbox{{\bf a)}}}
 \put(0,90){\includegraphics[height=8.5cm,width=8.5cm]{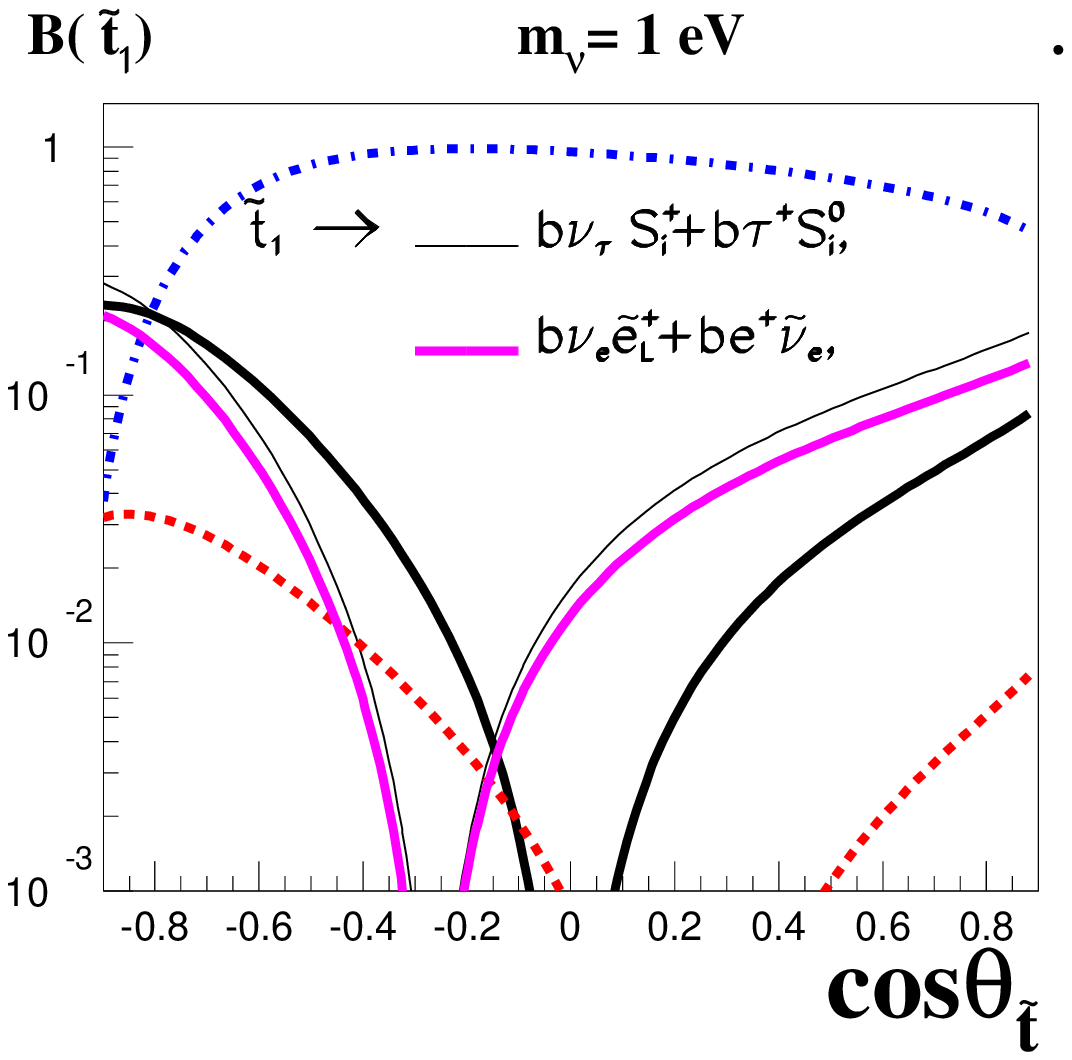}}
 \put(75,175){\mbox{{\bf b)}}}
 \put(70,90){\includegraphics[height=8.5cm,width=8.5cm]{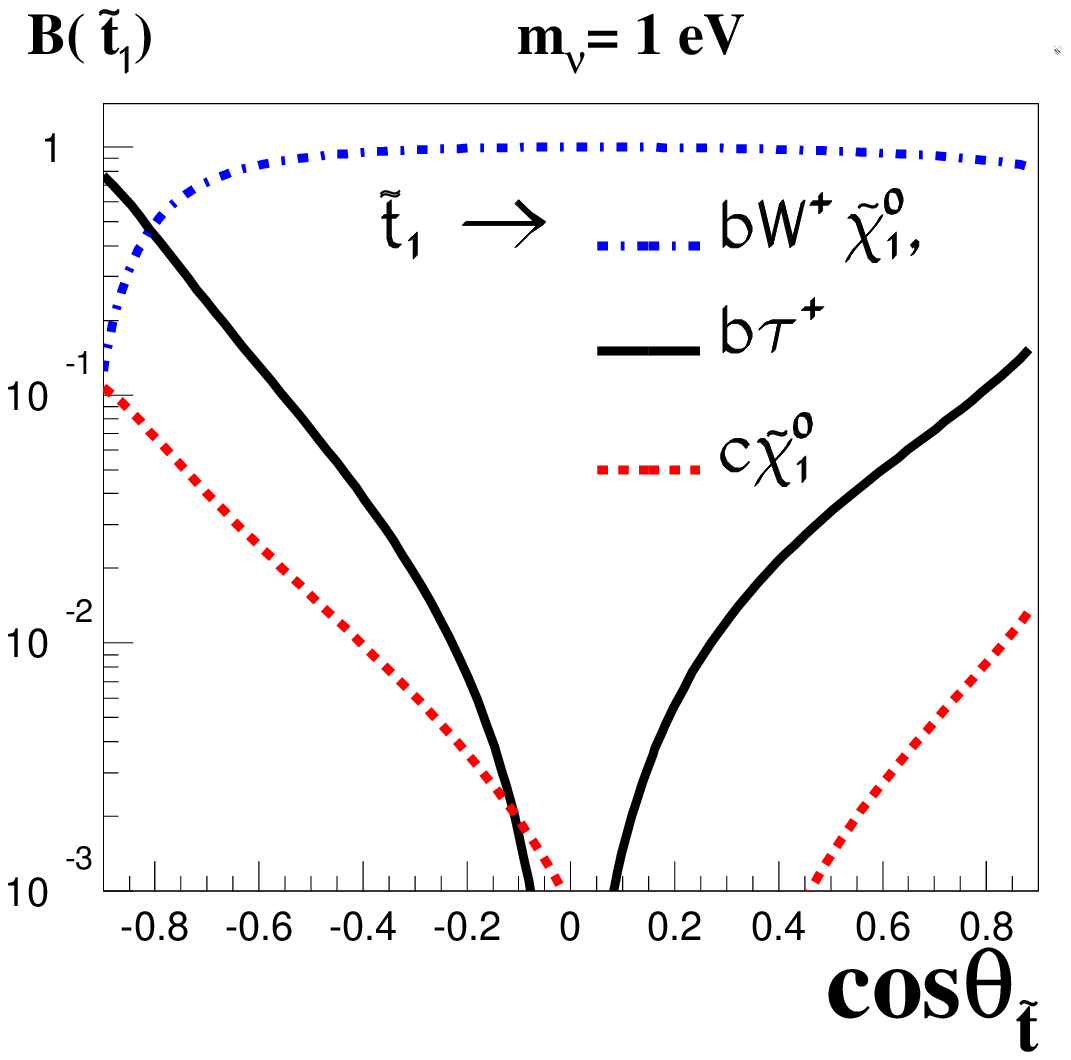}}
 \put(5,85){\mbox{{\bf c)}}}
 \put(0,0){\includegraphics[height=8.5cm,width=8.5cm]{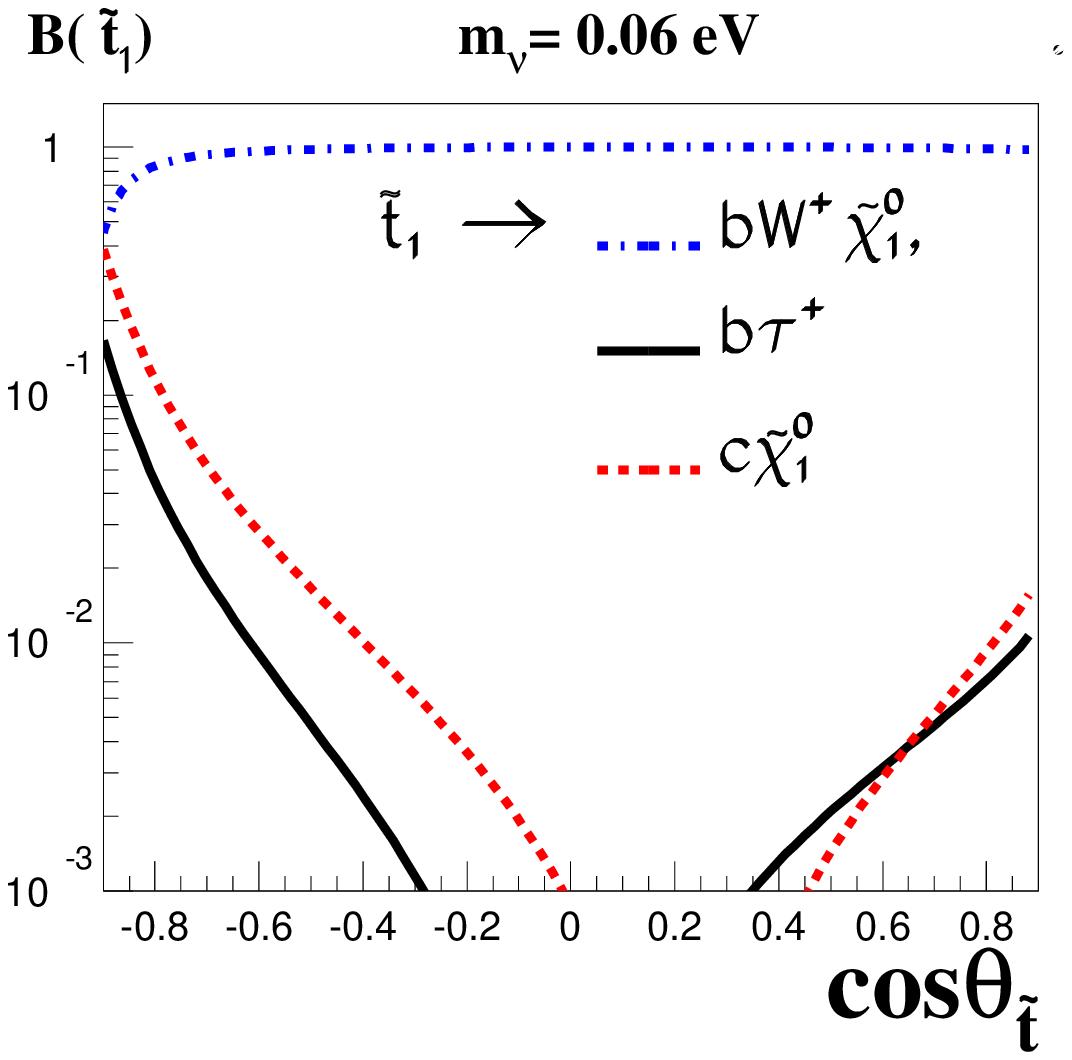}}
 \put(75,85){\mbox{{\bf d)}}}
 \put(70,0){\includegraphics[height=8.5cm,width=8.5cm]{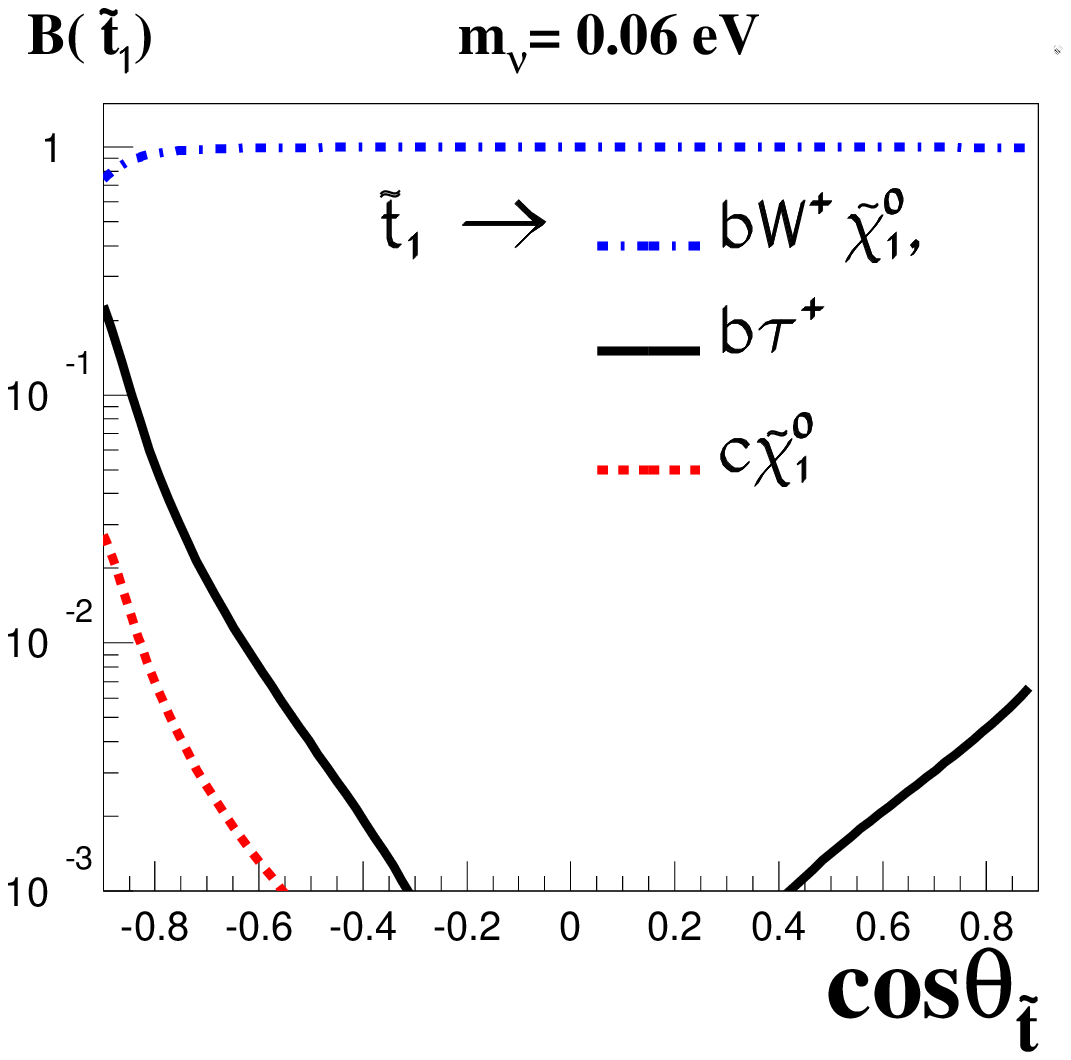}}
\end{picture}
    \caption{Branching ratios for the $\tilde t_1$ as a function of 
     $\cos \theta_{\tilde t}$ for different scenarios. We have fixed in
    a) $m_{\nu_3}=1\,$eV,
    b) $m_{\nu_3}=1\,$eV, $M_{\tilde
        E}>225\,$GeV, $M_{\tilde L}>225\,$GeV,
    c) $m_{\nu_3}=0.06\,$eV, $M_{\tilde E}>225\,$GeV,
      $M_{\tilde L}>225\,$GeV,
    d) Branching ratios for the $\tilde t_1$ as a function of $\cos
      \theta_{\tilde t}$ for $\tan \beta = 3$. $m_{\nu_3}=0.06\,$eV,
      $M_{\tilde E}>225\,$GeV, $M_{\tilde L}>225\,$GeV.
All the other inputs are given in Table~\protect{\ref{tab:1}}.
 }
\label{fig:1}
\end{figure}

In Fig.~\ref{fig:1} we show the branching ratios for the $\tilde t_1$
as a function of $\cos \theta_{\tilde t}$ in different scenarios.  In
order to calculate the partial width for the decay $\tilde t_1\to
c\tilde\chi_1^0$ we have taken the formula given in ref.
\cite{Hikasa:1987db}. According to the analysis performed
in~\cite{Diaz:1999ge}, where the full calculation was done in the
mSUGRA scenario, the result obtained with the present approximation
should be taken as one upper bound. This implies also that the shown
branching ratio for $\tilde t_1 \to b \tau^+$ can be viewed as a lower
bound. 
The parameters and physical quantities used in Fig.~\ref{fig:1} are
given in Tab.~\ref{tab:1}. For the case of Fig.~\ref{fig:1}(a) we have
fixed in addition the R-parity violating parameters such that $m_{\nu}
\approx 1$~eV. With this choice of parameters $S^0_1$, $S^0_3$,
$P^0_3$, and $S^-_4$ are mainly the MSSM Higgs-bosons whereas $S^0_2$,
$P^0_2$, $S^-_2$, and $S^-_3$ are mainly the MSSM sleptons of the
third generation.
In the plot we show the various branching ratios of the lighter stop
summing up those branching ratios for the decays into sleptons that
give the same final state, for example:
\begin{equation*}
{\tilde t}_1 \to b \, \nu_e \, {\tilde e}^+_L \,
          \to \, b \, e^+ \, \nu_e \, {\tilde \chi}^0_1 \;, \hspace{5mm}
{\tilde t}_1 \to b \, e^+ \, {\tilde \nu}_e \,
          \to \, b \, e^+ \, \nu_e \, {\tilde \chi}^0_1 \, .
\end{equation*}
The branching ratios for decays into $\tilde{\mu}_{L}$ or
$\tilde{\nu}_{\mu}$ are practically the same as those into
$\tilde{e}_{L}$ or $\tilde{\nu}_{e}$.  Note that the energy spectrum
of the leptons in the final will be somewhat different depending on
whether the scalar in the intermediate step is charged or neutral.
This offers in principle the possibility of determining the branching
ratios of the different decay chains even if the final state topology
is common.  Note, that states containing scalars or neutralinos will
lead to additional jet and/or lepton multiplicities absent in the
MSSM.

In Fig.~\ref{fig:1}(b) the slepton mass parameters are chosen such
that decays into scalars are kinematically forbidden.  Here we display
the channels ${\tilde t}_1 \to b \, W^+ {\tilde \chi}^0_1$, ${\tilde
  t}_1 \to b\, \tau^+$ and ${\tilde t}_1 \to c \, {\tilde \chi}^0_1$.
The remaining modes, such as ${\tilde t}_1 \to b \, S^0_1 \, \tau^+$,
turn out to be completely negligible.  In both cases, with and without
sleptons in the final state, one can see that in general the three
body mode ${\tilde t}_1 \to b \, W^+ {\tilde \chi}^0_1$, dominates
except for a somewhat narrow range of negative $\cos \theta_{\tilde
  t}$.  However, the branching ratio for $\tilde t_1\to b\,\tau^+$ is
above 0.1\% for most values of $|\cos \theta_{\tilde t}|$ implying the
observability of this mode.  Most importantly, note that even in the
parameter ranges where the three-body decay mode is dominant, its
resulting signature is rather different from that of the MSSM due to
the fact the lightest neutralino decays into SM-fermions, leading to
enhanced jet and/or lepton multiplicities, as discussed in detail in
\cite{Bartl2000rp,Porod:2000hv}. In the remaining part of this section
we assume that 3-body decays into scalars are kinematically forbidden.

\begin{table}
  \begin{center}
    \begin{tabular}{|l|lll|}\hline
      Input: & $\tan \beta = 6$ & $\mu = 500$ GeV & $M = 250$ GeV \\
      & $M_{\tilde D}=370$ GeV & $M_{\tilde Q}=340$ GeV & $A_b=150$ GeV \\
      & $M_{\tilde E}=210$ GeV & $M_{\tilde L}= 210$ GeV & $A_\tau=150$ GeV \\ 
      & $m_{{\tilde t}_1}=220$ GeV & $\cos \theta_{\tilde t}=-0.8$ &
      $m_{P^0_3}=300$ GeV \\ \hline
      Calculated & $m_{{\tilde \chi}^0_1}=122$ GeV &
      $m_{{\tilde \chi}^+_1}=234$ GeV & $m_{{\tilde \chi}^+_2}=519$ GeV \\
      &  $m_{{\tilde b}_1}=334$ GeV & $m_{{\tilde b}_2}=381$ GeV & 
      $\cos \theta_{\tilde b}=0.879$ \\
      & $m_{S^0_1}=107$ GeV & $m_{S^0_2}=200$ GeV & $m_{S^0_3}=302$ GeV \\
      &  $m_{P^0_2}=200$ GeV & $m_{P^0_3}=300$ GeV  & \\
      & $m_{S^-_2}=203$ GeV & $m_{S^-_3}=226$ GeV & $m_{S^-_4}=311$ GeV  \\
      & $m_{{\tilde e}_L}=215$ GeV &
      $m_{{\tilde \nu}_e}=m_{{\tilde \nu}_\mu }=200$ GeV  & \\ \hline
    \end{tabular}
  \end{center}
  \caption[]{Input parameters and resulting quantities used in 
Fig.~\protect{~\ref{fig:1}}.}
\label{tab:1}
\end{table}

In Fig.~\ref{fig:1}(c) the R-parity violating parameters are fixed in
such a way that the heaviest neutrino mass is in the range suggested
by the oscillation interpretation of the atmospheric neutrino anomaly
\cite{Gonzalez-Garcia:2001sq}.  

\begin{figure}[htp]
 \begin{center}
    \begin{picture}(232,190) 
      \put(0,-25){\includegraphics[height=8.0cm,width=8.5cm]{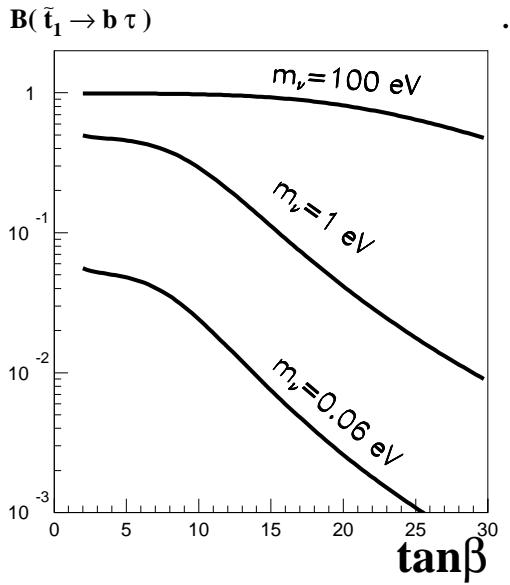}}
    \end{picture}
 \end{center}
    \caption{\small Branching ratios for ${\tilde t}_1$ decays
  for $m_{{\tilde t}_1} = 220$~GeV, $\mu = 500$~GeV, $M =
  240$~GeV, and $m_\nu = 100$, 1 and $0.06\,$eV.
  The branching ratios are shown as a function of
  $\tan \beta$. ($\cos\theta_{\tilde t}=-0.8$)}
   \label{fig:2}
\end{figure}

In Fig.~\ref{fig:1}(d) we show the same scenario as in
Fig.~\ref{fig:1}(c) but for $\tan \beta = 3$. The branching ratio into
$b \tau$ now increases, whereas the branching ratio into $c \tilde
\chi^0_1$ decreases.  This is easily understood by inspecting
Eqs.~(\ref{eq:6}) and (\ref{eq:7}). Indeed for the $b \, \tau$ case
the partial width is proportional to $h_b^2$, whereas for
$c\tilde\chi_1^0$ it is proportional to $h_b^4$.  This implies that
the partial width for $\tilde t_1\to c\tilde\chi_1^0$ grows faster
with $\tan \beta$ than the width for $\tilde t_1\to b \, \tau$.
This is also demonstrated in Fig.~\ref{fig:2} where we show the $\tan
\beta$ dependence of the branching ratio for the decay of $\tilde t_1$
into $b\tau^+$ for several values of the neutrino mass. For
$m_{\nu_3}=0.06\,$eV the $B(\tilde t_1\to b\,\tau)$ is still above
$0.1\%$ if $\tan\beta$ is not too large, as favored by the explanation
of the neutrino anomalies in this model~\cite{Romao:2000up}. As seen
from the figure, the the $\tilde t_1 \to b\tau^+$ branching ratio is
also somewhat correlated to the $\nu_3$ mass.  Should one add a
sterile neutrino to the model~\cite{Hirsch:2000xe}, then the neutrino
state $\nu_3$ could in principle be heavier than assumed above,
favoring ${\tilde t}_1 \to \tau^+ \, b $ decay mode.

Let us now turn to the general three neutrinos case. There are new
features that arise in this case, as opposed to the 1-generation case
considered so far.  In this model the solution to the present neutrino
anomalies implies that all the $\epsilon_i$ are of the same order of
magnitude~\cite{Romao:2000up}.  

Two furhter important results of \cite{Romao:2000up} are that the
atmospheric neutrino angle is controlled by the ratio $(\epsilon_2 v_d
+\mu v_2)/(\epsilon_3 v_d +\mu v_3)$ and that the solar mixing angle
is controlled by $(\epsilon_1/\epsilon_2)^2$.  One can get approximate
formulas for the decay widths ${\tilde t}_1 \to b \, e^+$ and ${\tilde
  t}_1 \to b \, \mu^+$ similar to Eq.~(\ref{eq:6}) by replacing
$\epsilon_3$ by $\epsilon_{1,2}$. This implies that {\it (i)} The
decays into $ b \, e^+$ and $ b \, \mu^+$ are as important as the
decay into $ b \, \tau^+$.  {\it (ii)} The decays ${\tilde t}_1 \to b
\, e^+$ and ${\tilde t}_1 \to b \, \mu^+$ are related with the solar
mixing angle.  Moreover, we find that $\sum_{l=e,\mu,\tau}
\Gamma(\tilde t_1 \to b \, l^+)$ in the 3-generation model is nearly
equal to $\Gamma(\tilde t_1 \to b \, \tau^+)$ in the 1-generation
model provided that $\sum_{i=1}^3 \epsilon^2_i$ is identified to
$\epsilon^2$ in the 1-generation model.

\begin{figure}
\setlength{\unitlength}{1mm}
\begin{center}
\begin{picture}(70,70)
\put(-5,-5){\includegraphics[height=7.0cm,width=7.cm]{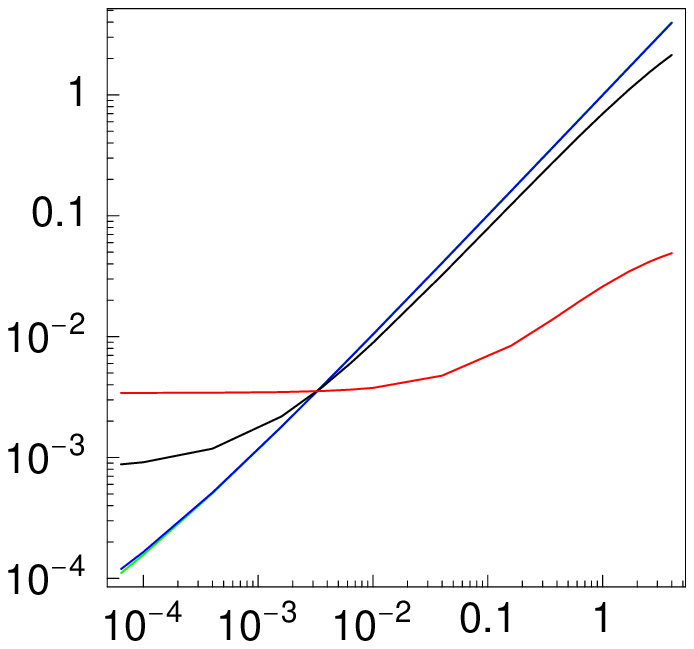}}
\put(0,63){\makebox(0,0)[bl]{{\small $\mbox{B}({\tilde t}_1 \to b \, e^+)/
                                    \mbox{B}({\tilde t}_1 \to b \, \mu^+) $}}}
\put(68,-8){\makebox(0,0)[br]{{\small $(\epsilon_1 / \epsilon_2)^2$}}}
\put(42,50){{\small $\geq 0.1$}}
\put(49,38){{\small $ 10^{-2}$}}
\put(52,41){\vector(0,1){6}}
\put(8,18){{\small $10^{-2}$}}
\put(52,28){{\small $10^{-3}$}}
\put(15,25){{\small $10^{-3}$}}
\put(12,7){{\small $\geq 0.1$}}
\end{picture}
\end{center}
\caption[]{Ratio of branching ratios: $\mbox{B}({\tilde t}_1 \to b e^+)/
  \mbox{B}({\tilde t}_1 \to b \mu^+) $ as a function of $(\epsilon_1 /
  \epsilon_2)^2$ for $m_{{\tilde t}_1} = 220$~GeV, $\mu = 500$~GeV, $M
  = 240$~GeV; $|\cos \theta_{\tilde t}| \geq 0.1, 0.01, 10^{-3}$,
  $m_{\nu_3} = 0.6$~eV.}
\label{fig:3}
\end{figure}

In Fig.~\ref{fig:3} we show the ratio of B$({\tilde t}_1 \to b \,
e^+)/$B$({\tilde t}_1 \to b \, \mu^+)$ versus
$(\epsilon_1/\epsilon_2)^2$ for different values of $\cos
\theta_{\tilde t}$. For definiteness we have fixed the heaviest
neutrino mass at the best-fit value indicated by the atmospheric
neutrino anomaly.
One can see that the dependence is nearly linear even for rather small
$\cos \theta_{\tilde t}$.  For $|\cos \theta_{\tilde t}| \lesssim
10^{-2}$ the approximation in Eq.~(\ref{eq:6}) breaks down and
additional pieces dependent on $\sin \theta_{\tilde t}$
\cite{Bartl:1996gz,Diaz:1999ge} become important, leading to the non-linear
dependence.  One sees from the figure that, as long as $\cos
\theta_{\tilde t} \gtrsim 10^{-2}$ there is a good degree of
correlation between the branching ratios into $B(\tilde t_1 \to
b\,e^+)$ and $B(\tilde t_1 \to b\,\mu^+)$ and the ratio
$(\epsilon_1/\epsilon_2)^2$. Thus by measuring these branchings one
will get information on the solar neutrino mixing, since $\tan^2
\theta_{sol}$ is proportional to
$(\epsilon_1/\epsilon_2)^2$~\cite{Romao:2000up} which makes it a
rather important quantity.
For the so--called small mixing angle or SMA solution of the solar
neutrino problem we expect ${\tilde t}_1 \to e^+ \, b$ to be
negligible. In contrast, for the large mixing angle type solutions
(LMA, LOW and QVAC, see ref.~\cite{Gonzalez-Garcia:2001sq} and
references therein) we expect \texttt{all} ${\tilde t}_1 \to l^+ \, b$
decays to have comparable rates.
As a result in this model one can directly test the solution to the
solar neutrino problem against the lighter stop decay pattern. 
This is also complementary to the case of neutralino decays considered
in \cite{Porod:2000hv}. In that case the sensitivity is mainly to
atmospheric mixing, as opposed to solar mixing. Testing the latter in
neutralino decays at a collider experiment requires more detailed
information on the complete spectrum to test the solar angle
\cite{Porod:2000hv}. In contrast we have obtained here a rather neat
connection of stop decays with the solar neutrino physics.

Note, that this result is much more general than the scenarios
discussed in this paper. It is of particular importance in scenarios
where only the R-parity violating decays and the decay into $\tilde
\chi^0_1 \, c$ are present \cite{Bartl:1996gz,Diaz:1999ge}. Similarly,
the other ratios of the final states $b \, l^+$ are proportional to
the square of the ratio of corresponding $\epsilon_i$ provided that
$\cos \theta_{\tilde t}$ is not too small.

\section{Conclusions}
\label{sec:con}

We have studied the phenomenology of the lightest stop in scenarios
where R-parity violating decays such as ${\tilde t}_1 \to b \, \tau^+$
compete with three--body decays. We have found that for $m_{{\tilde
    t}_1} \lesssim 250$~GeV there are regions of parameter where
${\tilde t}_1 \to b \, \tau^+$ is an important decay mode if not the
most important one. This implies that there exists the possibility of
full stop mass reconstruction from $\tau^+ \, \tau^- \, b \, \bar{b}$
final states, favoring the prospects for its discovery. In contrast, in
the MSSM the discovery of the lightest stop might not be possible at
the LHC within this mass range.
This implies that it is important to take into account this new decay
mode when designing the stop search strategies at a future $e^+ e^-$
Linear Collider.  Spontaneously and bilinearly broken R-parity
violation also imply additional leptons and/or jets in stop cascade
decays.  Looking at the three generation model the decays into
${\tilde t}_1 \to b \, l^+$ imply the possibility of probing
$\epsilon^2_1 /\epsilon^2_2$ and thus the solar mixing angle.
This complements information which can be obtained using neutralino
decays. In the latter case the sensitivity is mainly to the
atmospheric mixing, as opposed to solar mixing. In this model
neutralino decays is ideal to test the atmospheric anomaly at a
collider experiment, while stop decays provide neat complementary
information on the solar mixing angle.
Obtaining solar mixing information from neutralino decays would
require more detailed knowledge on the supersymmetric spectrum, since
it would be involved in the relevant loop calculations of the solar
neutrino mass scale and mixing angle.
By combining the two one can probe the parameters associated with both
solar and atmospheric neutrino anomalies at collider experiments.

\section*{Acknowledgments}

This work was supported by Spanish DGICYT under grant PB98-0693 and by
the European Commission TMR network HPRN-CT-2000-00148. D.~R.~was
supported by Colombian COLCIENCIAS fellowship, while W.~P.~was
supported by the Spanish ``Ministerio de Educaci{\'o}n y Cultura'' under
the contract SB97-BU0475382.

\newpage
\appendix

\section*{Appendix}

In this set of appendixes we present the formulas for the lighter stop
decay widths and couplings used in the paper and omitted in previous
sections.

\section{Formulas for the three-body decay widths}
\label{sec:appA}

\subsection{The width $\Gamma({\tilde t}_1 \to W^+ \, b \, {\tilde \chi}^0_i)$}

\noindent
\begin{eqnarray}
 \Gamma({\tilde t}_1 \to W^+ \, b \, {\tilde \chi}^0_i)  &=& \nonumber \\
 & & \hspace{-30mm} 
   =  \frac{\alpha^2}{16 \, \pi m^3_{{\tilde t}_1} \sin^4 \theta_W}
  \int\limits^{(m_{{\tilde t}_1}-m_{\scriptscriptstyle{W}})^2}_{
           (m_b + m_{{\tilde \chi}^0_i})^2} \hspace{-8mm}
     d \, s \,
   \left( G^W_{{\tilde \chi}^+ {\tilde \chi}^+} +
   G^W_{{\tilde \chi}^+ t} +
   G^W_{{\tilde \chi}^+ {\tilde b}} +
   G^W_{t t} +
   G^W_{t {\tilde b}} +
   G^W_{{\tilde b} {\tilde b}} \right) \nonumber
\end{eqnarray}
with
\begin{eqnarray} 
   G^W_{{\tilde \chi}^+ {\tilde \chi}^+} &=&
       \sum^3_{j=1} \Big[  \sum_{k=0}^3a_{ijk}s^k
          J^0_t(m^2_{{\tilde t}_1} + m^2_{W} + m^2_b
               + m^2_{{\tilde \chi}^0_i} - m^2_{{\tilde \chi}^+_j} - s
             ,\Gamma_{{\tilde \chi}^+_j} m_{{\tilde \chi}^+_j}) \nonumber \\
    & & \hspace{5mm} + \sum_{k=0}^2a_{ij,k+5}s^k
       J^1_t(m^2_{{\tilde t}_1} + m^2_{W} + m^2_b
            + m^2_{{\tilde \chi}^0_i} - m^2_{{\tilde \chi}^+_j} - s
             ,\Gamma_{{\tilde \chi}^+_j} m_{{\tilde \chi}^+_j}) \nonumber \\
    & & \hspace{5mm} + \, (a_{ij8}+a_{ij9}s) \,
          J^2_t(m^2_{{\tilde t}_1} + m^2_{W} + m^2_b
               + m^2_{{\tilde \chi}^0_i} - m^2_{{\tilde \chi}^+_j} - s
       ,\Gamma_{{\tilde \chi}^+_j} m_{{\tilde \chi}^+_j}) \Big] \nonumber \\
    & &  +  \sum_{k=0}^3a_{i4k}s^k
          J^0_{tt}(m^2_{{\tilde t}_1} + m^2_{W} + m^2_b
               + m^2_{{\tilde \chi}^0_i} - m^2_{{\tilde \chi}^+_1} - s
             ,\Gamma_{{\tilde \chi}^+_1} m_{{\tilde \chi}^+_1}  \nonumber \\
       & & \hspace{33mm} ,m^2_{{\tilde t}_1} + m^2_{W} + m^2_b
               + m^2_{{\tilde \chi}^0_i} - m^2_{{\tilde \chi}^+_2} - s
             ,\Gamma_{{\tilde \chi}^+_2} m_{{\tilde \chi}^+_2}) \nonumber \\
    & &  + \sum_{k=0}^2a_{i4,k+5}s^k
       J^1_{tt}(m^2_{{\tilde t}_1} + m^2_{W} + m^2_b
               + m^2_{{\tilde \chi}^0_i} - m^2_{{\tilde \chi}^+_1} - s
             ,\Gamma_{{\tilde \chi}^+_1} m_{{\tilde \chi}^+_1}  \nonumber \\
       & & \hspace{33mm} ,m^2_{{\tilde t}_1} + m^2_{W} + m^2_b
               + m^2_{{\tilde \chi}^0_i} - m^2_{{\tilde \chi}^+_2} - s
             ,\Gamma_{{\tilde \chi}^+_2} m_{{\tilde \chi}^+_2}) \nonumber \\
    & &  + \,( a_{i48}+a_{i49}s) \,
       J^2_{tt}(m^2_{{\tilde t}_1} + m^2_{W} + m^2_b
               + m^2_{{\tilde \chi}^0_i} - m^2_{{\tilde \chi}^+_1} - s
             ,\Gamma_{{\tilde \chi}^+_1} m_{{\tilde \chi}^+_1}  \nonumber \\
       & & \hspace{17mm} ,m^2_{{\tilde t}_1} + m^2_{W} + m^2_b
               + m^2_{{\tilde \chi}^0_i} - m^2_{{\tilde \chi}^+_2} - s
             ,\Gamma_{{\tilde \chi}^+_2} m_{{\tilde \chi}^+_2}) 
             \nonumber\\
    & &  +  \sum_{k=0}^3a_{i5k}s^k
          J^0_{tt}(m^2_{{\tilde t}_1} + m^2_{W} + m^2_b
               + m^2_{{\tilde \chi}^0_i} - m^2_{{\tilde \chi}^+_1} - s
             ,\Gamma_{{\tilde \chi}^+_1} m_{{\tilde \chi}^+_1}  \nonumber \\
       & & \hspace{33mm} ,m^2_{{\tilde t}_1} + m^2_{W} + m^2_b
               + m^2_{{\tilde \chi}^0_i} - m^2_{{\tilde \chi}^+_3} - s
             ,\Gamma_{{\tilde \chi}^+_3} m_{{\tilde \chi}^+_3}) \nonumber \\
    & &  + \sum_{k=0}^2a_{i5,k+5}s^k
       J^1_{tt}(m^2_{{\tilde t}_1} + m^2_{W} + m^2_b
               + m^2_{{\tilde \chi}^0_i} - m^2_{{\tilde \chi}^+_1} - s
             ,\Gamma_{{\tilde \chi}^+_1} m_{{\tilde \chi}^+_1}  \nonumber \\
       & & \hspace{33mm} ,m^2_{{\tilde t}_1} + m^2_{W} + m^2_b
               + m^2_{{\tilde \chi}^0_i} - m^2_{{\tilde \chi}^+_3} - s
             ,\Gamma_{{\tilde \chi}^+_3} m_{{\tilde \chi}^+_3}) \nonumber \\
    & &  + \, ( a_{i58}+a_{i59}s) \,
       J^2_{tt}(m^2_{{\tilde t}_1} + m^2_{W} + m^2_b
               + m^2_{{\tilde \chi}^0_i} - m^2_{{\tilde \chi}^+_1} - s
             ,\Gamma_{{\tilde \chi}^+_1} m_{{\tilde \chi}^+_1}  \nonumber \\
       & & \hspace{17mm} ,m^2_{{\tilde t}_1} + m^2_{W} + m^2_b
               + m^2_{{\tilde \chi}^0_i} - m^2_{{\tilde \chi}^+_3} - s
             ,\Gamma_{{\tilde \chi}^+_3} m_{{\tilde \chi}^+_3})\nonumber \\
    & &  +   \sum_{k=0}^3a_{i6k}s^k
          J^0_{tt}(m^2_{{\tilde t}_1} + m^2_{W} + m^2_b
               + m^2_{{\tilde \chi}^0_i} - m^2_{{\tilde \chi}^+_3} - s
             ,\Gamma_{{\tilde \chi}^+_3} m_{{\tilde \chi}^+_3}  \nonumber \\
       & & \hspace{33mm} ,m^2_{{\tilde t}_1} + m^2_{W} + m^2_b
               + m^2_{{\tilde \chi}^0_i} - m^2_{{\tilde \chi}^+_2} - s
             ,\Gamma_{{\tilde \chi}^+_2} m_{{\tilde \chi}^+_2}) \nonumber 
\end{eqnarray}
\begin{eqnarray}
      & &  + \sum_{k=0}^2a_{i6,k+5}s^k
       J^1_{tt}(m^2_{{\tilde t}_1} + m^2_{W} + m^2_b
               + m^2_{{\tilde \chi}^0_i} - m^2_{{\tilde \chi}^+_3} - s
             ,\Gamma_{{\tilde \chi}^+_3} m_{{\tilde \chi}^+_3}  \nonumber \\
       & & \hspace{33mm} ,m^2_{{\tilde t}_1} + m^2_{W} + m^2_b
               + m^2_{{\tilde \chi}^0_i} - m^2_{{\tilde \chi}^+_2} - s
             ,\Gamma_{{\tilde \chi}^+_2} m_{{\tilde \chi}^+_2}) \nonumber \\
    & &  + \, ( a_{i68}+a_{i69}s) \,
       J^2_{tt}(m^2_{{\tilde t}_1} + m^2_{W} + m^2_b
               + m^2_{{\tilde \chi}^0_i} - m^2_{{\tilde \chi}^+_3} - s
             ,\Gamma_{{\tilde \chi}^+_3} m_{{\tilde \chi}^+_3}  \nonumber \\
       & & \hspace{17mm} ,m^2_{{\tilde t}_1} + m^2_{W} + m^2_b
               + m^2_{{\tilde \chi}^0_i} - m^2_{{\tilde \chi}^+_2} - s
             ,\Gamma_{{\tilde \chi}^+_2} m_{{\tilde \chi}^+_2}) \, ,\nonumber
\end{eqnarray}
\begin{eqnarray}
   G^W_{{\tilde \chi}^+ t} &=&
       \sum^3_{j=1} \Big[ \sum_{k=0}^2b_{ijk}s^k 
 J^0_{tt}(m^2_{{\tilde t}_1} + m^2_{W} + m^2_b
               + m^2_{{\tilde \chi}^0_i} - m^2_{{\tilde \chi}^+_j} - s
       ,-\Gamma_{{\tilde \chi}^+_j} m_{{\tilde \chi}^+_j},m^2_t,
        \Gamma_t m_t) \nonumber \\
    & & \hspace{5mm} + \sum_{k=0}^2b_{ij,k+4}s^k 
 J^1_{tt}(m^2_{{\tilde t}_1} + m^2_{W} + m^2_b
               + m^2_{{\tilde \chi}^0_i} - m^2_{{\tilde \chi}^+_j} - s
       ,-\Gamma_{{\tilde \chi}^+_j} m_{{\tilde \chi}^+_j} ,m^2_t,
        \Gamma_t m_t) \nonumber \\
    & & \hspace{5mm} + \, (b_{ij7}+b_{ij8}s) \,
        J^2_{tt}(m^2_{{\tilde t}_1} + m^2_{W} + m^2_b
               + m^2_{{\tilde \chi}^0_i} - m^2_{{\tilde \chi}^+_j} - s
       ,-\Gamma_{{\tilde \chi}^+_j} m_{{\tilde \chi}^+_j} ,m^2_t,
         \Gamma_t m_t)  \Big] \, ,\nonumber 
\end{eqnarray}
\begin{eqnarray}
   G^W_{{\tilde \chi}^+ {\tilde b}} &=&
       -\sum^2_{k=1} \Big[
          \sum_{l=0}^3c_{ijkl}s^l J^0_{st}(
       m^2_{{\tilde b}_k}, \Gamma_ {{\tilde b}_k} m_{{\tilde b}_k}
       ,m^2_{{\tilde t}_1} + m^2_{W} + m^2_b
               + m^2_{{\tilde \chi}^0_i} - m^2_{{\tilde \chi}^+_j} - s
       ,-\Gamma_{{\tilde \chi}^+_j} m_{{\tilde \chi}^+_j}) \nonumber \\
    & &\hspace{-1.5cm}  + \sum_{l=0}^2c_{ijk,l+5}s^l  \, J^1_{st}(
       m^2_{{\tilde b}_k}, \Gamma_ {{\tilde b}_k} m_{{\tilde b}_k}
       ,m^2_{{\tilde t}_1} + m^2_{W} + m^2_b
               + m^2_{{\tilde \chi}^0_i} - m^2_{{\tilde \chi}^+_j} - s
       ,-\Gamma_{{\tilde \chi}^+_j} m_{{\tilde \chi}^+_j}) \Big] \, ,\nonumber
\end{eqnarray}
\begin{equation*}
   G^W_{tt} =
          ( d_{i1} + d_{i2} s) J^0_t(m^2_t, \Gamma_t m_t) +
          ( d_{i3} + d_{i4} s) J^1_t(m^2_t, \Gamma_t m_t)
 + ( d_{i5}+ d_{i6} s ) J^2_t(m^2_t, \Gamma_t m_t) \, ,
\end{equation*}
\begin{eqnarray}
   G^W_{t {\tilde b}} &=& \sum^2_{k=1} \Big[
   ( e_{ik1} + e_{ik2} s+ e_{ik3} s^2 )
   J^0_{st}(m^2_{{\tilde b}_k},
  \Gamma_ {{\tilde b}_k} m_{{\tilde b}_k},m^2_t, \Gamma_t m_t)\nonumber \\
&& + ( e_{ik4} + e_{ik5} s+ e_{ik6} s^2 ) J^1_{st}(m^2_{{\tilde b}_k},
    \Gamma_ {{\tilde b}_k} m_{{\tilde b}_k},m^2_t, \Gamma_t m_t)
   \Big] \, , \nonumber
\end{eqnarray}
\begin{eqnarray}
G^W_{{\tilde b} {\tilde b}} &=&
 \frac{\sqrt{\lambda(s,m^2_{{\tilde t}_1},m^2_{W})
                 \lambda(s,m^2_{{\tilde \chi}^0_i},m^2_b)}}{s} \nonumber \\
  & & \hspace{-12mm} \times \left\{ \sum^2_{k=1}
      \frac{(f_{ik1} + f_{ik2} s)}
   {(s-m^2_{{\tilde b}_k})^2 + \Gamma^2_{{\tilde b}_k} m^2_{{\tilde b}_k}}
 + \mbox{Re} \left[ \frac{(f_{i31} + f_{i31} s)}
      { (s-m^2_{{\tilde b}_1} + i \Gamma_{{\tilde b}_1} m_{{\tilde b}_1})
         (s-m^2_{{\tilde b}_2} - i \Gamma_{{\tilde b}_2} m_{{\tilde b}_2})}
          \right]  \right\} \, .\nonumber
\end{eqnarray}
The integrals   $J^{0,1,2}_{t,tt,st}$ are:
\begin{eqnarray}
 J^i_{t}(m^2_1, m_1 \Gamma_1)
        &=& \int\limits^{t_{max}}_{t_{min}}
     \hspace{-2mm} d \, t \frac{t^i}{(t-m^2_1)^2 \, + \,  m_1^2 \Gamma_1^2}
  \, , \nonumber\\
 J^i_{tt}(m^2_1,m_1 \, \Gamma_1,m^2_2,m_2 \, \Gamma_2)
        &=& \mbox{Re} \int\limits^{t_{max}}_{t_{min}}
     \hspace{-2mm} d \, t
          \frac{t^i}{(t-m^2_1 \, + \,i m_1 \Gamma_1)
                     (t-m^2_2 \, - \,i m_2 \Gamma_2)} \, , \nonumber\\
 J^i_{st}(m^2_1,m_1 \, \Gamma_1,m^2_2,m_2 \, \Gamma_2)
        &=& \mbox{Re} \frac1{s-m^2_1 \, + \,i  m_1 \Gamma_1}
         \int\limits^{t_{max}}_{t_{min}} \hspace{-2mm} d \, t
          \frac{t^i}{(t-m^2_2 \, - \,i m_2 \Gamma_2)} \nonumber
\end{eqnarray}
with $i=0,1,2$. 
Their integration range is given by
\begin{eqnarray}
t_{max \atop min} &=&
 \frac{m^2_{{\tilde t}_1} + m^2_b + m^2_{W}
        + m^2_{{\tilde \chi}^0_i} -s}2
       - \frac{(m^2_{{\tilde t}_1}-m^2_{W})
               (m^2_{{\tilde \chi}^0_i}-m^2_b)}{2 s} \nonumber \\
  & &  \pm  \frac{\sqrt{\lambda(s,m^2_{{\tilde t}_1},m^2_{W})
                  \lambda(s,m^2_{{\tilde \chi}^0_i},m^2_b)}}{2 s} \, ,
\nonumber
\end{eqnarray}
where $s = (p_{{\tilde t}_1} - p_{W})^2$
 and $t = (p_{{\tilde t}_1} - p_{t})^2$
are the usual Mandelstam variables.
Note, that $-\Gamma_{{\tilde \chi}^+_j} m_{{\tilde \chi}^+_j}$
appears in the entries of the integrals 
$G^W_{{\tilde \chi}^+ {\tilde b}_j}$ and
$G^W_{{\tilde \chi}^+ t}$ because the chargino is exchanged 
 in the $u$-channel in our convention.
The coefficients are given by (no sum upon repeated index):
\begin{eqnarray}
  a_{ij1}&=& 6 O^L_{ij} O^R_{ij} \left( (k^{\tilde t}_{1j})^2 + 
(l^{\tilde t}_{1j})^2 \right)   m_{\tilde{\chi}^+_j}m_{{\tilde \chi}^0_i}   
(2   m_b^2 + m^2_{{\tilde \chi}^0_i} + m_{W}^2)
\nonumber \\
&&
     +2 k^{\tilde t}_{1j}l^{\tilde t}_{1j}   m_b m_{{\tilde \chi}^0_i}   
(m_b^2+m^2_{{\tilde \chi}^0_i}+m^2_{{\tilde t}_1}+m^2_{\tilde{\chi}^+_j}
+m_{W}^2)
\nonumber \\
&&
- 2k^{\tilde t}_{1j}l^{\tilde t}_{1j}\left( (O^L_{ij})^2 
+ (O^R_{ij})^2 \right)
m_b m_{\tilde{\chi}^+_j}  
\left[
  3 m_b^2 + 2 m^2_{{\tilde \chi}^0_i} + 3 m^2_{{\tilde t}_1} 
  + (m^2_{{\tilde t}_1}+m_b^2)^2   /m_{W}^2
\right]
\nonumber \\
&&
- \left( (k^{\tilde t}_{1j})^2
  (O^R_{ij})^2 + (l^{\tilde t}_{1j})^2 (O^L_{ij})^2 \right)  
 \Big[ (4 m_b^2+m_{W}^2) (m^2_{{\tilde \chi}^0_i}+m_b^2) 
+ m^4_{{\tilde \chi}^0_i}
\nonumber \\
&&
      + m^2_{{\tilde t}_1} (2 m_{W}^2+4 m_b^2+m^2_{{\tilde \chi}^0_i})
   + (m_b^2+m^2_{{\tilde t}_1}) (m_b^4+m_b^2 m^2_{{\tilde t}_1}
+m^2_{{\tilde \chi}^0_i}m^2_{{\tilde t}_1}) /m_{W}^2 \Big]
\nonumber \\
&&
 - \left( (k^{\tilde t}_{1j})^2(O^L_{ij})^2 + 
(l^{\tilde t}_{1j})^2 (O^R_{ij})^2 \right)   m^2_{\tilde{\chi}^+_j}  
\left(
  m^2_{{\tilde \chi}^0_i} + m^2_{{\tilde t}_1} +2   m_b^2 
+ (m_b^4+m_b^2 m^2_{{\tilde t}_1}) /m_{W}^2
\right)
\nonumber 
\end{eqnarray}
\begin{eqnarray}
a_{ij2}&=& -12   k^{\tilde t}_{1j}l^{\tilde t}_{1j} O^L_{ij} O^R_{ij} 
m_b m_{{\tilde \chi}^0_i}-6\left( (k^{\tilde t}_{1j})^2 
+ (l^{\tilde t}_{1j})^2 \right)O^L_{ij} O^R_{ij}m_{\tilde{\chi}^+_j}
m_{{\tilde \chi}^0_i}
\nonumber \\
&&
+2   k^{\tilde t}_{1j}l^{\tilde t}_{1j}   \left( (O^L_{ij})^2 
+ (O^R_{ij})^2 \right)   m_b m_{\tilde{\chi}^+_j}  
(3+2 (m_b^2+m^2_{{\tilde t}_1}) /m_{W}^2)
\nonumber \\
&&
+\left( (k^{\tilde t}_{1j})^2(O^R_{ij})^2 + (l^{\tilde t}_{1j})^2 
(O^L_{ij})^2 \right)   \Big[6 m_b^2+2 m^2_{{\tilde \chi}^0_i}
+2 m^2_{{\tilde t}_1}+m_{W}^2
\nonumber \\
&&
+(3 m_b^4+m_b^2 m^2_{{\tilde \chi}^0_i}
+4 m_b^2 m^2_{{\tilde t}_1}+2 m^2_{{\tilde \chi}^0_i}
m^2_{{\tilde t}_1}+m^4_{{\tilde t}_1}) /m_{W}^2 \Big]
\nonumber \\
&&
+ \left( (k^{\tilde t}_{1j})^2(O^L_{ij})^2 + (l^{\tilde t}_{1j})^2 
(O^R_{ij})^2 \right)   m^2_{\tilde{\chi}^+_j}   (2+m_b^2 /m_{W}^2)    
\nonumber
\end{eqnarray}
\begin{eqnarray}
       a_{ij3}&=& -2k^{\tilde t}_{1j}l^{\tilde t}_{1j}   
\left( (O^L_{ij})^2 + (O^R_{ij})^2 \right)   
m_b m_{\tilde{\chi}^+_j}   /m_{W}^2
\nonumber \\
&&
              - \left( (k^{\tilde t}_{1j})^2
  (O^R_{ij})^2 + (l^{\tilde t}_{1j})^2 (O^L_{ij})^2 \right)   
\left[
  2+(m^2_{{\tilde \chi}^0_i}+3 m_b^2+2 m^2_{{\tilde t}_1}) /m_{W}^2
\right]
\nonumber
\end{eqnarray}
\begin{eqnarray}
       a_{ij4}&=& \left( (k^{\tilde t}_{1j})^2
  (O^R_{ij})^2 + (l^{\tilde t}_{1j})^2 (O^L_{ij})^2 \right)   /m_{W}^2
\nonumber
\end{eqnarray}
\begin{eqnarray}
       a_{ij5}&=& -12   k^{\tilde t}_{1j}l^{\tilde t}_{1j}   
O^L_{ij} O^R_{ij} m_b m_{{\tilde \chi}^0_i}
-6\left( (k^{\tilde t}_{1j})^2 +(l^{\tilde t}_{1j})^2 \right) 
O^L_{ij} O^R_{ij}   m_{\tilde{\chi}^+_j}m_{{\tilde \chi}^0_i}
\nonumber \\
&&            
+2 k^{\tilde t}_{1j}l^{\tilde t}_{1j}   
\left( (O^L_{ij})^2 + (O^R_{ij})^2 \right)  
m_b m_{\tilde{\chi}^+_j}
\left[
  3+2 (m_b^2+m^2_{{\tilde t}_1})/m_{W}^2
\right]
\nonumber \\
&&  
             +\left( (k^{\tilde t}_{1j})^2
  (O^R_{ij})^2 + (l^{\tilde t}_{1j})^2 (O^L_{ij})^2 \right)   
\left[
  6 m_b^2+3 m^2_{{\tilde \chi}^0_i}+2 m^2_{{\tilde t}_1}+2 m_{W}^2
\right.
\nonumber \\
&& 
\left.
  +(2 m_b^4+2 m_b^2 m^2_{{\tilde t}_1}+m^2_{{\tilde \chi}^0_i}
  m^2_{{\tilde t}_1}) /m_{W}^2
\right]
\nonumber \\
&&  
+\left( (k^{\tilde t}_{1j})^2(O^L_{ij})^2 + (l^{\tilde t}_{1j})^2 
(O^R_{ij})^2 \right)   m^2_{\tilde{\chi}^+_j}  
\left[
  1+(2 m_b^2+m^2_{{\tilde t}_1}) /m_{W}^2
\right]
\nonumber
\end{eqnarray}
\begin{eqnarray}
a_{ij6}&=& -
\left\{
  4  k^{\tilde t}_{1j}l^{\tilde t}_{1j}   
  \left( (O^L_{ij})^2 + (O^R_{ij})^2 \right) m_b m_{\tilde{\chi}^+_j}/m_{W}^2
\right.
\nonumber \\
&&  
+\left( (k^{\tilde t}_{1j})^2
  (O^R_{ij})^2 + (l^{\tilde t}_{1j})^2 (O^L_{ij})^2 \right)   
\left[
  4+(m^2_{{\tilde \chi}^0_i}+4 m_b^2+2 m^2_{{\tilde t}_1}) /m_{W}^2
\right]
\nonumber \\
&&
\left.
  +\left( (k^{\tilde t}_{1j})^2(O^L_{ij})^2 + 
(l^{\tilde t}_{1j})^2 (O^R_{ij})^2 \right)   m^2_{\tilde{\chi}^+_j}/m_{W}^2
\right\}
\nonumber
\end{eqnarray}
\begin{eqnarray}
       a_{ij7}&=& 2   \left( (k^{\tilde t}_{1j})^2
  (O^R_{ij})^2 + (l^{\tilde t}_{1j})^2 (O^L_{ij})^2 \right)   /m_{W}^2
\nonumber
\end{eqnarray}
\begin{eqnarray}
a_{ij8}&=& -
\left[
  2   k^{\tilde t}_{1j}l^{\tilde t}_{1j}   
  \left( (O^L_{ij})^2 + (O^R_{ij})^2 \right)   m_b m_{\tilde{\chi}^+_j}/m_{W}^2
\right.
\nonumber \\
&&
+ \left( (k^{\tilde t}_{1j})^2
  (O^R_{ij})^2 + (l^{\tilde t}_{1j})^2 (O^L_{ij})^2 \right)   
(2 + m_b^2 /m_{W}^2)
\nonumber \\
&&
\left.
  + \left( (k^{\tilde t}_{1j})^2(O^L_{ij})^2 + 
    (l^{\tilde t}_{1j})^2 (O^R_{ij})^2 \right)   
  m^2_{\tilde{\chi}^+_j}   /m_{W}^2 
\right]
\nonumber
\end{eqnarray}
\begin{eqnarray}
a_{ij9}&=&  \left( (k^{\tilde t}_{1j})^2
  (O^R_{ij})^2 + (l^{\tilde t}_{1j})^2 (O^L_{ij})^2 \right)   /m_{W}^2
\nonumber
\end{eqnarray}                
\begin{eqnarray}
 a_{i41}&=& 2\Big\{ 3\left(l^{\tilde t}_{11} l^{\tilde t}_{12} O^L_{i2}
  O^R_{i1} + k^{\tilde t}_{11} k^{\tilde t}_{12} 
O^L_{i1} O^R_{i2}\right)m_{\tilde{\chi}^+_j}m_{{\tilde \chi}^0_i}
(2 m_b^2+m^2_{{\tilde \chi}^0_i}+m_{W}^2)
\nonumber
\\ 
&&
+ 6\left(k^{\tilde t}_{11} l^{\tilde t}_{12}
O^L_{i2} O^R_{i1} +  k^{\tilde t}_{12} l^{\tilde t}_{11}
 O^L_{i1} O^R_{i2}\right)m_b m_{{\tilde \chi}^0_i}(m_b^2+m^2_{{\tilde \chi}^0_i}+m^2_{{\tilde t}_1}+m_{W}^2)
\nonumber
\\
&& 
+ 6\left(k^{\tilde t}_{12} l^{\tilde t}_{11}
 O^L_{i2} O^R_{i1} +  k^{\tilde t}_{11} l^{\tilde t}_{12}
 O^L_{i1} O^R_{i2}\right)m_b m_{{\tilde \chi}^0_i}m_{{\tilde \chi}^+_1} 
m_{{\tilde \chi}^+_2}
\nonumber
\\
&&
  + 3\left(k^{\tilde t}_{11} k^{\tilde t}_{12}
O^L_{i2} O^R_{i1} +l^{\tilde t}_{11} l^{\tilde t}_{12}
O^L_{i1} O^R_{i2}\right)m_{\tilde{\chi}^+_2} m_{{\tilde \chi}^0_i} 
(2 m_b^2+m^2_{{\tilde \chi}^0_i}+m_{W}^2)
\nonumber
\\
&&
  - \left(l^{\tilde t}_{11} l^{\tilde t}_{12}
 O^L_{i1} O^L_{i2} +  k^{\tilde t}_{11} k^{\tilde t}_{12} 
 O^R_{i1} O^R_{i2}\right)
 \Big[ (m^2_{{\tilde \chi}^0_i}+m_b^2) (m^2_{{\tilde \chi}^0_i}+
m^2_{{\tilde t}_1}+m_{W}^2+m_b^2)
\nonumber
\\
&&
  + m_b^2 (3 m_b^2+2 m^2_{{\tilde \chi}^0_i}+3 m^2_{{\tilde t}_1}) + 
2 m^2_{{\tilde t}_1} m_{W}^2
  + (m_b^2+m^2_{{\tilde t}_1}) (m_b^4+m^2_{{\tilde t}_1} 
(m^2_{{\tilde \chi}^0_i}+m_b^2)) /m_{W}^2 \Big]
\nonumber
\\
&&  
- \left(k^{\tilde t}_{11} l^{\tilde t}_{12}
  O^L_{i1} O^L_{i2} +  k^{\tilde t}_{12} l^{\tilde t}_{11}
  O^R_{i1} O^R_{i2}\right) m_b m_{\tilde{\chi}^+_j}
 \Big[3 m_b^2+2 m^2_{{\tilde \chi}^0_i}+3 m^2_{{\tilde t}_1}
\nonumber
\\
&&
  + (m_b^2+m^2_{{\tilde t}_1}) (m_b^2+m^2_{{\tilde t}_1}) /m_{W}^2 \Big]
\nonumber
\end{eqnarray}                
\begin{eqnarray}                
&&
  - \left(k^{\tilde t}_{12} l^{\tilde t}_{11}
O^L_{i1} O^L_{i2} +  k^{\tilde t}_{11} l^{\tilde t}_{12}
O^R_{i1} O^R_{i2}\right) m_b m_{\tilde{\chi}^+_2}
\Big[3 m_b^2+2 m^2_{{\tilde \chi}^0_i}+3 m^2_{{\tilde t}_1}
\nonumber
\\
&&
 + (m_b^2+m^2_{{\tilde t}_1}) (m_b^2+m^2_{{\tilde t}_1}) /m_{W}^2 \Big]
\nonumber
\\
&&
 - \left(k^{\tilde t}_{11} k^{\tilde t}_{12}
  O^L_{i1} O^L_{i2} +  l^{\tilde t}_{11} l^{\tilde t}_{12} 
  O^R_{i1}
  O^R_{i2}\right) m_{{\tilde \chi}^+_1} m_{{\tilde \chi}^+_2}
\Big[2 m_b^2+m^2_{{\tilde \chi}^0_i}+m^2_{{\tilde t}_1}
\nonumber
\\
&&
 + (m_b^2+m^2_{{\tilde t}_1}) m_b^2/m_W^2 \Big]\Big\}
\nonumber
\end{eqnarray}
\begin{eqnarray}
 a_{i42}&=& 2\Big\{-3 \left(l^{\tilde t}_{11} l^{\tilde t}_{12} O^L_{i2}
  O^R_{i1} + k^{\tilde t}_{11} k^{\tilde t}_{12} 
 O^L_{i1} O^R_{i2}\right) m_{\tilde{\chi}^+_j}m_{{\tilde \chi}^0_i}
\nonumber
\\
&&
 -6 \left(k^{\tilde t}_{11} l^{\tilde t}_{12}
 O^L_{i2} O^R_{i1} +  k^{\tilde t}_{12} l^{\tilde t}_{11}
 O^L_{i1} O^R_{i2}\right) m_b m_{{\tilde \chi}^0_i}
\nonumber
\\
&&
 -3 \left(k^{\tilde t}_{11} k^{\tilde t}_{12}
O^L_{i2} O^R_{i1} + l^{\tilde t}_{11} l^{\tilde t}_{12}
O^L_{i1} O^R_{i2}\right) m_{\tilde{\chi}^+_2} m_{{\tilde \chi}^0_i}
\nonumber
\\
&&
 + \left(l^{\tilde t}_{11} l^{\tilde t}_{12}
 O^L_{i1} O^L_{i2} +  k^{\tilde t}_{11} k^{\tilde t}_{12} 
 O^R_{i1} O^R_{i2}\right)
 \Big[6 m_b^2 + 2 m^2_{{\tilde \chi}^0_i} + 2 m^2_{{\tilde t}_1} + m_{W}^2
\nonumber
\\
&&
  + ((2 m_b^2+m^2_{{\tilde \chi}^0_i}) (m_b^2+2 m^2_{{\tilde t}_1})+
m_b^4+m^4_{{\tilde t}_1}) /m_{W}^2 \Big]
\nonumber
\\
&&
 + \left(k^{\tilde t}_{11} l^{\tilde t}_{12}
  O^L_{i1} O^L_{i2} +  k^{\tilde t}_{12} l^{\tilde t}_{11}
  O^R_{i1} O^R_{i2}\right) m_b m_{\tilde{\chi}^+_j} \Big[3+
2 (m_b^2+m^2_{{\tilde t}_1}) /m_{W}^2\Big]
\nonumber
\\
&&
 + \left(k^{\tilde t}_{12} l^{\tilde t}_{11}
O^L_{i1} O^L_{i2} +  k^{\tilde t}_{11} l^{\tilde t}_{12}
O^R_{i1} O^R_{i2}\right) m_b m_{\tilde{\chi}^+_2} \Big[3+
2 (m_b^2+m^2_{{\tilde t}_1}) /m_{W}^2\Big]
\nonumber
\\
&&
 + \left(k^{\tilde t}_{11} k^{\tilde t}_{12}
  O^L_{i1} O^L_{i2} +  l^{\tilde t}_{11} l^{\tilde t}_{12} 
  O^R_{i1}
  O^R_{i2}\right) m_{{\tilde \chi}^+_1} m_{{\tilde \chi}^+_2} 
(2+m_b^2/m_W^2) \Big\}
\nonumber
\end{eqnarray}
\begin{eqnarray}
 a_{i43}&=& -2\Big\{ \left(l^{\tilde t}_{11} l^{\tilde t}_{12}
 O^L_{i1} O^L_{i2} +  k^{\tilde t}_{11} k^{\tilde t}_{12} 
 O^R_{i1} O^R_{i2}\right)
 \Big[2+(m^2_{{\tilde \chi}^0_i}+3 m_b^2+2 m^2_{{\tilde t}_1}) /m_{W}^2\Big]
\nonumber
\\
&&
  + \left(k^{\tilde t}_{11} l^{\tilde t}_{12}
  O^L_{i1} O^L_{i2} +  k^{\tilde t}_{12} l^{\tilde t}_{11}
  O^R_{i1} O^R_{i2}\right) m_b m_{\tilde{\chi}^+_j} /m_{W}^2
\nonumber
\\
&&
  + \left(k^{\tilde t}_{12} l^{\tilde t}_{11}
O^L_{i1} O^L_{i2} +  k^{\tilde t}_{11} l^{\tilde t}_{12}
O^R_{i1} O^R_{i2}\right) m_b m_{\tilde{\chi}^+_2} /m_{W}^2 \Big\}
\nonumber
\end{eqnarray}
\begin{eqnarray}
 a_{i44}&=& 2\left(l^{\tilde t}_{11} l^{\tilde t}_{12}
 O^L_{i1} O^L_{i2} +  k^{\tilde t}_{11} k^{\tilde t}_{12} 
 O^R_{i1} O^R_{i2}\right) /m_{W}^2
\nonumber
\end{eqnarray}
\begin{eqnarray}
 a_{i45}&=& 2\Big\{- 3 \left(l^{\tilde t}_{11} l^{\tilde t}_{12} O^L_{i2}
  O^R_{i1} + k^{\tilde t}_{11} k^{\tilde t}_{12} 
 O^L_{i1} O^R_{i2}\right) m_{\tilde{\chi}^+_j}m_{{\tilde \chi}^0_i}
\nonumber
\\
&&
  - 6 \left(k^{\tilde t}_{11} l^{\tilde t}_{12}
 O^L_{i2} O^R_{i1} +  k^{\tilde t}_{12} l^{\tilde t}_{11}
 O^L_{i1} O^R_{i2}\right) m_b m_{{\tilde \chi}^0_i}
\nonumber
\\
&&
  - 3 \left(k^{\tilde t}_{11} k^{\tilde t}_{12}
O^L_{i2} O^R_{i1} + l^{\tilde t}_{11} l^{\tilde t}_{12}
O^L_{i1} O^R_{i2}\right) m_{\tilde{\chi}^+_2} m_{{\tilde \chi}^0_i}
\nonumber
\\
&&
  + \left(l^{\tilde t}_{11} l^{\tilde t}_{12}
 O^L_{i1} O^L_{i2} +  k^{\tilde t}_{11} k^{\tilde t}_{12} 
 O^R_{i1} O^R_{i2}\right)
 \Big[6 m_b^2+3 m^2_{{\tilde \chi}^0_i}+2 m^2_{{\tilde t}_1}+2 m_{W}^2
\nonumber
\\
&&
+(2m_b^4+m^2_{{\tilde t}_1}(m^2_{{\tilde \chi}^0_i}+2 m_b^2))/m_{W}^2\Big]
\nonumber
\\
&&
  + \left(k^{\tilde t}_{11} l^{\tilde t}_{12}
  O^L_{i1} O^L_{i2} +  k^{\tilde t}_{12} l^{\tilde t}_{11}
  O^R_{i1} O^R_{i2}\right) m_b m_{\tilde{\chi}^+_j}
 \Big[3+2 (m_b^2+m^2_{{\tilde t}_1}) /m_{W}^2 \Big]
\nonumber
\\
&&
  + \left(k^{\tilde t}_{12} l^{\tilde t}_{11}
O^L_{i1} O^L_{i2} +  k^{\tilde t}_{11} l^{\tilde t}_{12}
O^R_{i1} O^R_{i2}\right) m_b m_{\tilde{\chi}^+_2}
 \Big[3+2 (m_b^2+m^2_{{\tilde t}_1}) /m_{W}^2 \Big]
\nonumber
\\
&&
  + \left(k^{\tilde t}_{11} k^{\tilde t}_{12}
  O^L_{i1} O^L_{i2} +  l^{\tilde t}_{11} l^{\tilde t}_{12} 
  O^R_{i1}
  O^R_{i2}\right) m_{{\tilde \chi}^+_1} m_{{\tilde \chi}^+_2}
 \Big[1+(2 m_b^2+m^2_{{\tilde t}_1}) /m_{W}^2 \Big]\Big\}
\nonumber
\end{eqnarray}                
\begin{eqnarray}                
 a_{i46}&=& -2\Big\{ \left(l^{\tilde t}_{11} l^{\tilde t}_{12}
 O^L_{i1} O^L_{i2} +  k^{\tilde t}_{11} k^{\tilde t}_{12} 
 O^R_{i1} O^R_{i2}\right)
 \Big[4+(m^2_{{\tilde \chi}^0_i}+4 m_b^2+2 m^2_{{\tilde t}_1}) /m_{W}^2\Big]
\nonumber
\\
&&
  + 2 \left(k^{\tilde t}_{11} l^{\tilde t}_{12}
  O^L_{i1} O^L_{i2} +  k^{\tilde t}_{12} l^{\tilde t}_{11}
  O^R_{i1} O^R_{i2}\right) m_b m_{\tilde{\chi}^+_j} /m_{W}^2
\nonumber
\\
&&
  + 2 \left(k^{\tilde t}_{12} l^{\tilde t}_{11}
O^L_{i1} O^L_{i2} +  k^{\tilde t}_{11} l^{\tilde t}_{12}
O^R_{i1} O^R_{i2}\right) m_b m_{\tilde{\chi}^+_2} /m_{W}^2
\nonumber
\\
&&
  + \left(k^{\tilde t}_{11} k^{\tilde t}_{12}
  O^L_{i1} O^L_{i2} +  l^{\tilde t}_{11} l^{\tilde t}_{12} 
  O^R_{i1}
  O^R_{i2}\right)m_{{\tilde \chi}^+_1} m_{{\tilde \chi}^+_2}/m_{W}^2\Big\}
\nonumber
\end{eqnarray}
\begin{eqnarray}
 a_{i47}&=& 4 \left(l^{\tilde t}_{11} l^{\tilde t}_{12}
 O^L_{i1} O^L_{i2} +  k^{\tilde t}_{11} k^{\tilde t}_{12} 
 O^R_{i1} O^R_{i2}\right) /m_{W}^2
\nonumber
\end{eqnarray}
\begin{eqnarray}
 a_{i48}&=& -2\Big[ \left(l^{\tilde t}_{11} l^{\tilde t}_{12}
 O^L_{i1} O^L_{i2} +  k^{\tilde t}_{11} k^{\tilde t}_{12} 
 O^R_{i1} O^R_{i2}\right) (2+m_b^2/m_W^2)
\nonumber
\\
&&
  + \left(k^{\tilde t}_{11} l^{\tilde t}_{12}
  O^L_{i1} O^L_{i2} +  k^{\tilde t}_{12} l^{\tilde t}_{11}
  O^R_{i1} O^R_{i2}\right) m_b m_{\tilde{\chi}^+_j} /m_{W}^2
\nonumber
\\
&&
  + \left(k^{\tilde t}_{12} l^{\tilde t}_{11}
O^L_{i1} O^L_{i2} +  k^{\tilde t}_{11} l^{\tilde t}_{12}
O^R_{i1} O^R_{i2}\right) m_b m_{\tilde{\chi}^+_2} /m_{W}^2
\nonumber
\\
&&
  + \left(k^{\tilde t}_{11} k^{\tilde t}_{12}
  O^L_{i1} O^L_{i2} +  l^{\tilde t}_{11} l^{\tilde t}_{12} 
  O^R_{i1}
  O^R_{i2}\right)m_{{\tilde \chi}^+_1} m_{{\tilde \chi}^+_2}/m_{W}^2\Big]
\nonumber
\end{eqnarray}
\begin{eqnarray}
 a_{i49}&=& 2\left(l^{\tilde t}_{11} l^{\tilde t}_{12}
 O^L_{i1} O^L_{i2} +  k^{\tilde t}_{11} k^{\tilde t}_{12} 
 O^R_{i1} O^R_{i2}\right) /m_{W}^2
\nonumber
\end{eqnarray}
\noindent
The coefficients $a_{i5l}$ are obtained from $a_{i4l}$ by replacing:
$ l^{\tilde t}_{12}\to l^{\tilde t}_{13}
\quad k^{\tilde t}_{12}\to k^{\tilde t}_{13}
\quad  O^L_{i2}\to  O^L_{i3}
\quad  O^R_{i2}\to  O^R_{i3}
\quad m_{\tilde{\chi}^+_2}\to m_{\tilde{\chi}^+_3}$
and the coefficients $a_{i6l}$ are obtained from $a_{i4l}$ by replacing:
$l^{\tilde t}_{11}\to l^{\tilde t}_{13}
\quad k^{\tilde t}_{11}\to k^{\tilde t}_{13}
\quad  O^L_{i1}\to  O^L_{i3}
\quad  O^R_{i1}\to  O^R_{i3}
\quad m_{\tilde{\chi}^+_1}\to m_{\tilde{\chi}^+_3}$
%
\begin{eqnarray}
b_{ij1}&=& \sqrt{2}\Big\{ a^{\tilde{t}}_{1i}l^{\tilde t}_{1j}O^L_{ij} 
\Big[2 (m^2_{{\tilde \chi}^0_i}-m^2_{{\tilde t}_1}) 
(m_{W}^2+m^2_{{\tilde \chi}^0_i}+m_b^2)
\nonumber
\\
&&
 -m_b^2 m^2_{{\tilde t}_1} (1+(m^2_{{\tilde \chi}^0_i}-
m_b^2-m^2_{{\tilde t}_1}) /m_{W}^2) \Big]
 + 3 a^{\tilde{t}}_{1i} k^{\tilde t}_{1j} O^L_{ij} 
m_b m_{\tilde{\chi}^+_j} (m^2_{{\tilde \chi}^0_i}-m^2_{{\tilde t}_1})
\nonumber
\\
&&
 + 3 b^{\tilde{t}}_{1i} l^{\tilde t}_{1j} O^L_{ij} m_{{\tilde \chi}^0_i}m_t (2 m_b^2+m^2_{{\tilde \chi}^0_i}+m_{W}^2)
 + 6  b^{\tilde{t}}_{1i} k^{\tilde t}_{1j} O^L_{ij} m_{{\tilde \chi}^0_i}m_t m_b m_{\tilde{\chi}^+_j}
\nonumber
\\
&&
 + a^{\tilde{t}}_{1i} k^{\tilde t}_{1j} O^R_{ij}
 m_{\tilde{\chi}^+_j}m_{{\tilde \chi}^0_i} 
\Big[2 m_{W}^2-m_b^2 (1+m_b^2/m_W^2)\Big]
\nonumber
\\
&&
 + a^{\tilde{t}}_{1i} k^{\tilde t}_{1j} O^R_{ij} m_b m_{{\tilde \chi}^0_i} 
 (3 m_{W}^2+m^2_{{\tilde t}_1} (2-m_b^2/m_W^2)-m_b^2 (2+m_b^2/m_W^2)-
m^2_{{\tilde \chi}^0_i})
\nonumber
\\
&&
 - b^{\tilde{t}}_{1i} l^{\tilde t}_{1j} O^R_{ij} m_{\tilde{\chi}^+_j}m_t 
 (m^2_{{\tilde \chi}^0_i}+m^2_{{\tilde t}_1} (1+m_b^2/m_W^2)+
m_b^2 (2+m_b^2/m_W^2))
\nonumber
\\
&&
 - b^{\tilde{t}}_{1i} k^{\tilde t}_{1j} O^R_{ij} m_b m_t
 (2 m^2_{{\tilde \chi}^0_i} + (m_b^2+m^2_{{\tilde t}_1}) 
(3+(m_b^2+m^2_{{\tilde t}_1}) /m_{W}^2)) \Big\}
\nonumber
\end{eqnarray}
\begin{eqnarray}
b_{ij2}&=& \sqrt{2}\Big[ a^{\tilde{t}}_{1i}l^{\tilde t}_{1j}O^L_{ij} 
(m^2_{{\tilde t}_1}-m^2_{{\tilde \chi}^0_i}) (2-m_b^2/m_W^2)
 - 3  b^{\tilde{t}}_{1i} l^{\tilde t}_{1j} O^L_{ij} m_{{\tilde \chi}^0_i}m_t
\nonumber
\\
&& 
+ a^{\tilde{t}}_{1i} k^{\tilde t}_{1j} O^R_{ij} 
m_b m_{{\tilde \chi}^0_i} (m_b^2/m_W^2-1)
 + b^{\tilde{t}}_{1i} l^{\tilde t}_{1j} O^R_{ij} 
m_{\tilde{\chi}^+_j}m_t (2+m_b^2/m_W^2)
\nonumber
\\
&&
 + b^{\tilde{t}}_{1i} k^{\tilde t}_{1j} O^R_{ij} m_b m_t 
(3+2 (m_b^2+m^2_{{\tilde t}_1}) /m_{W}^2)  \Big]
\nonumber
\end{eqnarray}
\begin{eqnarray}
b_{ij3}&=& -\sqrt{2}\, b^{\tilde{t}}_{1i} k^{\tilde t}_{1j} O^R_{ij} 
m_b m_t /m_{W}^2 
\nonumber
\end{eqnarray}                
\begin{eqnarray}                
b_{ij4}&=& \sqrt{2}\Big\{ a^{\tilde{t}}_{1i}l^{\tilde t}_{1j}O^L_{ij} 
\Big[4 m_b^2+2 m_{W}^2+m^2_{{\tilde \chi}^0_i} m^2_{{\tilde t}_1} /m_{W}^2
 +m^2_{{\tilde t}_1} (2-m_b^2/m_W^2) \Big]
 + 3 a^{\tilde{t}}_{1i} k^{\tilde t}_{1j} O^L_{ij} m_b m_{\tilde{\chi}^+_j}
\nonumber
\\
&& 
 - 3 b^{\tilde{t}}_{1i} l^{\tilde t}_{1j} O^L_{ij} m_{{\tilde \chi}^0_i}m_t
 - a^{\tilde{t}}_{1i} k^{\tilde t}_{1j} O^R_{ij} 
m_{\tilde{\chi}^+_j}m_{{\tilde \chi}^0_i} (1-2 m_b^2/m_W^2)
\nonumber
\\
&&
 + a^{\tilde{t}}_{1i} k^{\tilde t}_{1j} O^R_{ij} 
m_b m_{{\tilde \chi}^0_i} (1+2 m_b^2/m_W^2+m^2_{{\tilde t}_1} /m_{W}^2)
\nonumber
\\
&&
 + b^{\tilde{t}}_{1i} l^{\tilde t}_{1j} O^R_{ij} 
m_{\tilde{\chi}^+_j}m_t (1+2 m_b^2/m_W^2+m^2_{{\tilde t}_1} /m_{W}^2)
\nonumber
\\
&&
 + b^{\tilde{t}}_{1i} k^{\tilde t}_{1j} O^R_{ij} m_b m_t 
\Big[3+2 (m_b^2+m^2_{{\tilde t}_1}) /m_{W}^2\Big] \Big\}
\nonumber
\end{eqnarray}
\begin{eqnarray}
b_{ij5}&=& -\sqrt{2}\Big\{ a^{\tilde{t}}_{1i}l^{\tilde t}_{1j}O^L_{ij} 
\Big[3+m_b^2/m_W^2+(m^2_{{\tilde t}_1}+m^2_{{\tilde \chi}^0_i}) /m_{W}^2\Big]
  + a^{\tilde{t}}_{1i} k^{\tilde t}_{1j} O^R_{ij} m
_b m_{{\tilde \chi}^0_i} /m_{W}^2
\nonumber
\\
&& 
  + b^{\tilde{t}}_{1i} l^{\tilde t}_{1j} O^R_{ij} 
m_{\tilde{\chi}^+_j}m_t /m_{W}^2
  + 2 b^{\tilde{t}}_{1i} k^{\tilde t}_{1j} O^R_{ij} m_b m_t /m_{W}^2 \Big\}
\nonumber
\end{eqnarray}
\begin{eqnarray}
b_{ij6}&=& \sqrt{2}\,a^{\tilde{t}}_{1i}l^{\tilde t}_{1j}O^L_{ij} /m_{W}^2
\nonumber\\
b_{ij7}&=& -\sqrt{2}\Big[ 2 a^{\tilde{t}}_{1i}l^{\tilde t}_{1j}O^L_{ij} + 
a^{\tilde{t}}_{1i} k^{\tilde t}_{1j} O^R_{ij} 
m_{\tilde{\chi}^+_j}m_{{\tilde \chi}^0_i} /m_{W}^2
  + a^{\tilde{t}}_{1i} k^{\tilde t}_{1j} O^R_{ij} 
m_b m_{{\tilde \chi}^0_i} /m_{W}^2 
\nonumber
\\
&& 
+ b^{\tilde{t}}_{1i} l^{\tilde t}_{1j} O^R_{ij} 
m_{\tilde{\chi}^+_j}m_t /m_{W}^2
  + b^{\tilde{t}}_{1i} k^{\tilde t}_{1j} O^R_{ij} m_b m_t /m_{W}^2 \Big]
\nonumber
\end{eqnarray}
\begin{eqnarray}
b_{ij8}&=&\sqrt{2}\, a^{\tilde{t}}_{1i}l^{\tilde t}_{1j}O^L_{ij} /m_{W}^2
\nonumber
\end{eqnarray}
\begin{eqnarray}
 c_{ijk1}&=& 2 \, A^{W}_{{\tilde t}_1 {\tilde b}_k} \Big\{ 
b^{\tilde b}_{ki} l^{\tilde t}_{1j} O^L_{ij} m_b m_{{\tilde \chi}^0_i}
  \big[m_b^2-m^2_{{\tilde t}_1}-2 m^2_{{\tilde \chi}^0_i}+
m^2_{{\tilde t}_1} (m_b^2/m_W^2+m^2_{{\tilde t}_1}/m_W^2)\Big]
\nonumber
\\
&&
 + a^{\tilde b}_{ki} l^{\tilde t}_{1j} O^L_{ij}
  \big[m_b^2 m_{W}^2+m^2_{{\tilde \chi}^0_i} m^2_{{\tilde t}_1} 
(m^2_{{\tilde t}_1}/m_W^2-1)+
 m_b^2 m^2_{{\tilde t}_1} (m_b^2/m_W^2-2)
\nonumber
\\
&&
+m_b^4-2 m_b^2 m^2_{{\tilde \chi}^0_i}+
 m_b^2 m^2_{{\tilde t}_1} m^2_{{\tilde t}_1}/m_W^2 \Big]
\nonumber
\\
&&
 + b^{\tilde b}_{ki} k^{\tilde t}_{1j} O^L_{ij} m_{\tilde{\chi}^+_j}
m_{{\tilde \chi}^0_i}
  (m_b^2-m_{W}^2-2 m^2_{{\tilde \chi}^0_i}+
m^2_{{\tilde t}_1} (1+m_b^2/m_W^2))
\nonumber
\\
&&
 + a^{\tilde b}_{ki} k^{\tilde t}_{1j} O^L_{ij} m_b m_{\tilde{\chi}^+_j}
  \big[m_b^2-2 m^2_{{\tilde \chi}^0_i}+m^2_{{\tilde t}_1} 
(m_b^2/m_W^2+m^2_{{\tilde t}_1}/m_W^2-1)\Big] \Big\}
\nonumber
\end{eqnarray}
\begin{eqnarray}
 c_{ijk2}&=& - 2 \, A^{W}_{{\tilde t}_1 {\tilde b}_k} \Big\{ 
b^{\tilde b}_{ki} l^{\tilde t}_{1j} O^L_{ij} m_b m_{{\tilde \chi}^0_i} 
(1+m_b^2/m_W^2+2 m^2_{{\tilde t}_1}/m_W^2)
\nonumber
\\
&&
+ a^{\tilde b}_{ki} l^{\tilde t}_{1j} O^L_{ij} \Big[m^2_{{\tilde \chi}^0_i} 
(1+2 m^2_{{\tilde t}_1}/m_W^2)+m^2_{{\tilde t}_1} m^2_{{\tilde t}_1}/m_W^2
  +m_{W}^2
\nonumber
\\
&&
+m_b^2 (3+m_b^2/m_W^2+3 m^2_{{\tilde t}_1}/m_W^2) \Big]
+ b^{\tilde b}_{ki} k^{\tilde t}_{1j} O^L_{ij} 
m_{\tilde{\chi}^+_j}m_{{\tilde \chi}^0_i} (m_b^2/m_W^2-1)
\nonumber
\\
&&
+ a^{\tilde b}_{ki} k^{\tilde t}_{1j} O^L_{ij} m_b 
m_{\tilde{\chi}^+_j} (1+m_b^2/m_W^2+2 m^2_{{\tilde t}_1}/m_W^2)  \Big\}
\nonumber
\end{eqnarray}
\begin{eqnarray}
 c_{ijk3}&=&2 \, A^{W}_{{\tilde t}_1 {\tilde b}_k} \Big[ 
b^{\tilde b}_{ki} l^{\tilde t}_{1j} O^L_{ij} m_b 
m_{{\tilde \chi}^0_i}/m_W^2
 + a^{\tilde b}_{ki} l^{\tilde t}_{1j} O^L_{ij} (2+2 m_b^2/m_W^2
+m^2_{{\tilde \chi}^0_i}/m_W^2+2 m^2_{{\tilde t}_1}/m_W^2)
\nonumber
\\
&&
 + a^{\tilde b}_{ki} k^{\tilde t}_{1j} O^L_{ij} m_b 
m_{\tilde{\chi}^+_j}/m_W^2  \Big]
\nonumber
\end{eqnarray}
\begin{eqnarray}
 c_{ijk4}&=& -2 \, A^{W}_{{\tilde t}_1 {\tilde b}_k}  a^{\tilde b}_{ki} 
l^{\tilde t}_{1j} O^L_{ij} /m_{W}^2
\nonumber\\
 c_{ijk5}&=&2 \, A^{W}_{{\tilde t}_1 {\tilde b}_k} \Big( 
b^{\tilde b}_{ki} k^{\tilde t}_{1j} O^L_{ij} m_{\tilde{\chi}^+_j}
m_{{\tilde \chi}^0_i} 
+ a^{\tilde b}_{ki} k^{\tilde t}_{1j} O^L_{ij} m_b m_{\tilde{\chi}^+_j}
 + b^{\tilde b}_{ki} l^{\tilde t}_{1j} O^L_{ij} m_b m_{{\tilde \chi}^0_i} 
\nonumber
\\
&&
+ a^{\tilde b}_{ki} l^{\tilde t}_{1j} O^L_{ij} m_b^2 \Big)
\Big(1-m^2_{{\tilde t}_1}/m_W^2\Big)
\nonumber
\end{eqnarray}                
\begin{eqnarray}                
 c_{ijk6}&=&2 \, A^{W}_{{\tilde t}_1 {\tilde b}_k} \Big[
b^{\tilde b}_{ki} l^{\tilde t}_{1j} O^L_{ij} m_b 
m_{{\tilde \chi}^0_i}/m_W^2
+ a^{\tilde b}_{ki} l^{\tilde t}_{1j} O^L_{ij} (1+
m_b^2/m_W^2+m^2_{{\tilde t}_1}/m_W^2)
\nonumber
\\
&&
 + b^{\tilde b}_{ki} k^{\tilde t}_{1j} O^L_{ij} 
m_{\tilde{\chi}^+_j}m_{{\tilde \chi}^0_i}/m_W^2
 + a^{\tilde b}_{ki} k^{\tilde t}_{1j} O^L_{ij} 
m_b m_{\tilde{\chi}^+_j}/m_W^2 \Big]
\nonumber
\end{eqnarray}
\begin{eqnarray}
 c_{ijk7}&=& c_{ijk4}
\nonumber
\end{eqnarray}
\begin{eqnarray}
d_{i1}&=& \hbox{$\frac12$}\Big\{(a^{\tilde{t}}_{1i})^2 
m^2_{{\tilde \chi}^0_i}-m^2_{{\tilde t}_1}) 
\Big[2 m_{W}^2-m_b^2 (1+m_b^2/m_W^2)\Big]
\nonumber
\\
&&
  - (b^{\tilde{t}}_{1i})^2 m_t^2 \Big[m^2_{{\tilde \chi}^0_i}
+m^2_{{\tilde t}_1}+m_b^2 (2+m_b^2/m_W^2+m^2_{{\tilde t}_1} /m_{W}^2)\Big]
\nonumber
\\
&&
  + 2 a^{\tilde{t}}_{1i} b^{\tilde{t}}_{1i} m_{{\tilde \chi}^0_i}m_t 
\Big[2 m_{W}^2-m_b^2 (1+m_b^2/m_W^2)\Big] \Big\}
\nonumber
\end{eqnarray}
\begin{eqnarray}
  d_{i2}&=& \hbox{$\frac12$}(b^{\tilde{t}}_{1i})^2 m_t^2 (2+m_b^2/m_W^2)
\nonumber\\
  d_{i3}&=& \hbox{$\frac12$}\Big\{ (a^{\tilde{t}}_{1i})^2 
\Big[m_b^2+2 m^2_{{\tilde t}_1}+2 m_{W}^2+(2 m^2_{{\tilde \chi}^0_i}-
m^2_{{\tilde t}_1}) m_b^2/m_W^2\Big]
\nonumber
\\
&&
  + (b^{\tilde{t}}_{1i})^2 m_t^2  \Big[1+(2 m_b^2
+m^2_{{\tilde t}_1})/m_{W}^2\Big]
  - 2 a^{\tilde{t}}_{1i} b^{\tilde{t}}_{1i} m_{{\tilde \chi}^0_i}
m_t (1-2 m_b^2/m_W^2) \Big\}
\nonumber
\end{eqnarray}
\begin{eqnarray}
  d_{i4}&=& - \hbox{$\frac12$}\Big[a^{\tilde{t}}_{1i})^2 (2+m_b^2/m_W^2) 
- (b^{\tilde{t}}_{1i})^2 m_t^2 /m_{W}^2\Big]
\nonumber
\end{eqnarray}
\begin{eqnarray}
  d_{i5}&=& -\hbox{$\frac12$}\Big[ (a^{\tilde{t}}_{1i})^2 
(2+m^2_{{\tilde \chi}^0_i} /m_{W}^2) + 2 a^{\tilde{t}}_{1i} 
b^{\tilde{t}}_{1i} m_{{\tilde \chi}^0_i}m_t /m_{W}^2
 + (b^{\tilde{t}}_{1i})^2 m_t^2 /m_{W}^2 \Big] 
\nonumber
\end{eqnarray}
\begin{eqnarray}
  d_{i6}&=& \hbox{$\frac12$}(a^{\tilde{t}}_{1i})^2 /m_{W}^2
\nonumber
\end{eqnarray}
\begin{eqnarray}
e_{ik1}&=& \sqrt2 \, A^{W}_{{\tilde t}_1 {\tilde b}_k}\Big\{ 
a^{\tilde{b}}_{ki} a^{\tilde{t}}_{1i} \Big[
m_b^2 (m^2_{{\tilde \chi}^0_i}+m^2_{{\tilde t}_1})+2 
m^2_{{\tilde \chi}^0_i} (m^2_{{\tilde t}_1}-m^2_{{\tilde \chi}^0_i}-m_{W}^2)
\nonumber
\\
&& 
+(m^2_{{\tilde \chi}^0_i}-m^2_{{\tilde t}_1}) 
m_b^2 m^2_{{\tilde t}_1}/m_W^2 \Big]
\nonumber
\\
&&
+ (a^{\tilde{b}}_{ki} b^{\tilde{t}}_{1i} m_{{\tilde \chi}^0_i}m_t + 
b^{\tilde{b}}_{ki} a^{\tilde{t}}_{1i} m_b m_{{\tilde \chi}^0_i})
(m_b^2+m^2_{{\tilde t}_1}-2 m^2_{{\tilde \chi}^0_i}-
m_{W}^2+m_b^2 m^2_{{\tilde t}_1}/m_W^2)
\nonumber
\\
&&
+ b^{\tilde{b}}_{ki} b^{\tilde{t}}_{1i} m_b m_t \Big[
m_b^2-2 m^2_{{\tilde \chi}^0_i}-m^2_{{\tilde t}_1}+
(m_b^2+m^2_{{\tilde t}_1}) m^2_{{\tilde t}_1}/m_W^2\Big] \Big\}
\nonumber
\end{eqnarray}
\begin{eqnarray}
e_{ik2}&=& \sqrt2 \, A^{W}_{{\tilde t}_1 {\tilde b}_k}\Big[ 
a^{\tilde{b}}_{ki} a^{\tilde{t}}_{1i} (m^2_{{\tilde \chi}^0_i}-
m^2_{{\tilde t}_1}) (2-m_b^2/m_W^2)
\nonumber
\\
&&
+ (a^{\tilde{b}}_{ki} b^{\tilde{t}}_{1i} m_{{\tilde \chi}^0_i}m_t 
+ b^{\tilde{b}}_{ki} a^{\tilde{t}}_{1i} m_b 
m_{{\tilde \chi}^0_i}) (1-m_b^2/m_W^2)
\nonumber
\\
&&
- b^{\tilde{b}}_{ki} b^{\tilde{t}}_{1i} m_b 
m_t (1+m_b^2/m_W^2+2 m^2_{{\tilde t}_1}/m_W^2) \Big]
\nonumber
\end{eqnarray}
\begin{eqnarray}
e_{ik3}&=& \sqrt2 \, A^{W}_{{\tilde t}_1 {\tilde b}_k}b^{\tilde{b}}_{ki} 
b^{\tilde{t}}_{1i} m_b m_t /m_{W}^2
\nonumber\\
e_{ik4}&=& \sqrt2 \, A^{W}_{{\tilde t}_1 {\tilde b}_k}\Big( 
a^{\tilde{b}}_{ki} a^{\tilde{t}}_{1i} m^2_{{\tilde \chi}^0_i} 
+ a^{\tilde{b}}_{ki} b^{\tilde{t}}_{1i} m_{{\tilde \chi}^0_i}m_t 
+ b^{\tilde{b}}_{ki} a^{\tilde{t}}_{1i} m_b m_{{\tilde \chi}^0_i}
\nonumber
\\
&&
 + b^{\tilde{b}}_{ki} b^{\tilde{t}}_{1i} m_b m_t\Big)\Big(
1-m^2_{{\tilde t}_1}/m_W^2\Big)
\nonumber
\end{eqnarray}
\begin{eqnarray}
e_{ik5}&=& \sqrt2 \, A^{W}_{{\tilde t}_1 {\tilde b}_k}\Big[ 
a^{\tilde{b}}_{ki} a^{\tilde{t}}_{1i} 
(1+m^2_{{\tilde \chi}^0_i}/m_W^2+m^2_{{\tilde t}_1}/m_W^2)
+ a^{\tilde{b}}_{ki} b^{\tilde{t}}_{1i}  
m_{{\tilde \chi}^0_i}m_t/m_W^2 
\nonumber
\\
&&
+ b^{\tilde{b}}_{ki} a^{\tilde{t}}_{1i} m_b m_{{\tilde \chi}^0_i}/m_W^2
+ b^{\tilde{b}}_{ki} b^{\tilde{t}}_{1i} m_b m_t/m_W^2\Big]
\nonumber
\end{eqnarray}
\begin{eqnarray}
e_{ik6}&=& -\sqrt2 \, A^{W}_{{\tilde t}_1 {\tilde b}_k}a^{\tilde{b}}_{ki} 
a^{\tilde{t}}_{1i} /m_{W}^2
\nonumber
\end{eqnarray}
\begin{eqnarray}
f_{ik1} &=& - \sqrt{2}\,(A^{W}_{{\tilde t}_1 {\tilde b}_k})^2
         \left[ \left((a^{\tilde{b}}_{ki})^2+(b^{\tilde{b}}_{ki})^2 \right)
                 \left( m^2_b + m^2_{{\tilde \chi}^0_i} \right)
     + 4 \, a^{\tilde{b}}_{ki} b^{\tilde{b}}_{ki} m_b m_{{\tilde \chi}^0_i}
       \right]  \, ,\nonumber
\end{eqnarray}
\begin{eqnarray}
f_{ik2} &=& \sqrt{2}\,(A^{W}_{{\tilde t}_1 {\tilde b}_k})^2
        \left((a^{\tilde{b}}_{ki})^2+(b^{\tilde{b}}_{ki})^2 \right) \, ,
\nonumber
\end{eqnarray}
\begin{eqnarray}
f_{i31} &=& - 4 \, A^{W}_{{\tilde t}_1 {\tilde b}_1}
                  A^{W}_{{\tilde t}_1 {\tilde b}_2}
 \left[ \left( a^{\tilde{b}}_{1i} a^{\tilde{b}}_{2i}
     + b^{\tilde{b}}_{1i} b^{\tilde{b}}_{2i} \right)
               \left( m^2_b + m^2_{{\tilde \chi}^0_i} \right)
     + 2  \left( a^{\tilde{b}}_{1i} b^{\tilde{b}}_{2i}
         + b^{\tilde{b}}_{1i} a^{\tilde{b}}_{2i} \right)
         m_b m_{{\tilde \chi}^0_i} \right]  \, ,\nonumber 
\end{eqnarray}
\begin{eqnarray}
f_{i31} &=& 4 \, A^{W}_{{\tilde t}_1 {\tilde b}_1}
        A^{W}_{{\tilde t}_1 {\tilde b}_2}
       \left( a^{\tilde{b}}_{1i} a^{\tilde{b}}_{2i}
              + b^{\tilde{b}}_{1i} b^{\tilde{b}}_{2i} \right) \, .
\nonumber
\end{eqnarray}

\subsection{The width $\Gamma({\tilde t}_1 \to S_k^+ \, b \, {\tilde \chi}^0_i)$}

\noindent
The decay width is given by
\begin{eqnarray}
 \Gamma({\tilde t}_1 \to S_k^+ \, b \, {\tilde \chi}^0_i)  &=& \nonumber \\
 & & \hspace{-30mm} 
   =  \frac{\alpha^2}{16 \, \pi m^3_{{\tilde t}_1} \sin^4 \theta_W}
  \int\limits^{(m_{{\tilde t}_1}-m_{S_k^+})^2}_{
           (m_b + m_{{\tilde \chi}^0_i})^2} \hspace{-8mm}
     d \, s \,
   \left( G_{{\tilde \chi}^+ {\tilde \chi}^+} +
   G_{{\tilde \chi}^+ t} +
   G_{{\tilde \chi}^+ {\tilde b}} +
   G_{t t} +
   G_{t {\tilde b}} +
   G_{{\tilde b} {\tilde b}} \right) 
\nonumber
\end{eqnarray}
with
\begin{eqnarray} 
   G_{{\tilde \chi}^+ {\tilde \chi}^+} &=&
       \sum^3_{j=1} \Big[  ( a_{ijk1} + a_{ijk2} s )
          J^0_t(m^2_{{\tilde t}_1} + m^2_{S_k^+} + m^2_b
               + m^2_{{\tilde \chi}^0_i} - m^2_{{\tilde \chi}^+_j} - s
             ,\Gamma_{{\tilde \chi}^+_j} m_{{\tilde \chi}^+_j}) \nonumber \\
    & & \hspace{5mm} + ( a_{ijk3} + a_{i4jk} s )
       J^1_t(m^2_{{\tilde t}_1} + m^2_{S_k^+} + m^2_b
            + m^2_{{\tilde \chi}^0_i} - m^2_{{\tilde \chi}^+_j} - s
             ,\Gamma_{{\tilde \chi}^+_j} m_{{\tilde \chi}^+_j}) \nonumber \\
    & & \hspace{5mm} + \, a_{ijk4} \,
          J^2_t(m^2_{{\tilde t}_1} + m^2_{S_k^+} + m^2_b
               + m^2_{{\tilde \chi}^0_i} - m^2_{{\tilde \chi}^+_j} - s
       ,\Gamma_{{\tilde \chi}^+_j} m_{{\tilde \chi}^+_j}) \Big] \nonumber \\
    & &  +  ( a_{i4k1} + a_{i4k2} s )
          J^0_{tt}(m^2_{{\tilde t}_1} + m^2_{S_k^+} + m^2_b
               + m^2_{{\tilde \chi}^0_i} - m^2_{{\tilde \chi}^+_1} - s
             ,\Gamma_{{\tilde \chi}^+_1} m_{{\tilde \chi}^+_1}  \nonumber \\
       & & \hspace{33mm} ,m^2_{{\tilde t}_1} + m^2_{S_k^+} + m^2_b
               + m^2_{{\tilde \chi}^0_i} - m^2_{{\tilde \chi}^+_2} - s
             ,\Gamma_{{\tilde \chi}^+_2} m_{{\tilde \chi}^+_2}) \nonumber \\
    & &  + ( a_{i4k3} + a_{i4k4} s )
       J^1_{tt}(m^2_{{\tilde t}_1} + m^2_{S_k^+} + m^2_b
               + m^2_{{\tilde \chi}^0_i} - m^2_{{\tilde \chi}^+_1} - s
             ,\Gamma_{{\tilde \chi}^+_1} m_{{\tilde \chi}^+_1}  \nonumber \\
       & & \hspace{33mm} ,m^2_{{\tilde t}_1} + m^2_{S_k^+} + m^2_b
               + m^2_{{\tilde \chi}^0_i} - m^2_{{\tilde \chi}^+_2} - s
             ,\Gamma_{{\tilde \chi}^+_2} m_{{\tilde \chi}^+_2}) \nonumber \\
    & &  + \, a_{i4k4} \,
       J^2_{tt}(m^2_{{\tilde t}_1} + m^2_{S_k^+} + m^2_b
               + m^2_{{\tilde \chi}^0_i} - m^2_{{\tilde \chi}^+_1} - s
             ,\Gamma_{{\tilde \chi}^+_1} m_{{\tilde \chi}^+_1}  \nonumber \\
       & & \hspace{17mm} ,m^2_{{\tilde t}_1} + m^2_{S_k^+} + m^2_b
               + m^2_{{\tilde \chi}^0_i} - m^2_{{\tilde \chi}^+_2} - s
             ,\Gamma_{{\tilde \chi}^+_2} m_{{\tilde \chi}^+_2}) 
             \nonumber\\
    & &  +  ( a_{i5k1} + a_{i5k2} s )
          J^0_{tt}(m^2_{{\tilde t}_1} + m^2_{S_k^+} + m^2_b
               + m^2_{{\tilde \chi}^0_i} - m^2_{{\tilde \chi}^+_1} - s
             ,\Gamma_{{\tilde \chi}^+_1} m_{{\tilde \chi}^+_1}  \nonumber \\
       & & \hspace{33mm} ,m^2_{{\tilde t}_1} + m^2_{S_k^+} + m^2_b
               + m^2_{{\tilde \chi}^0_i} - m^2_{{\tilde \chi}^+_3} - s
             ,\Gamma_{{\tilde \chi}^+_3} m_{{\tilde \chi}^+_3}) \nonumber \\
    & &  + ( a_{i5k3} + a_{i5k4} s )
       J^1_{tt}(m^2_{{\tilde t}_1} + m^2_{S_k^+} + m^2_b
               + m^2_{{\tilde \chi}^0_i} - m^2_{{\tilde \chi}^+_1} - s
             ,\Gamma_{{\tilde \chi}^+_1} m_{{\tilde \chi}^+_1}  \nonumber \\
       & & \hspace{33mm} ,m^2_{{\tilde t}_1} + m^2_{S_k^+} + m^2_b
               + m^2_{{\tilde \chi}^0_i} - m^2_{{\tilde \chi}^+_3} - s
             ,\Gamma_{{\tilde \chi}^+_3} m_{{\tilde \chi}^+_3}) \nonumber \\
    & &  + \, a_{i5k4} \,
       J^2_{tt}(m^2_{{\tilde t}_1} + m^2_{S_k^+} + m^2_b
               + m^2_{{\tilde \chi}^0_i} - m^2_{{\tilde \chi}^+_1} - s
             ,\Gamma_{{\tilde \chi}^+_1} m_{{\tilde \chi}^+_1}  \nonumber \\
       & & \hspace{17mm} ,m^2_{{\tilde t}_1} + m^2_{S_k^+} + m^2_b
               + m^2_{{\tilde \chi}^0_i} - m^2_{{\tilde \chi}^+_3} - s
             ,\Gamma_{{\tilde \chi}^+_3} m_{{\tilde \chi}^+_3})\nonumber 
\end{eqnarray}                
\begin{eqnarray}                
    & &  +  ( a_{i6k1} + a_{i6k2} s )
          J^0_{tt}(m^2_{{\tilde t}_1} + m^2_{S_k^+} + m^2_b
               + m^2_{{\tilde \chi}^0_i} - m^2_{{\tilde \chi}^+_3} - s
             ,\Gamma_{{\tilde \chi}^+_3} m_{{\tilde \chi}^+_3}  \nonumber \\
       & & \hspace{33mm} ,m^2_{{\tilde t}_1} + m^2_{S_k^+} + m^2_b
               + m^2_{{\tilde \chi}^0_i} - m^2_{{\tilde \chi}^+_2} - s
             ,\Gamma_{{\tilde \chi}^+_2} m_{{\tilde \chi}^+_2}) \nonumber \\
    & &  + ( a_{i6k3} + a_{i6k4} s )
       J^1_{tt}(m^2_{{\tilde t}_1} + m^2_{S_k^+} + m^2_b
               + m^2_{{\tilde \chi}^0_i} - m^2_{{\tilde \chi}^+_3} - s
             ,\Gamma_{{\tilde \chi}^+_3} m_{{\tilde \chi}^+_3}  \nonumber \\
       & & \hspace{33mm} ,m^2_{{\tilde t}_1} + m^2_{S_k^+} + m^2_b
               + m^2_{{\tilde \chi}^0_i} - m^2_{{\tilde \chi}^+_2} - s
             ,\Gamma_{{\tilde \chi}^+_2} m_{{\tilde \chi}^+_2}) \nonumber \\
    & &  + \, a_{i6k4} \,
       J^2_{tt}(m^2_{{\tilde t}_1} + m^2_{S_k^+} + m^2_b
               + m^2_{{\tilde \chi}^0_i} - m^2_{{\tilde \chi}^+_3} - s
             ,\Gamma_{{\tilde \chi}^+_3} m_{{\tilde \chi}^+_3}  \nonumber \\
       & & \hspace{17mm} ,m^2_{{\tilde t}_1} + m^2_{S_k^+} + m^2_b
               + m^2_{{\tilde \chi}^0_i} - m^2_{{\tilde \chi}^+_2} - s
             ,\Gamma_{{\tilde \chi}^+_2} m_{{\tilde \chi}^+_2}) \, ,
\nonumber
\end{eqnarray}
\begin{eqnarray}
   G_{{\tilde \chi}^+ t} &=&
       \sum^3_{j=1} \Big[ ( b_{ijk1} + b_{ijk2} s ) 
 J^0_{tt}(m^2_{{\tilde t}_1} + m^2_{S_k^+} + m^2_b
               + m^2_{{\tilde \chi}^0_i} - m^2_{{\tilde \chi}^+_j} - s
       ,-\Gamma_{{\tilde \chi}^+_j} m_{{\tilde \chi}^+_j},m^2_t,
        \Gamma_t m_t) \nonumber \\
    & & \hspace{5mm} + ( b_{ijk3} + b_{ijk4} s )
 J^1_{tt}(m^2_{{\tilde t}_1} + m^2_{S_k^+} + m^2_b
               + m^2_{{\tilde \chi}^0_i} - m^2_{{\tilde \chi}^+_j} - s
       ,-\Gamma_{{\tilde \chi}^+_j} m_{{\tilde \chi}^+_j} ,m^2_t,
        \Gamma_t m_t) \nonumber \\
    & & \hspace{5mm} + \, b_{ijk4} \,
        J^2_{tt}(m^2_{{\tilde t}_1} + m^2_{S_k^+} + m^2_b
               + m^2_{{\tilde \chi}^0_i} - m^2_{{\tilde \chi}^+_j} - s
       ,-\Gamma_{{\tilde \chi}^+_j} m_{{\tilde \chi}^+_j} ,m^2_t,
         \Gamma_t m_t)  \Big] \, ,
\nonumber
\end{eqnarray}
\begin{eqnarray}
   G_{{\tilde \chi}^+ {\tilde b}} &=&
       \sum^2_{l=1} \Big[
          ( c_{ijkl1} + c_{ijkl2} s ) \nonumber \\
    & & \hspace{7mm} * J^0_{st}(
       m^2_{{\tilde b}_l}, \Gamma_ {{\tilde b}_l} m_{{\tilde b}_l}
       ,m^2_{{\tilde t}_1} + m^2_{S_k^+} + m^2_b
               + m^2_{{\tilde \chi}^0_i} - m^2_{{\tilde \chi}^+_j} - s
       ,-\Gamma_{{\tilde \chi}^+_j} m_{{\tilde \chi}^+_j}) \nonumber \\
    & &  + \, c_{ijkl3} \, J^1_{st}(
       m^2_{{\tilde b}_l}, \Gamma_ {{\tilde b}_l} m_{{\tilde b}_l}
       ,m^2_{{\tilde t}_1} + m^2_{S_k^+} + m^2_b
               + m^2_{{\tilde \chi}^0_i} - m^2_{{\tilde \chi}^+_j} - s
       ,-\Gamma_{{\tilde \chi}^+_j} m_{{\tilde \chi}^+_j}) \Big] \, ,
\nonumber
\end{eqnarray}
\begin{equation*}
   G_{tt} =
          ( d_{ik1} + d_{ik2} s) J^0_t(m^2_t, \Gamma_t m_t) +
          ( d_{ik3} + d_{ik4} s) J^1_t(m^2_t, \Gamma_t m_t)
 + \, d_{ik4} \, J^2_t(m^2_t, \Gamma_t m_t) \, ,
\end{equation*}
\begin{eqnarray}
   G_{t {\tilde b}} &=& \sum^2_{l=1} \Big[
   ( e_{ikl1} + e_{ikl2} s )
   J^0_{st}(m^2_{{\tilde b}_l},
  \Gamma_ {{\tilde b}_l} m_{{\tilde b}_l},m^2_t, \Gamma_t m_t)
 + \, e_{ikl3} \, J^1_{st}(m^2_{{\tilde b}_l},
    \Gamma_ {{\tilde b}_l} m_{{\tilde b}_l},m^2_t, \Gamma_t m_t)
   \Big] \, ,
\nonumber
\end{eqnarray}
\begin{eqnarray}
G_{{\tilde b} {\tilde b}} &=&
 \frac{\sqrt{\lambda(s,m^2_{{\tilde t}_1},m^2_{S_k^+})
                 \lambda(s,m^2_{{\tilde \chi}^0_i},m^2_b)}}{s} \nonumber \\
  & & \hspace{-12mm} \times \left\{ \sum^2_{l=1}
      \frac{(f_{ikl1} + f_{ikl2} s)}
   {(s-m^2_{{\tilde b}_l})^2 + \Gamma^2_{{\tilde b}_l} m^2_{{\tilde b}_l}}
 + \mbox{Re} \left[ \frac{(f_{ik31} + f_{ik32} s)}
      { (s-m^2_{{\tilde b}_1} + i \Gamma_{{\tilde b}_1} m_{{\tilde b}_1})
         (s-m^2_{{\tilde b}_2} - i \Gamma_{{\tilde b}_2} m_{{\tilde b}_2})}
          \right]  \right\} \, .
\nonumber
\end{eqnarray}
Their integration range is given by
\begin{eqnarray}
t_{max \atop min} &=&
 \frac{m^2_{{\tilde t}_1} + m^2_b + m^2_{S_k^+}
        + m^2_{{\tilde \chi}^0_i} -s}2
       - \frac{(m^2_{{\tilde t}_1}-m^2_{S_k^+})
               (m^2_{{\tilde \chi}^0_i}-m^2_b)}{2 s} \nonumber \\
  & &  \pm  \frac{\sqrt{\lambda(s,m^2_{{\tilde t}_1},m^2_{S_k^+})
                  \lambda(s,m^2_{{\tilde \chi}^0_i},m^2_b)}}{2 s} \, ,
\nonumber
\end{eqnarray}
where $s = (p_{{\tilde t}_1} - p_{S_k^+})^2$
 and $t = (p_{{\tilde t}_1} - p_{t})^2$
are the usual Mandelstam variables.
Note, that $-\Gamma_{{\tilde \chi}^+_j} m_{{\tilde \chi}^+_j}$
appears in the entries of the integrals 
$G_{{\tilde \chi}^+ {\tilde b}_j}$ and
$G_{{\tilde \chi}^+ t}$ because the chargino is exchanged 
 in the $u$-channel in our convention.
The coefficients are given by (no sum upon repeated index):
\begin{eqnarray}
a_{ijk1} &=& - 4 \, k^{\tilde t}_{1j} l^{\tilde t}_{1j} 
       Q^L_{ijk}{}' Q^R_{ijk}{}' m_b m_{{\tilde \chi}^0_i}
    \left( m^2_b +  m^2_{\tilde{\chi}^+_j}
   + m^2_{{\tilde \chi}^0_i} + m^2_{{\tilde t}_1} 
   + m^2_{S_k^+} \right) \nonumber \\
  & &  - 2 \, Q^L_{ijk}{}' Q^R_{ijk}{}'
    \left( (k^{\tilde t}_{1j})^2 + (l^{\tilde t}_{1j})^2 \right)
    m_{{\tilde \chi}^0_i} m_{\tilde{\chi}^+_j}
    \left( 2 \, m^2_b + m^2_{{\tilde \chi}^0_i}
      + m^2_{S_k^+} \right)  \nonumber \\
  & &  - 2 \, k^{\tilde t}_{1j} l^{\tilde t}_{1j} 
        \left( (Q^L_{ijk}{}')^2 + (Q^R_{ijk}{}')^2 \right)
     m_b m_{\tilde{\chi}^+_j} 
        \left( m^2_b + 2 \, m^2_{{\tilde \chi}^0_i}
            + m^2_{{\tilde t}_1} \right)  \nonumber \\
  & &  - \left( (k^{\tilde t}_{1j})^2
  (Q^R_{ijk}{}')^2 + (l^{\tilde t}_{1j})^2 (Q^L_{ijk}{}')^2 \right) \nonumber \\
   & & \hspace{5mm} \times \left[ \left( m^2_b+m^2_{{\tilde \chi}^0_i} \right)^2
    + \left( m^2_b+m^2_{S_k^+} \right)
 \left(m^2_{{\tilde \chi}^0_i}+m^2_{{\tilde t}_1}\right) \right] \nonumber \\
  & &  - \left( (k^{\tilde t}_{1j})^2
       (Q^L_{ijk}{}')^2 + (l^{\tilde t}_{1j})^2 (Q^R_{ijk}{}')^2 \right)
      m^2_{\tilde{\chi}^+_j}
           \left(m^2_b+m^2_{{\tilde \chi}^0_i} \right) \, ,\nonumber
\end{eqnarray}
\begin{eqnarray}
a_{ijk2} &=& 4 \, k^{\tilde t}_{1j} l^{\tilde t}_{1j}
       Q^L_{ijk}{}' Q^R_{ijk}{}' m_b m_{{\tilde \chi}^0_i}
  + \left( (k^{\tilde t}_{1j})^2 
       (Q^R_{ijk}{}')^2 + (l^{\tilde t}_{1j})^2 (Q^L_{ijk}{}')^2 \right)
           \left( m^2_b+m^2_{{\tilde \chi}^0_i} \right) \nonumber \\
  & & + 2 \, k^{\tilde t}_{1j} 
       l^{\tilde t}_{1j} \left( (Q^L_{ijk}{}')^2 + (Q^R_{ijk}{}')^2 \right)
                 m_b m_{\tilde{\chi}^+_j}
        + 2 \, Q^L_{ijk}{}' Q^R_{ijk}{}'
        \left( (k^{\tilde t}_{1j})^2 + (l^{\tilde t}_{1j})^2 \right)
         m_{{\tilde \chi}^0_i} m_{\tilde{\chi}^+_j}  \nonumber \\
  & &  + \left( (k^{\tilde t}_{1j})^2
      (Q^L_{ijk}{}')^2 + (l^{\tilde t}_{1j})^2 (Q^R_{ijk}{}')^2 \right)
              m^2_{\tilde{\chi}^+_j}   \, , \nonumber
\end{eqnarray}
\begin{eqnarray}
a_{ijk3} &=& 4 \, k^{\tilde t}_{1j} l^{\tilde t}_{1j}
           Q^L_{ijk}{}' Q^R_{ijk}{}' m_b m_{{\tilde \chi}^0_i}
      + 2 \, Q^L_{ijk}{}' Q^R_{ijk}{}' \left( (k^{\tilde t}_{1j})^2
              + (l^{\tilde t}_{1j})^2 \right)
              m_{{\tilde \chi}^0_i} m_{\tilde{\chi}^+_j} \nonumber \\
  & & + \left( (k^{\tilde t}_{1j})^2 (Q^R_{ijk}{}')^2
             + (l^{\tilde t}_{1j})^2 (Q^L_{ijk}{}')^2 \right)
         \left(2 \, m^2_b+2 \, m^2_{{\tilde \chi}^0_i}
             +m^2_{S_k^+}+m^2_{{\tilde t}_1}\right)  \nonumber \\
  & & + 2 \, k^{\tilde t}_{1j} l^{\tilde t}_{1j}
          \left( (Q^L_{ijk}{}')^2 + (Q^R_{ijk}{}')^2 \right)
               m_b m_{\tilde{\chi}^+_j} \, , \nonumber
\end{eqnarray}
\begin{eqnarray}
a_{ijk4} &=& - (k^{\tilde t}_{1j})^2 (Q^R_{ijk}{}')^2 -
                 (l^{\tilde t}_{1j})^2 (Q^L_{ijk}{}')^2  \, ,
\nonumber \\
a_{i4k1} &=& - 2  \left(l^{\tilde t}_{11} l^{\tilde t}_{12} Q^L_{i2k}{}'
            Q^R_{i1k}{}' + k^{\tilde t}_{11} k^{\tilde t}_{12} 
        Q^L_{i1k}{}' Q^R_{i2k}{}'\right) m_{{\tilde \chi}^0_i}
             m_{\tilde{\chi}^+_1}
         \left( m^2_{{\tilde \chi}^0_i} + m^2_{S_k^+} + 2 \, m^2_b \right)
        \nonumber \\
  & &   - 4 \, \left(k^{\tilde t}_{11} l^{\tilde t}_{12}
        Q^L_{i2k}{}' Q^R_{i1k}{}' +  k^{\tilde t}_{12} l^{\tilde t}_{11}
           Q^L_{i1k}{}' Q^R_{i2k}{}'\right) m_b m_{{\tilde \chi}^0_i}
      \left( m^2_{{\tilde \chi}^0_i} + m^2_{S_k^+} + m^2_b
            + m^2_{{\tilde t}_1} \right) \nonumber \\
  & &  - 4 \, \left(k^{\tilde t}_{12} l^{\tilde t}_{11}
           Q^L_{i2k}{}' Q^R_{i1k}{}' +  k^{\tilde t}_{11} l^{\tilde t}_{12}
           Q^L_{i1k}{}' Q^R_{i2k}{}'\right) m_b m_{{\tilde \chi}^0_i}
         m_{\tilde{\chi}^+_1} m_{\tilde{\chi}^+_2} \nonumber \\
  & &  - 2 \, \left(k^{\tilde t}_{11} k^{\tilde t}_{12}
     Q^L_{i2k}{}' Q^R_{i1k}{}' +   l^{\tilde t}_{11} l^{\tilde t}_{12}
     Q^L_{i1k}{}' Q^R_{i2k}{}'\right) m_{{\tilde \chi}^0_i}
      m_{\tilde{\chi}^+_2}
      \left( m^2_{{\tilde \chi}^0_i} + m^2_{S_k^+} + 2 \, m^2_b \right) 
      \nonumber \\
  & &  - 2 \, \left(l^{\tilde t}_{11} l^{\tilde t}_{12}
        Q^L_{i1k}{}' Q^L_{i2k}{}' +  k^{\tilde t}_{11} k^{\tilde t}_{12} 
        Q^R_{i1k}{}' Q^R_{i2k}{}'\right)     \nonumber \\
  & & \hspace{5mm} \times \left[ \left(m^2_b+m^2_{{\tilde \chi}^0_i}\right)^2
      + \left(m^2_{{\tilde \chi}^0_i}+m^2_{{\tilde t}_1}\right)
        \left(m^2_b+m^2_{S_k^+}\right) \right] \nonumber \\
  & &  - 2 \, \left(k^{\tilde t}_{11} l^{\tilde t}_{12}
       Q^L_{i1k}{}' Q^L_{i2k}{}' +  k^{\tilde t}_{12} l^{\tilde t}_{11}
       Q^R_{i1k}{}' Q^R_{i2k}{}'\right) m_b m_{\tilde{\chi}^+_1}
     \left(m^2_b+m^2_{{\tilde t}_1}+2 \, m^2_{{\tilde \chi}^0_i} \right)
     \nonumber \\
  & &  - 2 \, \left(k^{\tilde t}_{12} l^{\tilde t}_{11}
     Q^L_{i1k}{}' Q^L_{i2k}{}' +  k^{\tilde t}_{11} l^{\tilde t}_{12}
     Q^R_{i1k}{}' Q^R_{i2k}{}'\right) m_b m_{\tilde{\chi}^+_2}
     \left(m^2_b+m^2_{{\tilde t}_1}+2 \, m^2_{{\tilde \chi}^0_i} \right)
     \nonumber \\
  & &  - 2 \, \left(k^{\tilde t}_{11} k^{\tilde t}_{12}
    Q^L_{i1k}{}' Q^L_{i2k}{}' +  l^{\tilde t}_{11} l^{\tilde t}_{12} 
    Q^R_{i1k}{}'
    Q^R_{i2k}{}'\right) m_{\tilde{\chi}^+_1} m_{\tilde{\chi}^+_2}
    \left(m^2_{{\tilde \chi}^0_i} + m^2_b \right)  \, , \nonumber
\end{eqnarray}                
\begin{eqnarray}                
a_{i4k2} &=& 2 \left(l^{\tilde t}_{11} l^{\tilde t}_{12} Q^L_{i2k}{}'
        Q^R_{i1k}{}' +   k^{\tilde t}_{11} k^{\tilde t}_{12} Q^L_{i1k}{}'
        Q^R_{i2k}{}'\right) m_{{\tilde \chi}^0_i} m_{\tilde{\chi}^+_1}
      \nonumber \\
  & &  + 4 \, \left(k^{\tilde t}_{11} l^{\tilde t}_{12}
      Q^L_{i2k}{}' Q^R_{i1k}{}' +  k^{\tilde t}_{12} l^{\tilde t}_{11}
      Q^L_{i1k}{}' Q^R_{i2k}{}'\right) m_b m_{{\tilde \chi}^0_i} \nonumber \\
  & &  + 2 \, \left(k^{\tilde t}_{11} k^{\tilde t}_{12} 
       Q^L_{i2k}{}' Q^R_{i1k}{}' +  l^{\tilde t}_{11} l^{\tilde t}_{12}
       Q^L_{i1k}{}' Q^R_{i2k}{}'\right) m_{{\tilde \chi}^0_i}
       m_{\tilde{\chi}^+_2} \nonumber \\
  & &  + 2 \, \left(l^{\tilde t}_{11} l^{\tilde t}_{12}
      Q^L_{i1k}{}' Q^L_{i2k}{}' +   k^{\tilde t}_{11} k^{\tilde t}_{12} 
      Q^R_{i1k}{}' Q^R_{i2k}{}'\right)
    \left( m^2_b+m^2_{{\tilde \chi}^0_i} \right)  \nonumber \\
  & &  + 2 \, \left(k^{\tilde t}_{11} l^{\tilde t}_{12}
      Q^L_{i1k}{}' Q^L_{i2k}{}' +  k^{\tilde t}_{12} l^{\tilde t}_{11}
      Q^R_{i1k}{}' Q^R_{i2k}{}'\right) m_b m_{\tilde{\chi}^+_1} 
      \nonumber \\
  & &  + 2 \, \left(k^{\tilde t}_{12} l^{\tilde t}_{11}
        Q^L_{i1k}{}' Q^L_{i2k}{}' +    k^{\tilde t}_{11} l^{\tilde t}_{12}
        Q^R_{i1k}{}' Q^R_{i2k}{}'\right) m_b m_{\tilde{\chi}^+_2}
      \nonumber \\
  & &  + 2 \, \left(k^{\tilde t}_{11} k^{\tilde t}_{12}
         Q^L_{i1k}{}' Q^L_{i2k}{}' +   l^{\tilde t}_{11} l^{\tilde t}_{12} 
         Q^R_{i1k}{}' Q^R_{i2k}{}'\right) m_{\tilde{\chi}^+_1}
          m_{\tilde{\chi}^+_2}  \, ,\nonumber
\end{eqnarray}
\begin{eqnarray}
a_{i4k3} &=& 2 \left(l^{\tilde t}_{11} l^{\tilde t}_{12} Q^L_{i2k}{}'
          Q^R_{i1k}{}' +  k^{\tilde t}_{11} k^{\tilde t}_{12} Q^L_{i1k}{}'
          Q^R_{i2k}{}'\right) m_{{\tilde \chi}^0_i}
          m_{\tilde{\chi}^+_1} \nonumber \\
  & &  + 4 \, \left(k^{\tilde t}_{11} l^{\tilde t}_{12}
        Q^L_{i2k}{}' Q^R_{i1k}{}' +  k^{\tilde t}_{12} l^{\tilde t}_{11}
        Q^L_{i1k}{}' Q^R_{i2k}{}'\right) m_b m_{{\tilde \chi}^0_i} 
\nonumber \\
  & &  + 2 \, \left(k^{\tilde t}_{11} k^{\tilde t}_{12}
        Q^L_{i2k}{}' Q^R_{i1k}{}' +   l^{\tilde t}_{11} l^{\tilde t}_{12}
        Q^L_{i1k}{}' Q^R_{i2k}{}'\right) m_{{\tilde \chi}^0_i}
        m_{\tilde{\chi}^+_2} \nonumber \\
  & &  + 2 \, \left(l^{\tilde t}_{11} l^{\tilde t}_{12}
        Q^L_{i1k}{}' Q^L_{i2k}{}' +  k^{\tilde t}_{11} k^{\tilde t}_{12}
        Q^R_{i1k}{}' Q^R_{i2k}{}'\right)
   \left(2 \, m^2_b+2 \, m^2_{{\tilde \chi}^0_i}
         +m^2_{S_k^+}+m^2_{{\tilde t}_1} \right) \nonumber \\
  & &  + 2 \, \left(k^{\tilde t}_{11} l^{\tilde t}_{12}
      Q^L_{i1k}{}' Q^L_{i2k}{}' +  k^{\tilde t}_{12} l^{\tilde t}_{11}
     Q^R_{i1k}{}' Q^R_{i2k}{}'\right) m_b m_{\tilde{\chi}^+_1} \nonumber \\
  & &  + 2 \, \left(k^{\tilde t}_{12} l^{\tilde t}_{11}
      Q^L_{i1k}{}' Q^L_{i2k}{}' +  k^{\tilde t}_{11} l^{\tilde t}_{12}
      Q^R_{i1k}{}' Q^R_{i2k}{}'\right) m_b m_{\tilde{\chi}^+_2} \, ,
    \nonumber
\end{eqnarray}
\begin{eqnarray}
a_{i4k4} &=& - 2 \, \left(l^{\tilde t}_{11} l^{\tilde t}_{12} Q^L_{i1k}{}'
            Q^L_{i2k}{}' +  k^{\tilde t}_{11} k^{\tilde t}_{12} Q^R_{i1k}{}'
            Q^R_{i2k}{}'\right) \, ,
\nonumber
\end{eqnarray}
\noindent
The coefficients $a_{i5kl}$ are obtained from $a_{i4kl}$ by replacing:
$ l^{\tilde t}_{12}\to l^{\tilde t}_{13}
\quad k^{\tilde t}_{12}\to k^{\tilde t}_{13}
\quad  Q^L_{i2k}{}'\to  Q^L_{i3k}{}'
\quad  Q^R_{i2k}{}'\to  Q^R_{i3k}{}'
\quad m_{\tilde{\chi}^+_2}\to m_{\tilde{\chi}^+_3}$
and the coefficients $a_{i6kl}$ are obtained from $a_{i4kl}$ by replacing:
$a_{i4kl}\to a_{i6kl}:\quad l^{\tilde t}_{11}\to l^{\tilde t}_{13}
\quad k^{\tilde t}_{11}\to k^{\tilde t}_{13}
\quad  Q^L_{i1k}{}'\to  Q^L_{i3k}{}'
\quad  Q^R_{i1k}{}'\to  Q^R_{i3k}{}'
\quad m_{\tilde{\chi}^+_1}\to m_{\tilde{\chi}^+_3}$
\begin{eqnarray}
b_{ijk1} &=& \frac{2\sqrt2 \,}{g} \Bigg\{
       b^{\tilde{t}}_{1i} k^{\tilde t}_{1j} Q^L_{ijk}{}'
       m_{\tilde{\chi}^+_j} m_b m_t
       \left[ \left( m^2_{{\tilde t}_1}-m^2_{{\tilde \chi}^0_i} \right)
       \frac{R^{S^\pm}_{k2}}{v_2}  - \left( m^2_b + 
       m^2_{{\tilde \chi}^0_i} \right)
       \frac{R^{S^\pm}_{k1}}{v_1} \right] \nonumber \\
  & &  + b^{\tilde{t}}_{1i} l^{\tilde t}_{1j} Q^L_{ijk}{}' m_t
       \left[ \left( m^2_{S_k^+} m^2_{{\tilde t}_1} - m^2_b
              m^2_{{\tilde \chi}^0_i} \right) \frac{R^{S^\pm}_{k2}}{v_2}
            - m^2_b  \left( m^2_b + m^2_{{\tilde t}_1} + 2 \,
             m^2_{{\tilde \chi}^0_i} \right)
             \frac{R^{S^\pm}_{k1}}{v_1} \right] \nonumber \\
  & &  + a^{\tilde{t}}_{1i} l^{\tilde t}_{1j} Q^L_{ijk}{}'
           m_{{\tilde \chi}^0_i} \nonumber \\
  & & \hspace{2mm} \times \left[ m^2_b
          \left( m^2_{S_k^+}+m^2_{{\tilde t}_1}-m^2_{{\tilde \chi}^0_i}
                  -m^2_b \right) \frac{R^{S^\pm}_{k1}}{v_1}
          - m^2_t \left( 2 \, m^2_b +m^2_{{\tilde \chi}^0_i}
                          +m^2_{S_k^+} \right)
          \frac{R^{S^\pm}_{k2}}{v_2} \right] \nonumber \\
  & &  + a^{\tilde{t}}_{1i} k^{\tilde t}_{1j} Q^L_{ijk}{}'
            m_{\tilde{\chi}^+_j} m_b m_{{\tilde \chi}^0_i}
       \left[ \left(m^2_{S_k^+}-m^2_b \right) \frac{R^{S^\pm}_{k1}}{v_1}
          - 2 \, m^2_t \frac{R^{S^\pm}_{k2}}{v_2} \right] \nonumber 
\end{eqnarray}
\begin{eqnarray}
  & &  + b^{\tilde{t}}_{1i} k^{\tilde t}_{1j} Q^R_{ijk}{}'
                m_{{\tilde \chi}^0_i} m_b m_t \nonumber \\
  & &  \hspace{2mm}\times \left[ \left(m^2_{{\tilde t}_1}+m^2_{S_k^+}-m^2_b
                -m^2_{{\tilde \chi}^0_i} \right) 
       \frac{R^{S^\pm}_{k2}}{v_2}
         - \left( m^2_{S_k^+}+2 \, m^2_b +m^2_{{\tilde \chi}^0_i} \right)
             \frac{R^{S^\pm}_{k1}}{v_1} \right] \nonumber \\
  & &  + b^{\tilde{t}}_{1i} l^{\tilde t}_{1j} Q^R_{ijk}{}'
                m_t m_{{\tilde \chi}^0_i} m_{\tilde{\chi}^+_j}
      \left[ \left(m^2_{S_k^+}-m^2_b \right) \frac{R^{S^\pm}_{k2}}{v_2}
         - 2 \, m^2_b  \frac{R^{S^\pm}_{k1}}{v_1} \right] \nonumber \\
  & &  - a^{\tilde{t}}_{1i} k^{\tilde t}_{1j} Q^R_{ijk}{}'
            m_b \nonumber \\
  & & \hspace{2mm} \times   \left[ \left(m^2_b m^2_{{\tilde \chi}^0_i}
           - m^2_{{\tilde t}_1} m^2_{S_k^+} \right) 
     \frac{R^{S^\pm}_{k1}}{v_1}
         + m^2_t \left( m^2_b +m^2_{{\tilde t}_1}+2 \,
             m^2_{{\tilde \chi}^0_i} \right)
          \frac{R^{S^\pm}_{k2}}{v_2} \right]  \nonumber \\
  & &  - a^{\tilde{t}}_{1i} l^{\tilde t}_{1j} Q^R_{ijk}{}'
              m_{\tilde{\chi}^+_j}
      \left[ m^2_b  \left(m^2_{{\tilde \chi}^0_i}-m^2_{{\tilde t}_1} \right)
           \frac{R^{S^\pm}_{k1}}{v_1}
         + m^2_t \left(m^2_b +m^2_{{\tilde \chi}^0_i} \right) 
          \frac{R^{S^\pm}_{k2}}{v_2}
           \right] \Bigg\}   \, , \nonumber
\end{eqnarray}
\begin{eqnarray}
b_{ijk2} &=& \frac{2\sqrt2 \,}{g} \Bigg\{ b^{\tilde{t}}_{1i}
                k^{\tilde t}_{1j} Q^L_{ijk}{}'
              m_{\tilde{\chi}^+_j} m_b m_t \frac{R^{S^\pm}_{k1}}{v_1}
      + b^{\tilde{t}}_{1i} l^{\tilde t}_{1j} Q^L_{ijk}{}' m^2_b  m_t 
        \frac{R^{S^\pm}_{k1}}{v_1}
              \nonumber \\
  & &  + a^{\tilde{t}}_{1i} l^{\tilde t}_{1j} Q^L_{ijk}{}'
              m^2_t m_{{\tilde \chi}^0_i} \frac{R^{S^\pm}_{k2}}{v_2}
              + b^{\tilde{t}}_{1i} k^{\tilde t}_{1j} Q^R_{ijk}{}'
            m_{{\tilde \chi}^0_i} m_b m_t \frac{R^{S^\pm}_{k1}}{v_1} 
\nonumber \\
  & &  + a^{\tilde{t}}_{1i} l^{\tilde t}_{1j} Q^R_{ijk}{}'
             m_{\tilde{\chi}^+_j} m^2_t \frac{R^{S^\pm}_{k2}}{v_2}
               + a^{\tilde{t}}_{1i} k^{\tilde t}_{1j} Q^R_{ijk}{}' m_b m^2_t
                 \frac{R^{S^\pm}_{k2}}{v_2} \Bigg\}  \, ,\nonumber
\end{eqnarray}
\begin{eqnarray}
b_{ijk3} &=& - \frac{2\sqrt2 \,}{g} \Bigg\{ b^{\tilde{t}}_{1i}
             k^{\tilde t}_{1j} Q^L_{ijk}{}'
           m_{\tilde{\chi}^+_j} m_b m_t \frac{R^{S^\pm}_{k2}}{v_2}
    - a^{\tilde{t}}_{1i} l^{\tilde t}_{1j} Q^L_{ijk}{}'
           m^2_t m_{{\tilde \chi}^0_i} \frac{R^{S^\pm}_{k2}}{v_2} 
\nonumber \\
  & &   + b^{\tilde{t}}_{1i} l^{\tilde t}_{1j} Q^L_{ijk}{}' m_t
     \left[ \left( m^2_{{\tilde \chi}^0_i}
            +m^2_{S_k^+}+m^2_{{\tilde t}_1}+m^2_b \right) 
      \frac{R^{S^\pm}_{k2}}{v_2}
          - m^2_b  \frac{R^{S^\pm}_{k1}}{v_1} \right] \nonumber \\
  & &   + a^{\tilde{t}}_{1i} k^{\tilde t}_{1j}
  Q^L_{ijk}{}' m_{\tilde{\chi}^+_j} m_b m_{{\tilde \chi}^0_i} 
     \frac{R^{S^\pm}_{k1}}{v_1}
    - b^{\tilde{t}}_{1i} k^{\tilde t}_{1j} Q^R_{ijk}{}' m_{{\tilde \chi}^0_i}
              m_b m_t \frac{R^{S^\pm}_{k1}}{v_1} \nonumber \\
  & &   + b^{\tilde{t}}_{1i} l^{\tilde t}_{1j} Q^R_{ijk}{}'
         m_{\tilde{\chi}^+_j} m_t m_{{\tilde \chi}^0_i} 
       \frac{R^{S^\pm}_{k2}}{v_2}
   + a^{\tilde{t}}_{1i} l^{\tilde t}_{1j} Q^R_{ijk}{}' m^2_b 
            m_{\tilde{\chi}^+_j} \frac{R^{S^\pm}_{k1}}{v_1} \nonumber \\
  & &   + a^{\tilde{t}}_{1i} k^{\tilde t}_{1j} Q^R_{ijk}{}' m_b
     \left[ \left(m^2_b +m^2_{{\tilde \chi}^0_i} +m^2_{S_k^+}
                 +m^2_{{\tilde t}_1} \right) \frac{R^{S^\pm}_{k1}}{v_1}
          - m^2_t \frac{R^{S^\pm}_{k2}}{v_2} \right] \Bigg\}  \,
        ,\nonumber
\end{eqnarray}
\begin{eqnarray}
b_{ijk4} &=& \frac{2\sqrt2 \,}{g}
         \left( b^{\tilde{t}}_{1i} l^{\tilde t}_{1j}
                Q^L_{ijk}{}' m_t \frac{R^{S^\pm}_{k2}}{v_2}
               + a^{\tilde{t}}_{1i} k^{\tilde t}_{1j} Q^R_{ijk}{}'
                 m_b \frac{R^{S^\pm}_{k1}}{v_2} \right) \, ,
\nonumber
\end{eqnarray}
\begin{eqnarray}
c_{ijkl1} &=& 2 \, C^{S_k^\pm}_{{\tilde t}_1 {\tilde b}_l} \Bigg[
         \left(b^{\tilde b}_{li} l^{\tilde t}_{1j} Q^L_{ijk}{}'
                        + a^{\tilde b}_{li} k^{\tilde t}_{1j} Q^R_{ijk}{}'
         \right) m_b \left( m^2_b + m^2_{{\tilde t}_1} + 2 \,
             m^2_{{\tilde \chi}^0_i} \right) \nonumber \\
  & & \hspace{13mm} + \left(a^{\tilde b}_{li} l^{\tilde t}_{1j} Q^L_{ijk}{}'
                        + b^{\tilde b}_{li} k^{\tilde t}_{1j} Q^R_{ijk}{}'
                \right) m_{{\tilde \chi}^0_i}
           \left( m^2_{{\tilde \chi}^0_i} + m^2_{S_k^+} + 2 \, m^2_b
             \right) \nonumber \\
  & & \hspace{13mm} + \left(b^{\tilde b}_{li} k^{\tilde t}_{1j} Q^L_{ijk}{}'
                        + a^{\tilde b}_{li} l^{\tilde t}_{1j} Q^R_{ijk}{}'
     \right) m_{\tilde{\chi}^+_j} \left( m^2_b+m^2_{{\tilde \chi}^0_i}
        \right) \nonumber \\
  & & \hspace{13mm} + \left(a^{\tilde b}_{li} k^{\tilde t}_{1j} Q^L_{ijk}{}'
                        + b^{\tilde b}_{li} l^{\tilde t}_{1j} Q^R_{ijk}{}'
         \right) 2 \, m_b m_{{\tilde \chi}^0_i} m_{\tilde{\chi}^+_j}
      \Bigg]  \, ,\nonumber
\end{eqnarray}
\begin{eqnarray}
c_{ijkl2} &=& - 2 \, C^{S_k^\pm}_{{\tilde t}_1 {\tilde b}_l} \Bigg[
     \left(b^{\tilde b}_{li} l^{\tilde t}_{1j} Q^L_{ijk}{}'
    + a^{\tilde b}_{li} k^{\tilde t}_{1j} Q^R_{ijk}{}' \right) m_b
  + \left(a^{\tilde b}_{li} l^{\tilde t}_{1j} Q^L_{ijk}{}'
      + b^{\tilde b}_{li} k^{\tilde t}_{1j} Q^R_{ijk}{}' \right)
                m_{{\tilde \chi}^0_i} \nonumber \\
   & & \hspace{16mm} + \left(b^{\tilde b}_{li} k^{\tilde t}_{1j} Q^L_{ijk}{}'
                        + a^{\tilde b}_{li} l^{\tilde t}_{1j} Q^R_{ijk}{}' 
            \right) m_{\tilde{\chi}^+_j} \Bigg]  \, ,\nonumber
\end{eqnarray}
\begin{eqnarray}
c_{ijkl3} &=& -2 \, C^{S_k^\pm}_{{\tilde t}_1 {\tilde b}_l} \Bigg[
              \left(b^{\tilde b}_{li} l^{\tilde t}_{1j} Q^L_{ijk}{}'
      + a^{\tilde b}_{li} k^{\tilde t}_{1j} Q^R_{ijk}{}' \right) m_b
 + \left(a^{\tilde b}_{li} l^{\tilde t}_{1j} Q^L_{ijk}{}'
      + b^{\tilde b}_{li} k^{\tilde t}_{1j} Q^R_{ijk}{}' \right)
        m_{{\tilde \chi}^0_i} \Bigg] \, , 
\nonumber
\end{eqnarray}
\begin{eqnarray}
d_{ik1} &=& \frac2{g^2} \left\{
   - \left[ (a^{\tilde{t}}_{1i})^2 m^2_t 
   \left(\frac{R^{S^\pm}_{k2}}{v_2}\right)^2
             + (b^{\tilde{t}}_{1i})^2 m^2_b 
   \left(\frac{R^{S^\pm}_{k1}}{v_1}\right)^2\, \right]
       m^2_t \left( m^2_b + m^2_{{\tilde \chi}^0_i} \right)\right. 
   \nonumber \\
  & & \hspace{5mm} + \left[ (a^{\tilde{t}}_{1i})^2 m^2_b 
    \left(\frac{R^{S^\pm}_{k1}}{v_1}\right)^2
                     + (b^{\tilde{t}}_{1i})^2 m^2_t 
  \left(\frac{R^{S^\pm}_{k2}}{v_2}\right)^2\, \right]
          \left( m^2_{{\tilde \chi}^0_i} - m^2_{{\tilde t}_1} \right)
          \left( m^2_{S_k^+} - m^2_b \right)  \nonumber \\
  & & \hspace{5mm} + 2 \, a^{\tilde{t}}_{1i} b^{\tilde{t}}_{1i}
             m_t m_{{\tilde \chi}^0_i}
      \left[ \left(m^2_{S_k^+}-m^2_b \right)
             \left(m^2_b \left(\frac{R^{S^\pm}_{k1}}{v_1}\right)^2 + 
     m^2_t \left(\frac{R^{S^\pm}_{k2}}{v_2}\right)^2 \right)
        - 2 \, m^2_b m^2_t \right]  \nonumber \\
  & & \hspace{5mm} + 2 \, \left( (a^{\tilde{t}}_{1i})^2 
             + (b^{\tilde{t}}_{1i})^2 \right)
       m^2_b m^2_t \left( m^2_{{\tilde t}_1}
            - m^2_{{\tilde \chi}^0_i} \right) \Bigg\}  \, , \nonumber
\end{eqnarray}
\begin{eqnarray}
d_{ik2} &=& \frac{2m^2_t}{g^2}
 \left[ (a^{\tilde{t}}_{1i})^2 m^2_t 
  \left(\frac{R^{S^\pm}_{k2}}{v_2}\right)^2
           + (b^{\tilde{t}}_{1i})^2 m^2_b 
  \left(\frac{R^{S^\pm}_{k1}}{v_1}\right)^2\, \right]  \, , \nonumber
\end{eqnarray}
\begin{eqnarray}
d_{ik3} &=& \frac{-2}{g^2} \Bigg\{
       2  \left( (a^{\tilde{t}}_{1i})^2 + (b^{\tilde{t}}_{1i})^2 \right)
         m^2_b m^2_t \nonumber\\
 &&  2  a^{\tilde{t}}_{1i} b^{\tilde{t}}_{1i}
              m_t m_{{\tilde \chi}^0_i}
   \left[ m^2_b \left( 2 +
  \left(\frac{R^{S^\pm}_{k1}}{v_1}\right)^2 \right)+ m^2_t 
  \left(\frac{R^{S^\pm}_{k2}}{v_2}\right)^2\, \right] \nonumber \\
  & & \hspace{12mm}
 - \left.\left[ (a^{\tilde{t}}_{1i})^2 m^2_b 
   \left(\frac{R^{S^\pm}_{k1}}{v_1}\right)^2
                     + (b^{\tilde{t}}_{1i})^2 m^2_t 
   \left(\frac{R^{S^\pm}_{k2}}{v_2}\right)^2\, \right]
        \left(m^2_{S_k^+}+m^2_{{\tilde t}_1} \right)  \right\}  \, ,\nonumber
%
\end{eqnarray}
\begin{eqnarray}
d_{ik4} &=& - \frac2{g^2}
   \left[{(a^{\tilde{t}}_{1i})^2 m^2_b 
   \left(\frac{R^{S^\pm}_{k1}}{v_1}\right)^2
                    + (b^{\tilde{t}}_{1i})^2 m^2_t 
   \left(\frac{R^{S^\pm}_{k2}}{v_2}\right)^2}\,\right]\, ,
\nonumber 
\end{eqnarray}
\begin{eqnarray}
e_{ikl1} &=& - \frac{2\sqrt2}{g}\, C^{S_k^\pm}_{{\tilde t}_1 {\tilde b}_l}
                \Bigg\{
       a^{\tilde{b}}_{li} b^{\tilde{t}}_{1i} m_t m_{{\tilde \chi}^0_i}
       \left[ \left( m^2_{S_k^+} - m^2_b \right) 
       \frac{R^{S^\pm}_{k2}}{v_2}
          - 2 \, m^2_b \frac{R^{S^\pm}_{k1}}{v_1} \right] \nonumber \\
  & & \hspace{21mm} + b^{\tilde{b}}_{ki} b^{\tilde{t}}_{1i} m_b m_t
      \left[ \left( m^2_{{\tilde t}_1} - m^2_{{\tilde \chi}^0_i} \right)
             \frac{R^{S^\pm}_{k2}}{v_2}
           - \left(m^2_b+m^2_{{\tilde \chi}^0_i} \right) 
      \frac{R^{S^\pm}_{k1}}{v_1} \right]
         \nonumber \\
  & & \hspace{21mm} - a^{\tilde{b}}_{li} a^{\tilde{t}}_{1i}
       \left[ m^2_t \left( m^2_b+m^2_{{\tilde \chi}^0_i} \right) 
      \frac{R^{S^\pm}_{k2}}{v_2}
         + m^2_b \left(m^2_{{\tilde \chi}^0_i} -m^2_{{\tilde t}_1} \right)
              \frac{R^{S^\pm}_{k1}}{v_1} \right] \nonumber \\
  & & \hspace{21mm} - b^{\tilde{b}}_{ki} a^{\tilde{t}}_{1i}
              m_b m_{{\tilde \chi}^0_i}
       \left[ 2 \, m^2_t \frac{R^{S^\pm}_{k2}}{v_2}
             + \left( m^2_b-m^2_{S_k^+} \right) 
     \frac{R^{S^\pm}_{k1}}{v_1} \right] \Bigg\}\, , \nonumber
\end{eqnarray}
\begin{eqnarray}
   e_{ikl2} &=& - \frac{ 2\sqrt2}g \, C^{S_k^\pm}_{{\tilde t}_1 {\tilde b}_l}
       \left(
           a^{\tilde{b}}_{li} a^{\tilde{t}}_{1i} m^2_t 
        \frac{R^{S^\pm}_{k2}}{v_2}
      + b^{\tilde{b}}_{ki} b^{\tilde{t}}_{1i} m_b m_t 
       \frac{R^{S^\pm}_{k1}}{v_1} \right)\, , \nonumber
\end{eqnarray}
\begin{eqnarray}
e_{ikl3} &=& \frac{2\sqrt2}g \, C^{S_k^\pm}_{{\tilde t}_1 {\tilde b}_l}
    \bigg(  a^{\tilde{b}}_{li} b^{\tilde{t}}_{1i} m_t
       m_{{\tilde \chi}^0_i} \frac{R^{S^\pm}_{k2}}{v_2}
       + b^{\tilde{b}}_{ki} b^{\tilde{t}}_{1i} m_b m_t 
      \frac{R^{S^\pm}_{k2}}{v_2} \nonumber \\
  & & \hspace{17mm} + \, a^{\tilde{b}}_{li} a^{\tilde{t}}_{1i} m^2_b 
      \frac{R^{S^\pm}_{k1}}{v_1}
       + b^{\tilde{b}}_{ki} a^{\tilde{t}}_{1i} m_b
              m_{{\tilde \chi}^0_i} 
         \frac{R^{S^\pm}_{k1}}{v_1} \bigg)  \, ,
\nonumber
\end{eqnarray}
\begin{eqnarray}
f_{ikl1} &=& - (C^{S_k^\pm}_{{\tilde t}_1 {\tilde b}_l})^2
         \left[ \left((a^{\tilde{b}}_{li})^2+(b^{\tilde{b}}_{11})^2 \right)
                 \left( m^2_b + m^2_{{\tilde \chi}^0_i} \right)
     + 4 \, a^{\tilde{b}}_{li} b^{\tilde{b}}_{ki} m_b m_{{\tilde \chi}^0_i}
       \right]  \, ,\nonumber
\end{eqnarray}
\begin{eqnarray}
f_{ikl2} &=& (C^{S_k^\pm}_{{\tilde t}_1 {\tilde b}_l})^2
        \left((a^{\tilde{b}}_{li})^2+(b^{\tilde{b}}_{11})^2 \right) \, ,
\nonumber
\end{eqnarray}
\begin{eqnarray}
f_{ik31} &=& - 2 \, C^{S_k^\pm}_{{\tilde t}_1 {\tilde b}_1}
                  C^{S_k^\pm}_{{\tilde t}_1 {\tilde b}_2}
 \left[ \left( a^{\tilde{b}}_{1i} a^{\tilde{b}}_{2i}
     + b^{\tilde{b}}_{1i} b^{\tilde{b}}_{2i} \right)
               \left( m^2_b + m^2_{{\tilde \chi}^0_i} \right)
     + 2  \left( a^{\tilde{b}}_{1i} b^{\tilde{b}}_{2i}
         + b^{\tilde{b}}_{1i} a^{\tilde{b}}_{2i} \right)
         m_b m_{{\tilde \chi}^0_i} \right]  \, ,\nonumber 
\end{eqnarray}
\begin{eqnarray}
f_{ik32} &=& 2 \, C^{S_k^\pm}_{{\tilde t}_1 {\tilde b}_1}
        C^{S_k^\pm}_{{\tilde t}_1 {\tilde b}_2}
       \left( a^{\tilde{b}}_{1i} a^{\tilde{b}}_{2i}
              + b^{\tilde{b}}_{1i} b^{\tilde{b}}_{2i} \right) \, .
\nonumber
\end{eqnarray}

\subsection{The width $\Gamma({\tilde t}_1 \to S_k^0 \, b 
\, {\tilde \chi}^+_i)$}

\noindent
The decay width is given by
\begin{eqnarray}
 \Gamma({\tilde t}_1 \to S_k^0 \, b \, {\tilde \chi}^+_i)  &=& \nonumber \\
 & & \hspace{-30mm} 
   =  \frac{\alpha^2}{16 \, \pi m^3_{{\tilde t}_1} \sin^4 \theta_W}
  \int\limits^{(m_{{\tilde t}_1}-m_{S_k^0})^2}_{
           (m_b + m_{{\tilde \chi}^+_i})^2} \hspace{-8mm}
     d \, s \,
   \left( G^{S^0}_{{\tilde \chi}^+ {\tilde \chi}^+} +
   G^{S^0}_{{\tilde \chi}^+ t} +
   G^{S^0}_{{\tilde \chi}^+ {\tilde t}} +
   G^{S^0}_{t t} +
   G^{S^0}_{t {\tilde t}} +
   G^{S^0}_{{\tilde t} {\tilde t}} \right) 
\nonumber
\end{eqnarray}
with
\begin{eqnarray} 
   G^{S^0}_{{\tilde \chi}^+ {\tilde \chi}^+} &=&
       \sum^3_{j=1} \Big[  ( a_{ijk1} + a_{ijk2} s )
          J^0_b(m^2_{{\tilde t}_1} + m^2_{S_k^0} + m^2_b
               + m^2_{{\tilde \chi}^+_i} - m^2_{{\tilde \chi}^+_j} - s
             ,\Gamma_{{\tilde \chi}^+_j} m_{{\tilde \chi}^+_j}) \nonumber \\
    & & \hspace{5mm} + ( a_{ijk3} + a_{i4jk} s )
       J^1_b(m^2_{{\tilde t}_1} + m^2_{S_k^0} + m^2_b
            + m^2_{{\tilde \chi}^+_i} - m^2_{{\tilde \chi}^+_j} - s
             ,\Gamma_{{\tilde \chi}^+_j} m_{{\tilde \chi}^+_j}) \nonumber \\
    & & \hspace{5mm} + \, a_{ijk4} \,
          J^2_b(m^2_{{\tilde t}_1} + m^2_{S_k^0} + m^2_b
               + m^2_{{\tilde \chi}^+_i} - m^2_{{\tilde \chi}^+_j} - s
       ,\Gamma_{{\tilde \chi}^+_j} m_{{\tilde \chi}^+_j}) \Big] \nonumber \\
    & &  +  ( a_{i4k1} + a_{i4k2} s )
          J^0_{bb}(m^2_{{\tilde t}_1} + m^2_{S_k^0} + m^2_b
               + m^2_{{\tilde \chi}^+_i} - m^2_{{\tilde \chi}^+_1} - s
             ,\Gamma_{{\tilde \chi}^+_1} m_{{\tilde \chi}^+_1}  \nonumber \\
       & & \hspace{33mm} ,m^2_{{\tilde t}_1} + m^2_{S_k^0} + m^2_b
               + m^2_{{\tilde \chi}^+_i} - m^2_{{\tilde \chi}^+_2} - s
             ,\Gamma_{{\tilde \chi}^+_2} m_{{\tilde \chi}^+_2}) \nonumber \\
    & &  + ( a_{i4k3} + a_{i4k4} s )
       J^1_{bb}(m^2_{{\tilde t}_1} + m^2_{S_k^0} + m^2_b
               + m^2_{{\tilde \chi}^+_i} - m^2_{{\tilde \chi}^+_1} - s
             ,\Gamma_{{\tilde \chi}^+_1} m_{{\tilde \chi}^+_1}  \nonumber \\
       & & \hspace{33mm} ,m^2_{{\tilde t}_1} + m^2_{S_k^0} + m^2_b
               + m^2_{{\tilde \chi}^+_i} - m^2_{{\tilde \chi}^+_2} - s
             ,\Gamma_{{\tilde \chi}^+_2} m_{{\tilde \chi}^+_2}) \nonumber \\
    & &  + \, a_{i4k4} \,
       J^2_{bb}(m^2_{{\tilde t}_1} + m^2_{S_k^0} + m^2_b
               + m^2_{{\tilde \chi}^+_i} - m^2_{{\tilde \chi}^+_1} - s
             ,\Gamma_{{\tilde \chi}^+_1} m_{{\tilde \chi}^+_1}  \nonumber \\
       & & \hspace{17mm} ,m^2_{{\tilde t}_1} + m^2_{S_k^0} + m^2_b
               + m^2_{{\tilde \chi}^+_i} - m^2_{{\tilde \chi}^+_2} - s
             ,\Gamma_{{\tilde \chi}^+_2} m_{{\tilde \chi}^+_2}) 
             \nonumber\\
    & &  +  ( a_{i5k1} + a_{i5k2} s )
          J^0_{bb}(m^2_{{\tilde t}_1} + m^2_{S_k^0} + m^2_b
               + m^2_{{\tilde \chi}^+_i} - m^2_{{\tilde \chi}^+_1} - s
             ,\Gamma_{{\tilde \chi}^+_1} m_{{\tilde \chi}^+_1}  \nonumber \\
       & & \hspace{33mm} ,m^2_{{\tilde t}_1} + m^2_{S_k^0} + m^2_b
               + m^2_{{\tilde \chi}^+_i} - m^2_{{\tilde \chi}^+_3} - s
             ,\Gamma_{{\tilde \chi}^+_3} m_{{\tilde \chi}^+_3}) \nonumber \\
    & &  + ( a_{i5k3} + a_{i5k4} s )
       J^1_{bb}(m^2_{{\tilde t}_1} + m^2_{S_k^0} + m^2_b
               + m^2_{{\tilde \chi}^+_i} - m^2_{{\tilde \chi}^+_1} - s
             ,\Gamma_{{\tilde \chi}^+_1} m_{{\tilde \chi}^+_1}  \nonumber \\
       & & \hspace{33mm} ,m^2_{{\tilde t}_1} + m^2_{S_k^0} + m^2_b
               + m^2_{{\tilde \chi}^+_i} - m^2_{{\tilde \chi}^+_3} - s
             ,\Gamma_{{\tilde \chi}^+_3} m_{{\tilde \chi}^+_3}) \nonumber \\
    & &  + \, a_{i5k4} \,
       J^2_{bb}(m^2_{{\tilde t}_1} + m^2_{S_k^0} + m^2_b
               + m^2_{{\tilde \chi}^+_i} - m^2_{{\tilde \chi}^+_1} - s
             ,\Gamma_{{\tilde \chi}^+_1} m_{{\tilde \chi}^+_1}  \nonumber \\
       & & \hspace{17mm} ,m^2_{{\tilde t}_1} + m^2_{S_k^0} + m^2_b
               + m^2_{{\tilde \chi}^+_i} - m^2_{{\tilde \chi}^+_3} - s
             ,\Gamma_{{\tilde \chi}^+_3} m_{{\tilde \chi}^+_3})\nonumber \\
    & &  +  ( a_{i6k1} + a_{i6k2} s )
          J^0_{bb}(m^2_{{\tilde t}_1} + m^2_{S_k^0} + m^2_b
               + m^2_{{\tilde \chi}^+_i} - m^2_{{\tilde \chi}^+_3} - s
             ,\Gamma_{{\tilde \chi}^+_3} m_{{\tilde \chi}^+_3}  \nonumber \\
       & & \hspace{33mm} ,m^2_{{\tilde t}_1} + m^2_{S_k^0} + m^2_b
               + m^2_{{\tilde \chi}^+_i} - m^2_{{\tilde \chi}^+_2} - s
             ,\Gamma_{{\tilde \chi}^+_2} m_{{\tilde \chi}^+_2}) \nonumber \\
    & &  + ( a_{i6k3} + a_{i6k4} s )
       J^1_{bb}(m^2_{{\tilde t}_1} + m^2_{S_k^0} + m^2_b
               + m^2_{{\tilde \chi}^+_i} - m^2_{{\tilde \chi}^+_3} - s
             ,\Gamma_{{\tilde \chi}^+_3} m_{{\tilde \chi}^+_3}  \nonumber \\
       & & \hspace{33mm} ,m^2_{{\tilde t}_1} + m^2_{S_k^0} + m^2_b
               + m^2_{{\tilde \chi}^+_i} - m^2_{{\tilde \chi}^+_2} - s
             ,\Gamma_{{\tilde \chi}^+_2} m_{{\tilde \chi}^+_2}) \nonumber \\
    & &  + \, a_{i6k4} \,
       J^2_{bb}(m^2_{{\tilde t}_1} + m^2_{S_k^0} + m^2_b
               + m^2_{{\tilde \chi}^+_i} - m^2_{{\tilde \chi}^+_3} - s
             ,\Gamma_{{\tilde \chi}^+_3} m_{{\tilde \chi}^+_3}  \nonumber \\
       & & \hspace{17mm} ,m^2_{{\tilde t}_1} + m^2_{S_k^0} + m^2_b
               + m^2_{{\tilde \chi}^+_i} - m^2_{{\tilde \chi}^+_2} - s
             ,\Gamma_{{\tilde \chi}^+_2} m_{{\tilde \chi}^+_2}) \, ,
\nonumber
\end{eqnarray}
\begin{eqnarray}
   G^{S^0}_{{\tilde \chi}^+ t} &=&
       \sum^3_{j=1} \Big[ ( b_{ijk1} + b_{ijk2} s ) 
 J^0_{bb}(m^2_{{\tilde t}_1} + m^2_{S_k^0} + m^2_b
               + m^2_{{\tilde \chi}^+_i} - m^2_{{\tilde \chi}^+_j} - s
       ,-\Gamma_{{\tilde \chi}^+_j} m_{{\tilde \chi}^+_j},m^2_b,
        \Gamma_b m_b) \nonumber \\
    & & \hspace{5mm} + ( b_{ijk3} + b_{ijk4} s )
 J^1_{bb}(m^2_{{\tilde t}_1} + m^2_{S_k^0} + m^2_b
               + m^2_{{\tilde \chi}^+_i} - m^2_{{\tilde \chi}^+_j} - s
       ,-\Gamma_{{\tilde \chi}^+_j} m_{{\tilde \chi}^+_j} ,m^2_b,
        \Gamma_b m_b) \nonumber \\
    & & \hspace{5mm} + \, b_{ijk4} \,
        J^2_{bb}(m^2_{{\tilde t}_1} + m^2_{S_k^0} + m^2_b
               + m^2_{{\tilde \chi}^+_i} - m^2_{{\tilde \chi}^+_j} - s
       ,-\Gamma_{{\tilde \chi}^+_j} m_{{\tilde \chi}^+_j} ,m^2_b,
         \Gamma_b m_b)  \Big] \, ,
\nonumber
\end{eqnarray}
\begin{eqnarray}
   G^{S^0}_{{\tilde \chi}^+ {\tilde t}} &=&
       \sum^2_{l=1} \Big[
          ( c_{ijkl1} + c_{ijkl2} s ) \nonumber \\
    & & \hspace{7mm} * J^0_{st}(
       m^2_{{\tilde t}_l}, \Gamma_ {{\tilde t}_l} m_{{\tilde t}_l}
       ,m^2_{{\tilde t}_1} + m^2_{S_k^0} + m^2_b
               + m^2_{{\tilde \chi}^+_i} - m^2_{{\tilde \chi}^+_j} - s
       ,-\Gamma_{{\tilde \chi}^+_j} m_{{\tilde \chi}^+_j}) \nonumber \\
    & &  + \, c_{ijkl3} \, J^1_{st}(
       m^2_{{\tilde t}_l}, \Gamma_ {{\tilde t}_l} m_{{\tilde t}_l}
       ,m^2_{{\tilde t}_1} + m^2_{S_k^0} + m^2_b
               + m^2_{{\tilde \chi}^+_i} - m^2_{{\tilde \chi}^+_j} - s
       ,-\Gamma_{{\tilde \chi}^+_j} m_{{\tilde \chi}^+_j}) \Big] \, ,
\nonumber
\end{eqnarray}
\begin{equation*}
   G^{S^0}_{tt} =
          ( d_{ik1} + d_{ik2} s) J^0_b(m^2_b, \Gamma_b m_b) +
          ( d_{ik3} + d_{ik4} s) J^1_b(m^2_b, \Gamma_b m_b)
 + \, d_{ik4} \, J^2_b(m^2_b, \Gamma_b m_b) \, ,
\end{equation*}
\begin{eqnarray}
   G^{S^0}_{t {\tilde t}} &=& \sum^2_{l=1} \Big[
   ( e_{ikl1} + e_{ikl2} s )
   J^0_{st}(m^2_{{\tilde t}_l},
  \Gamma_ {{\tilde t}_l} m_{{\tilde t}_l},m^2_b, \Gamma_b m_b)
 + \, e_{ikl3} \, J^1_{st}(m^2_{{\tilde t}_l},
    \Gamma_ {{\tilde t}_l} m_{{\tilde t}_l},m^2_b, \Gamma_b m_b)
   \Big] \, , 
\nonumber
\end{eqnarray}
\begin{eqnarray}
G^{S^0}_{{\tilde t} {\tilde t}} &=&
 \frac{\sqrt{\lambda(s,m^2_{{\tilde t}_1},m^2_{S_k^0})
                 \lambda(s,m^2_{{\tilde \chi}^+_i},m^2_b)}}{s} \nonumber \\
  & & \hspace{-12mm} \times \left\{ \sum^2_{l=1}
      \frac{(f_{ikl1} + f_{ikl2} s)}
   {(s-m^2_{{\tilde t}_l})^2 + \Gamma^2_{{\tilde t}_l} m^2_{{\tilde t}_l}}
 + \mbox{Re} \left[ \frac{(f_{ik31} + f_{ik32} s)}
      { (s-m^2_{{\tilde t}_1} + i \Gamma_{{\tilde t}_1} m_{{\tilde t}_1})
         (s-m^2_{{\tilde t}_2} - i \Gamma_{{\tilde t}_2} m_{{\tilde t}_2})}
          \right]  \right\} \, .
\nonumber
\end{eqnarray}
The integrals   $J^{0,1,2}_{t,tt,st}$ are:
\begin{eqnarray}
 J^i_{t}(m^2_1, m_1 \Gamma_1)
        &=& \int\limits^{t_{max}}_{t_{min}}
     \hspace{-2mm} d \, t \frac{t^i}{(t-m^2_1)^2 \, + \,  m_1^2 \Gamma_1^2}
  \, , \nonumber\\
 J^i_{bb}(m^2_1,m_1 \, \Gamma_1,m^2_2,m_2 \, \Gamma_2)
        &=& \mbox{Re} \int\limits^{t_{max}}_{t_{min}}
     \hspace{-2mm} d \, t
          \frac{t^i}{(t-m^2_1 \, + \,i m_1 \Gamma_1)
                     (t-m^2_2 \, - \,i m_2 \Gamma_2)} \, , \nonumber\\
 J^i_{st}(m^2_1,m_1 \, \Gamma_1,m^2_2,m_2 \, \Gamma_2)
        &=& \mbox{Re} \frac1{s-m^2_1 \, + \,i  m_1 \Gamma_1}
         \int\limits^{t_{max}}_{t_{min}} \hspace{-2mm} d \, t
          \frac{t^i}{(t-m^2_2 \, - \,i m_2 \Gamma_2)} 
\nonumber
\end{eqnarray}
with $i=0,1,2$. Their integration range is given by
\begin{eqnarray}
t_{max \atop min} &=&
 \frac{m^2_{{\tilde t}_1} + m^2_b + m^2_{S_k^0}
        + m^2_{{\tilde \chi}^+_i} -s}2
       - \frac{(m^2_{{\tilde t}_1}-m^2_{S_k^0})
               (m^2_{{\tilde \chi}^+_i}-m^2_b)}{2 s} \nonumber \\
  & &  \pm  \frac{\sqrt{\lambda(s,m^2_{{\tilde t}_1},m^2_{S_k^0})
                  \lambda(s,m^2_{{\tilde \chi}^+_i},m^2_b)}}{2 s} \, ,
\nonumber
\end{eqnarray}
where $s = (p_{{\tilde t}_1} - p_{S_k^0})^2$
 and $t = (p_{{\tilde t}_1} - p_{t})^2$
are the usual Mandelstam variables.
Note, that $-\Gamma_{{\tilde \chi}^+_j} m_{{\tilde \chi}^+_j}$
appears in the entries of the integrals 
$G^{S^0}_{{\tilde \chi}^+ {\tilde t}_j}$ and
$G^{S^0}_{{\tilde \chi}^+ t}$ because the chargino is exchanged 
 in the $u$-channel in our convention.
The coefficients are given by (no sum upon repeated index):
\begin{eqnarray}
a_{ijk1} &=& - 4 \, k^{\tilde t}_{1j} l^{\tilde t}_{1j} 
       l^{S^0}_{ijk} k^{S^0}_{ijk} m_b m_{{\tilde \chi}^+_i}
    \left( m^2_b +  m^2_{\tilde{\chi}^+_j}
   + m^2_{{\tilde \chi}^+_i} + m^2_{{\tilde t}_1} 
   + m^2_{S_k^0} \right) \nonumber \\
  & &  - 2 \, l^{S^0}_{ijk} k^{S^0}_{ijk}
    \left( (k^{\tilde t}_{1j})^2 + (l^{\tilde t}_{1j})^2 \right)
    m_{{\tilde \chi}^+_i} m_{\tilde{\chi}^+_j}
    \left( 2 \, m^2_b + m^2_{{\tilde \chi}^+_i}
      + m^2_{S_k^0} \right)  \nonumber \\
  & &  - 2 \, k^{\tilde t}_{1j} l^{\tilde t}_{1j} 
        \left( (l^{S^0}_{ijk})^2 + (k^{S^0}_{ijk})^2 \right)
     m_b m_{\tilde{\chi}^+_j} 
        \left( m^2_b + 2 \, m^2_{{\tilde \chi}^+_i}
            + m^2_{{\tilde t}_1} \right)  \nonumber \\
  & &  - \left( (k^{\tilde t}_{1j})^2
  (k^{S^0}_{ijk})^2 + (l^{\tilde t}_{1j})^2 (l^{S^0}_{ijk})^2 \right) \nonumber \\
   & & \hspace{5mm} \times \left[ \left( m^2_b+m^2_{{\tilde \chi}^+_i} \right)^2
    + \left( m^2_b+m^2_{S_k^0} \right)
 \left(m^2_{{\tilde \chi}^+_i}+m^2_{{\tilde t}_1}\right) \right] \nonumber \\
  & &  - \left( (k^{\tilde t}_{1j})^2
       (l^{S^0}_{ijk})^2 + (l^{\tilde t}_{1j})^2 (k^{S^0}_{ijk})^2 \right)
      m^2_{\tilde{\chi}^+_j}
           \left(m^2_b+m^2_{{\tilde \chi}^+_i} \right) \, ,\nonumber
\end{eqnarray}
\begin{eqnarray}
a_{ijk2} &=& 4 \, k^{\tilde t}_{1j} l^{\tilde t}_{1j}
       l^{S^0}_{ijk} k^{S^0}_{ijk} m_b m_{{\tilde \chi}^+_i}
  + \left( (k^{\tilde t}_{1j})^2 
       (k^{S^0}_{ijk})^2 + (l^{\tilde t}_{1j})^2 (l^{S^0}_{ijk})^2 \right)
           \left( m^2_b+m^2_{{\tilde \chi}^+_i} \right) \nonumber \\
  & & + 2 \, k^{\tilde t}_{1j} 
       l^{\tilde t}_{1j} \left( (l^{S^0}_{ijk})^2 + (k^{S^0}_{ijk})^2 \right)
                 m_b m_{\tilde{\chi}^+_j}
        + 2 \, l^{S^0}_{ijk} k^{S^0}_{ijk}
        \left( (k^{\tilde t}_{1j})^2 + (l^{\tilde t}_{1j})^2 \right)
         m_{{\tilde \chi}^+_i} m_{\tilde{\chi}^+_j}  \nonumber \\
  & &  + \left( (k^{\tilde t}_{1j})^2
      (l^{S^0}_{ijk})^2 + (l^{\tilde t}_{1j})^2 (k^{S^0}_{ijk})^2 \right)
              m^2_{\tilde{\chi}^+_j}   \, , \nonumber
\end{eqnarray}
\begin{eqnarray}
a_{ijk3} &=& 4 \, k^{\tilde t}_{1j} l^{\tilde t}_{1j}
           l^{S^0}_{ijk} k^{S^0}_{ijk} m_b m_{{\tilde \chi}^+_i}
      + 2 \, l^{S^0}_{ijk} k^{S^0}_{ijk} \left( (k^{\tilde t}_{1j})^2
              + (l^{\tilde t}_{1j})^2 \right)
              m_{{\tilde \chi}^+_i} m_{\tilde{\chi}^+_j} \nonumber \\
  & & + \left( (k^{\tilde t}_{1j})^2 (k^{S^0}_{ijk})^2
             + (l^{\tilde t}_{1j})^2 (l^{S^0}_{ijk})^2 \right)
         \left(2 \, m^2_b+2 \, m^2_{{\tilde \chi}^+_i}
             +m^2_{S_k^0}+m^2_{{\tilde t}_1}\right)  \nonumber \\
  & & + 2 \, k^{\tilde t}_{1j} l^{\tilde t}_{1j}
          \left( (l^{S^0}_{ijk})^2 + (k^{S^0}_{ijk})^2 \right)
               m_b m_{\tilde{\chi}^+_j} \, , \nonumber
\end{eqnarray}
\begin{eqnarray}
a_{ijk4} &=& - (k^{\tilde t}_{1j})^2 (k^{S^0}_{ijk})^2 -
                 (l^{\tilde t}_{1j})^2 (l^{S^0}_{ijk})^2  \, ,
\nonumber
\end{eqnarray}
\begin{eqnarray}
a_{i4k1} &=& - 2  \left(l^{\tilde t}_{11} l^{\tilde t}_{12} k^{S^0}_{i2k}
            l^{S^0}_{i1k} + k^{\tilde t}_{11} k^{\tilde t}_{12} 
        k^{S^0}_{i1k} l^{S^0}_{i2k}\right) m_{{\tilde \chi}^+_i}
             m_{\tilde{\chi}^+_1}
         \left( m^2_{{\tilde \chi}^+_i} + m^2_{S_k^0} + 2 \, m^2_b \right)
        \nonumber \\
  & &   - 4 \, \left(k^{\tilde t}_{11} l^{\tilde t}_{12}
        k^{S^0}_{i2k} l^{S^0}_{i1k} +  k^{\tilde t}_{12} l^{\tilde t}_{11}
           k^{S^0}_{i1k} l^{S^0}_{i2k}\right) m_b m_{{\tilde \chi}^+_i}
      \left( m^2_{{\tilde \chi}^+_i} + m^2_{S_k^0} + m^2_b
            + m^2_{{\tilde t}_1} \right) \nonumber \\
  & &  - 4 \, \left(k^{\tilde t}_{12} l^{\tilde t}_{11}
           k^{S^0}_{i2k} l^{S^0}_{i1k} +  k^{\tilde t}_{11} l^{\tilde t}_{12}
           k^{S^0}_{i1k} l^{S^0}_{i2k}\right) m_b m_{{\tilde \chi}^+_i}
         m_{\tilde{\chi}^+_1} m_{\tilde{\chi}^+_2} \nonumber \\
  & &  - 2 \, \left(k^{\tilde t}_{11} k^{\tilde t}_{12}
     k^{S^0}_{i2k} l^{S^0}_{i1k} +   l^{\tilde t}_{11} l^{\tilde t}_{12}
     k^{S^0}_{i1k} l^{S^0}_{i2k}\right) m_{{\tilde \chi}^+_i}
      m_{\tilde{\chi}^+_2}
      \left( m^2_{{\tilde \chi}^+_i} + m^2_{S_k^0} + 2 \, m^2_b \right) 
      \nonumber \\
  & &  - 2 \, \left(l^{\tilde t}_{11} l^{\tilde t}_{12}
        k^{S^0}_{i1k} k^{S^0}_{i2k} +  k^{\tilde t}_{11} k^{\tilde t}_{12} 
        l^{S^0}_{i1k} l^{S^0}_{i2k}\right)     \nonumber \\
  & & \hspace{5mm} \times \left[ \left(m^2_b+m^2_{{\tilde \chi}^+_i}\right)^2
      + \left(m^2_{{\tilde \chi}^+_i}+m^2_{{\tilde t}_1}\right)
        \left(m^2_b+m^2_{S_k^0}\right) \right] \nonumber \\
  & &  - 2 \, \left(k^{\tilde t}_{11} l^{\tilde t}_{12}
       k^{S^0}_{i1k} k^{S^0}_{i2k} +  k^{\tilde t}_{12} l^{\tilde t}_{11}
       l^{S^0}_{i1k} l^{S^0}_{i2k}\right) m_b m_{\tilde{\chi}^+_1}
     \left(m^2_b+m^2_{{\tilde t}_1}+2 \, m^2_{{\tilde \chi}^+_i} \right)
     \nonumber \\
  & &  - 2 \, \left(k^{\tilde t}_{12} l^{\tilde t}_{11}
     k^{S^0}_{i1k} k^{S^0}_{i2k} +  k^{\tilde t}_{11} l^{\tilde t}_{12}
     l^{S^0}_{i1k} l^{S^0}_{i2k}\right) m_b m_{\tilde{\chi}^+_2}
     \left(m^2_b+m^2_{{\tilde t}_1}+2 \, m^2_{{\tilde \chi}^+_i} \right)
     \nonumber \\
  & &  - 2 \, \left(k^{\tilde t}_{11} k^{\tilde t}_{12}
    k^{S^0}_{i1k} k^{S^0}_{i2k} +  l^{\tilde t}_{11} l^{\tilde t}_{12} 
    l^{S^0}_{i1k}
    l^{S^0}_{i2k}\right) m_{\tilde{\chi}^+_1} m_{\tilde{\chi}^+_2}
    \left(m^2_{{\tilde \chi}^+_i} + m^2_b \right)  \, , \nonumber
\end{eqnarray}
\begin{eqnarray}
a_{i4k2} &=& 2 \left(l^{\tilde t}_{11} l^{\tilde t}_{12} k^{S^0}_{i2k}
        l^{S^0}_{i1k} +   k^{\tilde t}_{11} k^{\tilde t}_{12} k^{S^0}_{i1k}
        l^{S^0}_{i2k}\right) m_{{\tilde \chi}^+_i} m_{\tilde{\chi}^+_1}
      \nonumber \\
  & &  + 4 \, \left(k^{\tilde t}_{11} l^{\tilde t}_{12}
      k^{S^0}_{i2k} l^{S^0}_{i1k} +  k^{\tilde t}_{12} l^{\tilde t}_{11}
      k^{S^0}_{i1k} l^{S^0}_{i2k}\right) m_b m_{{\tilde \chi}^+_i} \nonumber \\
  & &  + 2 \, \left(k^{\tilde t}_{11} k^{\tilde t}_{12} 
       k^{S^0}_{i2k} l^{S^0}_{i1k} +  l^{\tilde t}_{11} l^{\tilde t}_{12}
       k^{S^0}_{i1k} l^{S^0}_{i2k}\right) m_{{\tilde \chi}^+_i}
       m_{\tilde{\chi}^+_2} \nonumber \\
  & &  + 2 \, \left(l^{\tilde t}_{11} l^{\tilde t}_{12}
      k^{S^0}_{i1k} k^{S^0}_{i2k} +   k^{\tilde t}_{11} k^{\tilde t}_{12} 
      l^{S^0}_{i1k} l^{S^0}_{i2k}\right)
    \left( m^2_b+m^2_{{\tilde \chi}^+_i} \right)  \nonumber \\
  & &  + 2 \, \left(k^{\tilde t}_{11} l^{\tilde t}_{12}
      k^{S^0}_{i1k} k^{S^0}_{i2k} +  k^{\tilde t}_{12} l^{\tilde t}_{11}
      l^{S^0}_{i1k} l^{S^0}_{i2k}\right) m_b m_{\tilde{\chi}^+_1} 
      \nonumber \\
  & &  + 2 \, \left(k^{\tilde t}_{12} l^{\tilde t}_{11}
        k^{S^0}_{i1k} k^{S^0}_{i2k} +    k^{\tilde t}_{11} l^{\tilde t}_{12}
        l^{S^0}_{i1k} l^{S^0}_{i2k}\right) m_b m_{\tilde{\chi}^+_2}
      \nonumber \\
  & &  + 2 \, \left(k^{\tilde t}_{11} k^{\tilde t}_{12}
         k^{S^0}_{i1k} k^{S^0}_{i2k} +   l^{\tilde t}_{11} l^{\tilde t}_{12} 
         l^{S^0}_{i1k} l^{S^0}_{i2k}\right) m_{\tilde{\chi}^+_1}
          m_{\tilde{\chi}^+_2}  \, ,\nonumber
\end{eqnarray}
\begin{eqnarray}
a_{i4k3} &=& 2 \left(l^{\tilde t}_{11} l^{\tilde t}_{12} k^{S^0}_{i2k}
          l^{S^0}_{i1k} +  k^{\tilde t}_{11} k^{\tilde t}_{12} k^{S^0}_{i1k}
          l^{S^0}_{i2k}\right) m_{{\tilde \chi}^+_i}
          m_{\tilde{\chi}^+_1} \nonumber \\
  & &  + 4 \, \left(k^{\tilde t}_{11} l^{\tilde t}_{12}
        k^{S^0}_{i2k} l^{S^0}_{i1k} +  k^{\tilde t}_{12} l^{\tilde t}_{11}
        k^{S^0}_{i1k} l^{S^0}_{i2k}\right) m_b m_{{\tilde \chi}^+_i} 
\nonumber \\
  & &  + 2 \, \left(k^{\tilde t}_{11} k^{\tilde t}_{12}
        k^{S^0}_{i2k} l^{S^0}_{i1k} +   l^{\tilde t}_{11} l^{\tilde t}_{12}
        k^{S^0}_{i1k} l^{S^0}_{i2k}\right) m_{{\tilde \chi}^+_i}
        m_{\tilde{\chi}^+_2} \nonumber \\
  & &  + 2 \, \left(l^{\tilde t}_{11} l^{\tilde t}_{12}
        k^{S^0}_{i1k} k^{S^0}_{i2k} +  k^{\tilde t}_{11} k^{\tilde t}_{12}
        l^{S^0}_{i1k} l^{S^0}_{i2k}\right)
   \left(2 \, m^2_b+2 \, m^2_{{\tilde \chi}^+_i}
         +m^2_{S_k^0}+m^2_{{\tilde t}_1} \right) \nonumber \\
  & &  + 2 \, \left(k^{\tilde t}_{11} l^{\tilde t}_{12}
      k^{S^0}_{i1k} k^{S^0}_{i2k} +  k^{\tilde t}_{12} l^{\tilde t}_{11}
     l^{S^0}_{i1k} l^{S^0}_{i2k}\right) m_b m_{\tilde{\chi}^+_1} \nonumber \\
  & &  + 2 \, \left(k^{\tilde t}_{12} l^{\tilde t}_{11}
      k^{S^0}_{i1k} k^{S^0}_{i2k} +  k^{\tilde t}_{11} l^{\tilde t}_{12}
      l^{S^0}_{i1k} l^{S^0}_{i2k}\right) m_b m_{\tilde{\chi}^+_2} \, ,
    \nonumber
\end{eqnarray}
\begin{eqnarray}
a_{i4k4} &=& - 2 \, \left(l^{\tilde t}_{11} l^{\tilde t}_{12} k^{S^0}_{i1k}
            k^{S^0}_{i2k} +  k^{\tilde t}_{11} k^{\tilde t}_{12} l^{S^0}_{i1k}
            l^{S^0}_{i2k}\right) \, ,
\nonumber
\end{eqnarray}
\noindent
The coefficients $a_{i5kl}$ are obtained from $a_{i4kl}$ by replacing:
$ l^{\tilde t}_{12}\to l^{\tilde t}_{13}
\quad k^{\tilde t}_{12}\to k^{\tilde t}_{13}
\quad  k^{S^0}_{i2k}\to  k^{S^0}_{i3k}
\quad  l^{S^0}_{i2k}\to  l^{S^0}_{i3k}
\quad m_{\tilde{\chi}^+_2}\to m_{\tilde{\chi}^+_3}$
and the coefficients $a_{i6kl}$ are obtained from $a_{i4kl}$ by replacing:
$a_{i4kl}\to a_{i6kl}:\quad l^{\tilde t}_{11}\to l^{\tilde t}_{13}
\quad k^{\tilde t}_{11}\to k^{\tilde t}_{13}
\quad  k^{S^0}_{i1k}\to  k^{S^0}_{i3k}
\quad  l^{S^0}_{i1k}\to  l^{S^0}_{i3k}
\quad m_{\tilde{\chi}^+_1}\to m_{\tilde{\chi}^+_3}$
\begin{eqnarray}
b_{ijk1} &=& 2  \Bigg\{
       k^{\tilde{t}}_{1i} k^{\tilde t}_{1j} l^{S^0}_{ijk}
       m_{\tilde{\chi}^+_j} m_b^2
       \left[ \left( m^2_{{\tilde t}_1}-m^2_{{\tilde \chi}^+_i} \right)
      l^{S^0}_{bb}  - \left( m^2_b + 
       m^2_{{\tilde \chi}^+_i} \right)
       k^{S^0}_{bb} \right] \nonumber \\
  & &  + k^{\tilde{t}}_{1i} l^{\tilde t}_{1j} l^{S^0}_{ijk} m_b
       \left[ \left( m^2_{S_k^0} m^2_{{\tilde t}_1} - m^2_b
              m^2_{{\tilde \chi}^+_i} \right)l^{S^0}_{bb}
            - m^2_b  \left( m^2_b + m^2_{{\tilde t}_1} + 2 \,
             m^2_{{\tilde \chi}^+_i} \right)
             k^{S^0}_{bb} \right] \nonumber \\
  & &  + l^{\tilde{t}}_{1i} l^{\tilde t}_{1j} l^{S^0}_{ijk}
           m_{{\tilde \chi}^+_i} \nonumber \\
  & & \hspace{2mm} \times \left[ m^2_b
          \left( m^2_{S_k^0}+m^2_{{\tilde t}_1}-m^2_{{\tilde \chi}^+_i}
                  -m^2_b \right) k^{S^0}_{bb}
          - m^2_b \left( 2 \, m^2_b +m^2_{{\tilde \chi}^+_i}
                          +m^2_{S_k^0} \right)
         l^{S^0}_{bb} \right] \nonumber \\
  & &  + l^{\tilde{t}}_{1i} k^{\tilde t}_{1j} l^{S^0}_{ijk}
            m_{\tilde{\chi}^+_j} m_b m_{{\tilde \chi}^+_i}
       \left[ \left(m^2_{S_k^0}-m^2_b \right) k^{S^0}_{bb}
          - 2 \, m^2_bl^{S^0}_{bb} \right] \nonumber \\
  & &  + k^{\tilde{t}}_{1i} k^{\tilde t}_{1j} k^{S^0}_{ijk}
                m_{{\tilde \chi}^+_i} m_b^2 \nonumber \\
  & &  \hspace{2mm}\times \left[ \left(m^2_{{\tilde t}_1}+m^2_{S_k^0}-m^2_b
                -m^2_{{\tilde \chi}^+_i} \right) 
      l^{S^0}_{bb}
         - \left( m^2_{S_k^0}+2 \, m^2_b +m^2_{{\tilde \chi}^+_i} \right)
             k^{S^0}_{bb} \right] \nonumber \\
  & &  + k^{\tilde{t}}_{1i} l^{\tilde t}_{1j} k^{S^0}_{ijk}
                m_b m_{{\tilde \chi}^+_i} m_{\tilde{\chi}^+_j}
      \left[ \left(m^2_{S_k^0}-m^2_b \right)l^{S^0}_{bb}
         - 2 \, m^2_b  k^{S^0}_{bb} \right] \nonumber 
\end{eqnarray}
\begin{eqnarray}
  & &  - l^{\tilde{t}}_{1i} k^{\tilde t}_{1j} k^{S^0}_{ijk}
            m_b \nonumber \\
  & & \hspace{2mm} \times   \left[ \left(m^2_b m^2_{{\tilde \chi}^+_i}
           - m^2_{{\tilde t}_1} m^2_{S_k^0} \right) 
     k^{S^0}_{bb}
         + m^2_b \left( m^2_b +m^2_{{\tilde t}_1}+2 \,
             m^2_{{\tilde \chi}^+_i} \right)
         l^{S^0}_{bb} \right]  \nonumber \\
  & &  - l^{\tilde{t}}_{1i} l^{\tilde t}_{1j} k^{S^0}_{ijk}
              m_{\tilde{\chi}^+_j}
      \left[ m^2_b  \left(m^2_{{\tilde \chi}^+_i}-m^2_{{\tilde t}_1} \right)
           k^{S^0}_{bb}
         + m^2_b \left(m^2_b +m^2_{{\tilde \chi}^+_i} \right) 
         l^{S^0}_{bb}
           \right] \Bigg\}   \, , \nonumber
\end{eqnarray}
\begin{eqnarray}
b_{ijk2} &=& 2  \Bigg\{ k^{\tilde{t}}_{1i}
                k^{\tilde t}_{1j} l^{S^0}_{ijk}
              m_{\tilde{\chi}^+_j} m_b^2 k^{S^0}_{bb}
      + k^{\tilde{t}}_{1i} l^{\tilde t}_{1j} l^{S^0}_{ijk} m^2_b  m_b 
        k^{S^0}_{bb}
              \nonumber \\
  & &  + l^{\tilde{t}}_{1i} l^{\tilde t}_{1j} l^{S^0}_{ijk}
              m^2_b m_{{\tilde \chi}^+_i}l^{S^0}_{bb}
              + k^{\tilde{t}}_{1i} k^{\tilde t}_{1j} k^{S^0}_{ijk}
            m_{{\tilde \chi}^+_i} m_b^2 k^{S^0}_{bb} 
\nonumber \\
  & &  + l^{\tilde{t}}_{1i} l^{\tilde t}_{1j} k^{S^0}_{ijk}
             m_{\tilde{\chi}^+_j} m^2_bl^{S^0}_{bb}
               + l^{\tilde{t}}_{1i} k^{\tilde t}_{1j} k^{S^0}_{ijk} m_b m^2_b
                l^{S^0}_{bb} \Bigg\}  \, ,\nonumber
\end{eqnarray}
\begin{eqnarray}
b_{ijk3} &=& - 2  \Bigg\{ k^{\tilde{t}}_{1i}
             k^{\tilde t}_{1j} l^{S^0}_{ijk}
           m_{\tilde{\chi}^+_j} m_b^2l^{S^0}_{bb}
    - l^{\tilde{t}}_{1i} l^{\tilde t}_{1j} l^{S^0}_{ijk}
           m^2_b m_{{\tilde \chi}^+_i}l^{S^0}_{bb} 
\nonumber \\
  & &   + k^{\tilde{t}}_{1i} l^{\tilde t}_{1j} l^{S^0}_{ijk} m_b
     \left[ \left( m^2_{{\tilde \chi}^+_i}
            +m^2_{S_k^0}+m^2_{{\tilde t}_1}+m^2_b \right) 
     l^{S^0}_{bb}
          - m^2_b  k^{S^0}_{bb} \right] \nonumber \\
  & &   + l^{\tilde{t}}_{1i} k^{\tilde t}_{1j}
  l^{S^0}_{ijk} m_{\tilde{\chi}^+_j} m_b m_{{\tilde \chi}^+_i} 
     k^{S^0}_{bb}
    - k^{\tilde{t}}_{1i} k^{\tilde t}_{1j} k^{S^0}_{ijk} m_{{\tilde \chi}^+_i}
              m_b^2 k^{S^0}_{bb} \nonumber \\
  & &   + k^{\tilde{t}}_{1i} l^{\tilde t}_{1j} k^{S^0}_{ijk}
         m_{\tilde{\chi}^+_j} m_b m_{{\tilde \chi}^+_i} 
      l^{S^0}_{bb}
   + l^{\tilde{t}}_{1i} l^{\tilde t}_{1j} k^{S^0}_{ijk} m^2_b 
            m_{\tilde{\chi}^+_j} k^{S^0}_{bb} \nonumber \\
  & &   + l^{\tilde{t}}_{1i} k^{\tilde t}_{1j} k^{S^0}_{ijk} m_b
     \left[ \left(m^2_b +m^2_{{\tilde \chi}^+_i} +m^2_{S_k^0}
                 +m^2_{{\tilde t}_1} \right) k^{S^0}_{bb}
          - m^2_bl^{S^0}_{bb} \right] \Bigg\}  \, ,\nonumber
\end{eqnarray}
\begin{eqnarray}
b_{ijk4} &=& 2 
         \left( k^{\tilde{t}}_{1i} l^{\tilde t}_{1j}
                l^{S^0}_{ijk} m_bl^{S^0}_{bb}
               + l^{\tilde{t}}_{1i} k^{\tilde t}_{1j} k^{S^0}_{ijk}
                 m_b k^{S^0}_{bb} \right) \, ,
\nonumber
\end{eqnarray}
\begin{eqnarray}
c_{ijkl1} &=& 2 \, C^{S_k^0}_{{\tilde t}_1 {\tilde t}_l} \Bigg[
         \left(k^{\tilde t}_{li} l^{\tilde t}_{1j} l^{S^0}_{ijk}
                        + a^{\tilde t}_{li} k^{\tilde t}_{1j} k^{S^0}_{ijk}
         \right) m_b \left( m^2_b + m^2_{{\tilde t}_1} + 2 \,
             m^2_{{\tilde \chi}^+_i} \right) \nonumber \\
  & & \hspace{13mm} + \left(l^{\tilde t}_{li} l^{\tilde t}_{1j} l^{S^0}_{ijk}
                        + k^{\tilde t}_{li} k^{\tilde t}_{1j} k^{S^0}_{ijk}
                \right) m_{{\tilde \chi}^+_i}
           \left( m^2_{{\tilde \chi}^+_i} + m^2_{S_k^0} + 2 \, m^2_b
             \right) \nonumber \\
  & & \hspace{13mm} + \left(k^{\tilde t}_{li} k^{\tilde t}_{1j} l^{S^0}_{ijk}
                        + l^{\tilde t}_{li} l^{\tilde t}_{1j} k^{S^0}_{ijk}
     \right) m_{\tilde{\chi}^+_j} \left( m^2_b+m^2_{{\tilde \chi}^+_i}
        \right) \nonumber \\
  & & \hspace{13mm} + \left(l^{\tilde t}_{li} k^{\tilde t}_{1j} l^{S^0}_{ijk}
                        + k^{\tilde t}_{li} l^{\tilde t}_{1j} k^{S^0}_{ijk}
         \right) 2 \, m_b m_{{\tilde \chi}^+_i} m_{\tilde{\chi}^+_j}
      \Bigg]  \, ,\nonumber
\end{eqnarray}
\begin{eqnarray}
c_{ijkl2} &=& - 2 \, C^{S_k^0}_{{\tilde t}_1 {\tilde t}_l} \Bigg[
     \left(k^{\tilde t}_{li} l^{\tilde t}_{1j} l^{S^0}_{ijk}
    + l^{\tilde t}_{li} k^{\tilde t}_{1j} k^{S^0}_{ijk} \right) m_b
  + \left(l^{\tilde t}_{li} l^{\tilde t}_{1j} l^{S^0}_{ijk}
      + k^{\tilde t}_{li} k^{\tilde t}_{1j} k^{S^0}_{ijk} \right)
                m_{{\tilde \chi}^+_i} \nonumber \\
   & & \hspace{16mm} + \left(k^{\tilde t}_{li} k^{\tilde t}_{1j} l^{S^0}_{ijk}
                        + l^{\tilde t}_{li} l^{\tilde t}_{1j} k^{S^0}_{ijk} 
            \right) m_{\tilde{\chi}^+_j} \Bigg]  \, ,\nonumber
\end{eqnarray}
\begin{eqnarray}
c_{ijkl3} &=& -2 \, C^{S_k^0}_{{\tilde t}_1 {\tilde t}_l} \Bigg[
              \left(k^{\tilde t}_{li} l^{\tilde t}_{1j} l^{S^0}_{ijk}
      + l^{\tilde t}_{li} k^{\tilde t}_{1j} k^{S^0}_{ijk} \right) m_b
 + \left(l^{\tilde t}_{li} l^{\tilde t}_{1j} l^{S^0}_{ijk}
      + k^{\tilde t}_{li} k^{\tilde t}_{1j} k^{S^0}_{ijk} \right)
        m_{{\tilde \chi}^+_i} \Bigg] \, , \nonumber \\
\nonumber
\end{eqnarray}
\begin{eqnarray}
d_{ik1} &=&  \left\{
   - \left[ (l^{\tilde{t}}_{1i})^2 m^2_b 
   \left(l^{S^0}_{bb}\right)^2
             + (k^{\tilde{t}}_{1i})^2 m^2_b 
   \left(k^{S^0}_{bb}\right)^2\, \right]
       m^2_b \left( m^2_b + m^2_{{\tilde \chi}^+_i} \right)\right. 
   \nonumber \\
  & & \hspace{5mm} + \left[ (l^{\tilde{t}}_{1i})^2 m^2_b 
    \left(k^{S^0}_{bb}\right)^2
                     + (k^{\tilde{t}}_{1i})^2 m^2_b 
  \left(l^{S^0}_{bb}\right)^2\, \right]
          \left( m^2_{{\tilde \chi}^+_i} - m^2_{{\tilde t}_1} \right)
          \left( m^2_{S_k^0} - m^2_b \right)  \nonumber \\
  & & \hspace{5mm} + 2 \, l^{\tilde{t}}_{1i} k^{\tilde{t}}_{1i}
             m_b m_{{\tilde \chi}^+_i}
      \left[ \left(m^2_{S_k^0}-m^2_b \right)
             \left(m^2_b \left(k^{S^0}_{bb}\right)^2 + 
     m^2_b \left(l^{S^0}_{bb}\right)^2 \right)
        - 2 \, m^2_b m^2_b \right]  \nonumber \\
  & & \hspace{5mm} + 2 \, \left( (l^{\tilde{t}}_{1i})^2 
             + (k^{\tilde{t}}_{1i})^2 \right)
       m^2_b m^2_b \left( m^2_{{\tilde t}_1}
            - m^2_{{\tilde \chi}^+_i} \right) \Bigg\}  \, , \nonumber
\end{eqnarray}
\begin{eqnarray}
d_{ik2} &=& {m^2_b}
 \left[ (l^{\tilde{t}}_{1i})^2 m^2_b 
  \left(l^{S^0}_{bb}\right)^2
           + (k^{\tilde{t}}_{1i})^2 m^2_b 
  \left(k^{S^0}_{bb}\right)^2\, \right]  \, , \nonumber
\end{eqnarray}
\begin{eqnarray}
d_{ik3} &=& - \left\{
       2  \left( (l^{\tilde{t}}_{1i})^2 + (k^{\tilde{t}}_{1i})^2 \right)
         m^2_b m^2_b 
 + 2  l^{\tilde{t}}_{1i} k^{\tilde{t}}_{1i}
              m_b m_{{\tilde \chi}^+_i}
   \left[ m^2_b \left( 2 +
  \left(k^{S^0}_{bb}\right)^2 \right)+ m^2_b 
  \left(l^{S^0}_{bb}\right)^2\, \right]\right. \nonumber \\
  & & \hspace{12mm}
 - \left.\left[ (l^{\tilde{t}}_{1i})^2 m^2_b 
   \left(k^{S^0}_{bb}\right)^2
                     + (k^{\tilde{t}}_{1i})^2 m^2_b 
   \left(l^{S^0}_{bb}\right)^2\, \right]
        \left(m^2_{S_k^0}+m^2_{{\tilde t}_1} \right)  \right\}  \,
      ,\nonumber
\end{eqnarray}
\begin{eqnarray}
d_{ik4} &=& - 
   \left[{(l^{\tilde{t}}_{1i})^2 m^2_b 
   \left(k^{S^0}_{bb}\right)^2
                    + (k^{\tilde{t}}_{1i})^2 m^2_b 
   \left(l^{S^0}_{bb}\right)^2}\,\right]\, ,
\nonumber
\end{eqnarray}
\begin{eqnarray}
e_{ikl1} &=& - { 2 \, C^{S_k^0}_{{\tilde t}_1 {\tilde t}_l}}
                \Bigg\{
       l^{\tilde{t}}_{li} k^{\tilde{t}}_{1i} m_b m_{{\tilde \chi}^+_i}
       \left[ \left( m^2_{S_k^0} - m^2_b \right) 
      l^{S^0}_{bb}
          - 2 \, m^2_b k^{S^0}_{bb} \right] \nonumber \\
  & & \hspace{21mm} + k^{\tilde{t}}_{11} k^{\tilde{t}}_{1i} m_b^2
      \left[ \left( m^2_{{\tilde t}_1} - m^2_{{\tilde \chi}^+_i} \right)
            l^{S^0}_{bb}
           - \left(m^2_b+m^2_{{\tilde \chi}^+_i} \right) 
      k^{S^0}_{bb} \right]
         \nonumber \\
  & & \hspace{21mm} - l^{\tilde{t}}_{li} l^{\tilde{t}}_{1i}
       \left[ m^2_b \left( m^2_b+m^2_{{\tilde \chi}^+_i} \right) 
     l^{S^0}_{bb}
         + m^2_b \left(m^2_{{\tilde \chi}^+_i} -m^2_{{\tilde t}_1} \right)
              k^{S^0}_{bb} \right] \nonumber \\
  & & \hspace{21mm} - k^{\tilde{t}}_{11} l^{\tilde{t}}_{1i}
              m_b m_{{\tilde \chi}^+_i}
       \left[ 2 \, m^2_bl^{S^0}_{bb}
             + \left( m^2_b-m^2_{S_k^0} \right) 
     k^{S^0}_{bb} \right] \Bigg\}\, , \nonumber
\end{eqnarray}
\begin{eqnarray}
e_{ikl2} &=& - { 2 \, C^{S_k^0}_{{\tilde t}_1 {\tilde t}_l}}
       \left(
           l^{\tilde{t}}_{li} l^{\tilde{t}}_{1i} m^2_b 
       l^{S^0}_{bb}
      + k^{\tilde{t}}_{ki} k^{\tilde{t}}_{1i} m_b^2 
       k^{S^0}_{bb} \right)\, , \nonumber
\end{eqnarray}
\begin{eqnarray}
e_{ikl3} &=& { 2 \, C^{S_k^0}_{{\tilde t}_1 {\tilde t}_l}}
    \bigg(  l^{\tilde{t}}_{li} k^{\tilde{t}}_{1i} m_b
       m_{{\tilde \chi}^+_i}l^{S^0}_{bb}
       + k^{\tilde{t}}_{ki} k^{\tilde{t}}_{1i} m_b^2 
     l^{S^0}_{bb} \nonumber \\
  & & \hspace{17mm} + \, l^{\tilde{t}}_{li} l^{\tilde{t}}_{1i} m^2_b 
      k^{S^0}_{bb}
       + k^{\tilde{t}}_{ki} l^{\tilde{t}}_{1i} m_b
              m_{{\tilde \chi}^+_i} 
         k^{S^0}_{bb} \bigg)  \, ,
\nonumber
\end{eqnarray}
\begin{eqnarray}
f_{ikl1} &=& - (C^{S_k^0}_{{\tilde t}_1 {\tilde t}_l})^2
         \left[ \left((l^{\tilde{t}}_{li})^2+(k^{\tilde{t}}_{ki})^2 \right)
                 \left( m^2_b + m^2_{{\tilde \chi}^+_i} \right)
     + 4 \, l^{\tilde{t}}_{li} k^{\tilde{t}}_{ki} m_b m_{{\tilde \chi}^+_i}
       \right]  \, ,\nonumber
\end{eqnarray}
\begin{eqnarray}
f_{ikl2} &=& (C^{S_k^0}_{{\tilde t}_1 {\tilde t}_l})^2
        \left((l^{\tilde{t}}_{li})^2+(k^{\tilde{t}}_{ki})^2 \right) \, ,
\nonumber
\end{eqnarray}
\begin{eqnarray}
f_{ik31} &=& - 2 \, C^{S_k^0}_{{\tilde t}_1 {\tilde t}_1}
                  C^{S_k^0}_{{\tilde t}_1 {\tilde t}_2}
 \left[ \left( l^{\tilde{t}}_{1i} l^{\tilde{t}}_{2i}
     + k^{\tilde{t}}_{1i} k^{\tilde{t}}_{2i} \right)
               \left( m^2_b + m^2_{{\tilde \chi}^+_i} \right)
     + 2  \left( l^{\tilde{t}}_{1i} k^{\tilde{t}}_{2i}
         + k^{\tilde{t}}_{1i} l^{\tilde{t}}_{2i} \right)
         m_b m_{{\tilde \chi}^+_i} \right]  \, ,\nonumber 
\end{eqnarray}
\begin{eqnarray}
f_{ik32} &=& 2 \, C^{S_k^0}_{{\tilde t}_1 {\tilde t}_1}
        C^{S_k^0}_{{\tilde t}_1 {\tilde t}_2}
       \left( l^{\tilde{t}}_{1i} l^{\tilde{t}}_{2i}
              + k^{\tilde{t}}_{1i} k^{\tilde{t}}_{2i} \right) \, .
\nonumber
\end{eqnarray}

\section{Couplings}
\label{sec:appB}

Here we give the couplings that were used in sec.~\ref{sec:appA}.

The ${\tilde q}_i$-$q'$-${\tilde \chi}^\pm_j$ couplings read then
\begin{equation*}
l^{\tilde q}_{ij} = {\cal{R}}^{\tilde q}_{in} {\cal{O}}^{q}_{jn}, \hspace{5mm}
k^{\tilde q}_{ij} = {\cal{R}}^{\tilde q}_{i1} {\cal{O}}^{q'}_{j2}
\end{equation*}
with
\begin{equation*}
 {\cal{O}}^{t}_j = \left( \begin{array}{c} -V_{j1}
                      \\ h_t V_{j2} \end{array} \right), \hspace{5mm}
 {\cal{O}}^{b}_j = \left( \begin{array}{c} -U_{j1}
                      \\ h_b U_{j2} \end{array} \right) .
\end{equation*}
where $U_{ij}$ and $V_{ij}$ are the mixing matrices of the
charginos~\cite{Akeroyd:1998iq}.

For the couplings $\tilde{\chi}^\pm-S^0/P^0-\tilde{\chi}^\pm$ we have
\begin{eqnarray}
  l^{S^0}_{ijk}&=&\Big[-R^{S^0}_{k3}V_{j1}U_{i3}-R^{S^0}_{k2}V_{j2}U_{i1}
  -R^{S^0}_{k1}V_{j1}U_{i2}+\hat h_\tau\big(R^{S^0}_{k3}V_{j3}U_{i2}-
  R^{S^0}_{k1}V_{j3}U_{i3}\big)\Big]
  \nonumber
  \\
  k^{S^0}_{ijk}&=&\Big[-R^{S^0}_{k3}U_{j3}V_{i1}-R^{S^0}_{k2}U_{j1}V_{i2}
  -R^{S^0}_{k1}U_{j2}V_{i1}+\hat h_\tau\big(R^{S^0}_{k3}U_{j2}V_{i3}-
  R^{S^0}_{k1}U_{j3}V_{i3}\big)\Big]
  \nonumber
  \\
  l^{P^0}_{ijk}&=&-i\Big[-R^{P^0}_{k3}V_{j1}U_{i3}-R^{P^0}_{k2}V_{j2}U_{i1}
  -R^{P^0}_{k1}V_{j1}U_{i2}-\hat h_\tau\big(R^{P^0}_{k3}V_{j3}U_{i2}-
  R^{P^0}_{k1}V_{j3}U_{i3}\big)\Big]
  \nonumber
  \\
  k^{P^0}_{ijk}&=&-i\Big[-R^{P^0}_{k3}U_{j3}V_{i1}-R^{P^0}_{k2}U_{j1}V_{i2}
  -R^{P^0}_{k1}U_{j2}V_{i1}-\hat h_\tau\big(R^{P^0}_{k3}U_{j2}V_{i3}-
  R^{P^0}_{k1}U_{j3}V_{i3}\big)\Big]
  \nonumber
\end{eqnarray}
The ${\tilde q}_i$-$q$-${\tilde \chi}^0_k$ couplings are given by
\begin{equation*}
a^{\tilde q}_{ik} = {\cal{R}}^{\tilde q}_{in} {\cal{A}}^{f}_{kn}, \hspace{5mm}
b^{\tilde q}_{ik} = {\cal{R}}^{\tilde q}_{in} {\cal{B}}^{f}_{kn}
\end{equation*}
with
\begin{equation*}
 {\cal{A}}^{f}_{k} = \left( \begin{array}{c} f^f_{Lk}
                      \\ h^f_{Rk} \end{array} \right), \hspace{5mm}
 {\cal{B}}^{f}_{k} = \left( \begin{array}{c} h^f_{Lk}
                      \\ f^f_{Rk} \end{array} \right),
\end{equation*}
and
\begin{equation*}
  \begin{array}{l}
    h^t_{Lk} =  - h_t N_{k4}  \\
    f^t_{Lk} = -\frac{\sqrt2}2
    \left[
      N_{k2} + \frac13\frac{g'}{g}N_{k1}
    \right]\\
    h^t_{Rk} =  - h_t N_{k4}  \\
    f^t_{Rk} =  \frac{-2\sqrt{2}}{3} N_{k1}
  \end{array}
\end{equation*}
\begin{equation*}
  \begin{array}{l}
    h^b_{Lk} = - h_b N_{k3} \\
    f^b_{Lk} = \frac{\sqrt{2}}{2} 
    \left(
      N_{k2} - \frac13\frac{g'}{g} N_{k2}
    \right)\\
    h^b_{Rk} = - h_b N_{k4}  \\
    f^b_{Rk} =  \frac{\sqrt{2}}{3} N_{k1}
  \end{array}
\end{equation*}
where $N_{ij}$ is the mixing matrix of the neutralinos.
The couplings $\overline{\tilde t}_i$-${\tilde b}_j$-$W^+$ read
\begin{equation*}
A^W_{{\tilde t}_i{\tilde b}_j} = (A^W_{{\tilde b}_i{\tilde t}_j})^T
= \frac1{\sqrt2 \,} \left( \begin{array}{rr}
 \cos \theta_{\tilde b} \cos \theta_{\tilde t} & - \sin \theta_{\tilde b}
            \cos \theta_{\tilde t} \\
 -\cos \theta_{\tilde b} \sin \theta_{\tilde t} & \sin \theta_{\tilde b}
               \sin \theta_{\tilde t} \end{array} \right) \, .
\end{equation*}
The couplings $\overline{\tilde t}_i$-${\tilde b}_j$-$S^+_k$ are given by
\begin{equation*} 
  \hspace*{-6mm}
  C^{S^\pm_k}_{{\tilde t}_i{\tilde b}_j} = 
  (C^{S^\pm_k}_{{\tilde b}_i{\tilde t}_j})^T 
  =  \frac1{\sqrt2 }
  {\cal{R}}^{{\tilde t}}
  \left( 
    \begin{array}{cc}
      \mathcal{A}_{11}
      &
      \mathcal{A}_{12}\\
      \mathcal{A}_{21} & \mathcal{A}_{22}
    \end{array}  
  \right)
  \left(  {\cal{R}}^{{\tilde b}} \right)^{\dagger}.
\end{equation*}
where
\begin{eqnarray}
  \mathcal{A}_{11}&=&
  v_1h_b^2R^{S^\pm}_{k1}+v_2h_t^2R^{S^\pm}_{k2}-\frac12 g^2
  \sum_{j=1}^3v_jR^{S^\pm}_{kj}
  \nonumber
  \\
  \mathcal{A}_{12}&=&
  \sqrt{2}h_b(A_b R^{S^\pm}_{k1}+\mu R^{S^\pm}_{k2}-v_3h_\tau  
  R^{S^\pm}_{k4})
  \nonumber
  \\
\mathcal{A}_{21}&=&
  \sqrt{2}h_t(A_t R^{S^\pm}_{k2}+\mu R^{S^\pm}_{k1}-\epsilon_3 R^{S^\pm}_{k3})
  \nonumber
  \\
  \mathcal{A}_{22}&=&
h_bh_t(v_2R^{S^\pm}_{k1}+v_1R^{S^\pm}_{k2})\nonumber
\end{eqnarray}
The couplings $\overline{t}$-${b}$-$S^+$ are given by
\begin{equation*}
  C^{S_k}_{{t}{b}}=\bar t(h_tR^{S^\pm}_{k2}P_L+h_b R^{S^\pm}_{k1}P_R)
\end{equation*}
The $W^+$-${\tilde \chi}^-_j$-${\tilde \chi}^0_k$ couplings read:
The $W^+$-${\tilde \chi}^-_j$-${\tilde \chi}^0_k$ couplings read:
\begin{eqnarray}
 O^L_{kj}{}' & = & -\frac{V_{j2}}{\sqrt2 \,}N_{k4} + V_{j1} N_{k2} \, ,\nonumber \\
 O^R_{kj}{}' & = & \frac{U_{j2}}{\sqrt2 \,} N_{k3} 
  + U_{j1} N_{j2} \, .\nonumber
\end{eqnarray}
The $S^+$-${\tilde \chi}^-_j$-${\tilde \chi}^0_k$ couplings are given by:
\begin{eqnarray}
  Q^L_{ijk}{}' & = &-g\Bigg[R^{S^\pm}_{k2}
  \left(
    V_{j1}N_{j4}+V_{j2}\frac{g'}{g}N_{i1}+N_{i2}
  \right)
  +\hat h_\tau
  \left(
    R^{S^\pm}_{k1}V_{j3}N_{i5}+R^{S^\pm}_{k3}V_{j3}N_{i3}
  \right)
  \nonumber
  \\
  &&+\sqrt{2}N_{i1}V_{j3}R^{S^\pm}_{k4}
  \Bigg]
\nonumber  \\  
  Q^R_{ikj}{}' & = & -g\Bigg\{ R^{S^\pm}_{k1}
  \left[
    U_{j1} N_{i3} - \frac{U_{j2}}{\sqrt2 \,} 
    \left(
      \frac{g'}{g} N_{i1}+ N_{i2}  
    \right)
  \right]
  -\hat h_\tau
  \left(
    R^{S^\pm}_{k4}U_{j2}N_{i5}+R^{S^\pm}_{k4}U_{j3} N_{i3}
  \right)
  \nonumber
  \\
  &&-\frac{1}{\sqrt{2}}R^{S^\pm}_{k3}U_{j3}
  \left(
    \frac{g'}{g}N_{i1}+N_{i2}
  \right)
 \Bigg\}\nonumber
\end{eqnarray}

\end{document}